\documentclass[11pt]{article}
\usepackage{amsfonts}
\usepackage{amssymb}
\usepackage{tikz}
\usepackage{amsmath,amssymb}
\usepackage{bbm}

\def\sfrac#1#2{{\textstyle\frac{#1}{#2}}}
\newcommand{\unit}{\mathbbm{1}}   

\newcommand{\CA}{\mathcal{A}}    
\newcommand{\CG}{\mathcal{G}} 
\newcommand{\CT}{\mathcal{T}} 
\newcommand{\CO}{\mathcal{O}}    
\newcommand{\CL}{\mathcal{L}}    
\newcommand{\CF}{\mathcal{F}}   
\newcommand{\CE}{\mathcal{E}}  

\newcommand{\CH}{\mathcal{H}}

\newcommand{\CR}{\mathcal{R}} 
\newcommand{\Z}{\mathbb{Z}}   
   
\newcommand{\R}{\mathbb{R}}     
\newcommand{\C}{\mathbb{C}}     
\newcommand{\CPP}{{\mathbb{C}P}}    
\newcommand{\Nbb}{{\mathbb{N}}}    

\newcommand{\im}{\mathrm{i}} 
 
\newcommand{\dd}{\mathrm{d}}     
\newcommand{\dpar}{\partial}     
\newcommand{\hra}{{\hookrightarrow}}     
\newcommand{\diag}{{\mathrm{diag}}}     

\newcommand{\+}{\dagger}

\newcommand{\Ad}{\mathrm{Ad}} 
\newcommand{\sU}{\mathrm{U}}     
\newcommand{\sSU}{\mathrm{SU}}     
\newcommand{\sSL}{\mathrm{SL}}     
\newcommand{\sSO}{\mathrm{SO}}     
\newcommand{\sGL}{\mathrm{GL}}  
\newcommand{\rsu}{\mathrm{su}} 
\newcommand{\ru}{\mathrm{u}}

\newcommand{\sO}{\mathrm{O}}  

\newcommand{\al}{{{\alpha}}} 
 \newcommand{\ga}{{{\gamma}}} 
\newcommand{\dal}{{{\dot\alpha}}}

\newcommand{\veps}{{\varepsilon}} 
\newcommand{\vph}{{{\varphi}}}

\newcommand{\zh}{{\hat{z}}}

\newcommand{\zb}{{\bar{z}}}
\newcommand{\zd}{{\dot{z}}}

\newcommand{\Av}{{A_{\sf{vac}}}}
\newcommand{\Lv}{{L_{\sf{v}}}}
\newcommand{\Jv}{{J_{\sf{v}}}}
\newcommand{\qv}{{q_{\sf{v}}}}
\newcommand{\Fv}{{F_{\sf{vac}}}}
\newcommand{\Qv}{{Q_{\sf{v}}}}

\newcommand{\und}{{\quad\mathrm{and}\quad}}
\newcommand{\with}{{\quad\mathrm{with}\quad}}
\newcommand{\for}{{\ \mathrm{for}\ }}

\makeatletter

\@addtoreset{equation}{section}
\makeatother

\begin{document}
\begin{titlepage}
\setcounter{page}{0}
.
\vskip 15mm
\begin{center}
{\LARGE \bf Dirac particles, spin and photons}\\
\vskip 2cm
{\Large Alexander D. Popov}
\vskip 1cm
{\em Institut f\"{u}r Theoretische Physik,
Leibniz Universit\"{a}t Hannover\\
Appelstra{\ss}e 2, 30167 Hannover, Germany}\\
{Email: alexander.popov@itp.uni-hannover.de}
\vskip 1.1cm
\end{center}
\begin{center}
{\bf Abstract}
\end{center}

We describe relativistic particles with spin as points moving in phase space 
$X=T^*{\mathbb R}^{1,3}\times{\mathbb C}^2_L\times{\mathbb C}^2_R$, where $T^*{\mathbb R}^{1,3}={\mathbb R}^{1,3}\times{\mathbb R}^{1,3}$ is the space of coordinates and momenta, and ${\mathbb C}^2_L$ and ${\mathbb C}^2_R$ are the spaces of representation of the Lorentz group of type $(\sfrac12 , 0)$ and $(0, \sfrac12)$. Passing from relativistic mechanics with a Lorentz invariant Hamiltonian function $H$ on the phase space $X$ to quantum mechanics with a Hamiltonian operator $\hat H$, we introduce two complex conjugate line bundles $L_{\mathbb C}^+$ and $L_{\mathbb C}^-$ over $X$. Quantum particles are introduced as sections $\Psi_+^{}$ of the bundle $L_{\mathbb C}^+$ holomorphic along the space ${\mathbb C}^2_L\times{\mathbb C}^2_R$, and antiparticles are sections $\Psi_-^{}$ of the bundle $L_{\mathbb C}^-$ antiholomorphic along the internal spin space ${\mathbb C}^2_L\times{\mathbb C}^2_R$. The wave functions $\Psi_\pm^{}$ are characterized by conserved charges $q^{}_{\sf{v}}=\pm 1$ associated with the structure group $\mathrm{U}(1)_{\sf{v}}^{}$ of the bundles $L_{\mathbb C}^\pm$. Wave functions $\Psi_\pm^{}$ are governed by relativistic analogue of the Schr\"odinger equation. We show how fields with spin $s=0$ (Klein-Gordon), spin $s=\sfrac12$ (Dirac) and spin $s=1$ (Proca fields) arise from these equations in the zeroth, first, and second order expansions of the functions $\Psi_\pm^{}$ in the coordinates of the spin space ${\mathbb C}^2_L\times{\mathbb C}^2_R$. The Klein-Gordon, Dirac and Proca equations for these fields follow from the Schr\"odinger equation on the extended phase space $T^*{\mathbb R}^{1,3}\times{\mathbb C}^2_L\times{\mathbb C}^2_R$. Using these results, we also introduce equations describing first quantized photons. We show that taking into account the charges $q^{}_{\sf{v}}=\pm 1$ of the fields $\Psi_\pm^{}$ changes the definitions of the inner products and currents, which eliminates negative energies and negative probabilities from relativistic quantum mechanics.

\end{titlepage}

\newpage
\setcounter{page}{1}

\tableofcontents

\newpage

\section{Introduction}

\noindent   Quantum field theory is based on the free Klein-Gordon and Dirac equations, the complex solutions of which decompose into a sum of positive frequency solutions, interpreted as particles, and negative frequency solutions associated with antiparticles. For real solutions of wave equations (for example, Maxwell's equations), antiparticle-type solutions are complex conjugate to particle-type solutions, which is interpreted as neutrality. It is believed that at the first quantized level, the negative frequency solutions of the Klein-Gordon and Dirac equations have negative probabilities and negative energies, which is corrected only when moving to operator-valued solutions at the second quantized level. Due to above difficulties, the point of view has become widespread that the equations of Klein-Gordon, Dirac, Maxwell, etc. should be considered as equations of classical fields to which the concepts of quantum mechanics should not be applied. Accordingly, the second quantization of particles should be considered as the first quantization of fields.

The above point of view is contradictory and untenable for at least two reasons. Firstly, the non-relativistic limits of the Klein-Gordon and Dirac equations coincide with the Schr\"odinger and Pauli equations, i.e. ``classical" fields are reduced to ``quantum" fields, which contradicts elementary logic. Secondly, the Klein-Gordon and Dirac equations can be derived using the same standard methods of transition from classical to quantum mechanics (see e.g. \cite{Wood1, BerMar, Bona}) as the non-relativistic Schr\"odinger equation \cite{Sour}-\cite{Wood}. In fact, Wigner's generally accepted approach to relativistic equations is nothing more than quantization of the coadjoint orbits of the Poincare group \cite{Wigner, BW}. The purpose of this paper is to analyze the reasons for the appearance of non-physical (from a quantum mechanical point of view) solutions of relativistic equations and to discuss their elimination at the level of classical mechanics, which will eliminate them at the first quantized level.

In fact, already at the level of classical mechanics one should understand what the spin of a particle is and what an antiparticle is. The situation with spin is as follows. The relativistic phase space $T^*\R^{1,3}$ of a massive particle with fixed parity extends to a phase space $T^*\R^{1,3}\times\C^2$, where $\C^2$ is the fundamental representation of the group $\sSL (2, \C)$ double covering the proper ortochronous Lorentz group $\sSO^+(1,3)$ \cite{Wood1}. Next, a formal transition is made to the orbit $T^*H_+^3\times\CPP^1$ of the Poincare group, over this orbit a complex line bundle $L_\C^+$ is introduced and the polarized sections of this bundle define fields with spin $s=0, \sfrac12, 1,...$ \cite{Wood1}. Here $H_+^3$ is one sheet of a two-sheeted hyperboloid $\eta^{\mu\nu}p_\mu p_\nu = m^2$ in the momentum space, $p_\mu\in\R^{1,3}$, $\mu, \nu , ... =0,...,3$, $(\eta_{\mu\nu})=\diag (1, -1, -1, -1)$. If the parity is not fixed, then the phase space should be extended to the space $T^*\R^{1,3}\times\C^2_L\times\C_R^2$, where $\C_L^2$ and $\C_R^2$ are the representation spaces of the Lorentz group of type $(\sfrac12, 0)$ and $(0, \sfrac12)$. This allows us to obtain a description of fields in arbitrary representations $(s, j)$ of the Lorentz group, and not only representations of type $(s, 0)$ and $(0,s)$ obtained for fixed parity. In the non-relativistic case, the phase space $\C^2_L\times\C_R^2$ is reduced to the fundamental representation $\C^2$ of the group SU(2) covering the rotation group SO(3), and the coset space $\CPP^1=\sSU(2)/\sU(1)\subset\C^2$ of this group is used to introduce the space $\C^{2s+1}$ of quantum spin. The space of quantum spin in both the relativistic and non-relativistic cases is defined as the space $\C^{2s+1}$ of sections of the bundle $\CO(2s)$ over the space $\CPP^1$.

When considering classical spin, they always talk about homogeneous space  $\CPP^1$ of the group SU(2) or SL(2, $\C$) and do not talk about the dynamics of a particle in space $\C^2$. In this paper we analyse the dynamics of a nonrelativistic particle in phase space $T^*\R^{3}\times\C^2$. We will show that a particle of spin $s$ moves in a circle in a lens space $S^3/\Z_n\subset \C^2$ with $n=2s=1,2,...$, where $\Z_n$ is the cyclic group of order $n$, generated by an element $\zeta$ with $\zeta^n=1$, i.e. $\zeta$ is the $n$-th root of unity.  The space $\CPP^1\subset S^3/\Z_n$ parametrizes $S^1$-orbits along which the particle moves, i.e. spin is indeed related to the rotation. Moreover, if we include the dynamics of the particle in the consideration, it becomes clear that spin is discrete already at the classical level. We will carry out a detailed differential-geometric analysis of the dynamics of classical spin variables and their quantization in the first part of the paper.

From the very beginning we introduce the concept of antiparticle and use it throughout the paper. We define the mapping of a particle into its antiparticle as a mapping $\tau\mapsto-\tau$ of the parameter $\tau$ on the particle's trajectory, i.e. a change in orientation on the trajectory. In the non-relativistic case, $\tau$ can be identified with time $t$. In the relativistic case, the coordinate time $x^0$ cannot be identified with the scalar $\tau$, and the mapping $\tau\mapsto -\tau$ is a charge conjugation transformation.  This mapping is antilinear, it maps complex structures on the phase manifold into the conjugate complex structures. In the quantum case, the map $\tau\mapsto-\tau$ also induces a map of any complex vector bundle to a complex conjugate vector bundle. The correlated signs $q_\pm =\pm 1$ of orientation on the particle tajectory and the signs of all the above mentioned complex structures distinguish particles ($q_+=1$)  from antiparticles ($q_-=-1$). This definition was analyzed for the non-relativistic classical and quantum oscillator in \cite{Popov1} and for the relativistic oscillator and its supersymmetric version in \cite{Popov2, Popov3}. In particular, it was shown that when taking into account the charges $q_\pm$, relativistic oscillator models are Lorentz covariant, unitary and does not contain non-physical states. In this paper we will show the absence of non-physical states for free classical and quantum relativistic particles with spin.

We introduce and study relativistic Hamiltonian mechanics in terms of Lorentz invariant functions $H$ (they are not energy) on the phase space $T^*\R^{1,3}\times\C^2_L\times\C^2_R$ that define both the dynamics of particles with spin and the space of initial data for their motion. We then introduce a relativistic analogue of the Schr\"odinger equation for evolution in $\tau$ with a Hamiltonian operator $\hat H$, using ideas from the geometric quantization approach \cite{Sour}-\cite{Wood}. This equation and wave function are defined on the space $T^*\R^{1,3}\times\C^2_L\times\C^2_R$. Expanding the wave function in terms of bosonic spin variables $z_\al\in\C^2_L$ and $y^{\dot\al}\in\C^2_R$,
we obtain fields in the representations of the Lorentz group of type $(s, j)$ for any half-integer $s$ and $j$. In particular, we obtain the Klein-Gordon equation (zero order in $z_\al$, $y^{\dot\al}$), the Dirac equation (first order), and derive the Proca equation (second order). From the consideration of relativistic classical mechanics of particles of spin $s=0, \sfrac12$ and 1, we will derive formulae for inner products and currents that take into account the charges $q_\pm =\pm 1$, which leads to the elimination of non-physical states at the first quantized level. We use all the obtained results to describe first quantized photons. In the final section, we present a concise overview of the main results of the paper.

\section{Classical mechanics: spin and antiparticles}

\subsection{Dynamical spin variables}

\noindent {\bf Lorentz group.} The Lorentz group O(1,3) with matrices $L=(L_\nu^\mu)$, $\mu , \nu =0,...,3$, has four connected components,
\begin{equation}\label{2.1}
\begin{split}
L_+^\uparrow:&\quad \det L=+1,\ {\mathrm{sign}}\, L_0^0=+1,\quad        
L_-^\uparrow:\quad \det L=-1,\ {\mathrm {sign}}\, L_0^0=+1,
\\[2pt]
L_+^\downarrow:&\quad \det L=+1,\ {\mathrm{sign}}\, L_0^0=-1,\quad        
L_-^\downarrow:\quad \det L=-1,\ {\mathrm{sign}}\, L_0^0=-1.
\end{split}
\end{equation}
The transformations $L\in L_+^\uparrow$ form a subgroup SO$^+$(1,3)$\subset$O(1,3), which is called the proper orthochronous Lorentz group. It is a connected component of the identity of the Lorentz group and is isomorphic to the group SL(2, $\C )/\Z_2$, where $\Z_2=\{1, -1\}$. The remaining components do not form subgroups in O(1,3) and are obtained from the space $L_+^\uparrow =\sSO^+(1,3)$ by the parity mapping $P: x^a\mapsto -x^a$, $a=1,2,3$ (the space $L_-^\uparrow$), by time reversal $T: x^0\mapsto -x^0$ (the space $L_+^\downarrow$) and the map $PT: x^\mu\mapsto -x^\mu$ (the space $L_-^\downarrow$).

\noindent {\bf Spinor notation.} The group SL(2, $\C$) has a representation of type $(\sfrac12 , 0)$ on the space $\C_L^2$ with coordinates $z_\al$ and representation of type $(0, \sfrac12)$  on the space $\C_R^2$ with coordinates $y^{\dot\al}$. These are the spaces of left and right spinors. Representations of type $(s,j)$ are obtained by the tensor product  of $2s$ spaces $\C_L^2$ and $2j$ spaces $\C_R^2$. In particular, the representation of type $(\sfrac12, \sfrac12)$ is given by the space $\C_L^2\otimes\C_R^2$ with elements
\begin{equation}\label{2.2}
A^{\al\dot\al}=A_\mu\sigma^{\mu\al\dot\al}\ ,
\end{equation}
where $\sigma^{\mu}=(\unit_2, \sigma^a)$, and $\sigma^a$ are Pauli matrices. Here $A_\mu$ is a complex vector from the $\C^4$-representation of the complexified Lorentz group $\sO(4,\C )$. If we impose the reality condition in the form $\C_R^2=\bar\C_L^2$, then the matrix $(A^{\al\dot\al})$ will be Hermitian and $A_\mu$ will be real, $A_\mu\in\R^{1,3}$.

\noindent {\bf Covariant phase space.} Free massive spinless particles move in phase space $T^*\R^{1,3}=\R^{1,3}\times\R^{1,3}$ of coordinates $x^\mu\in\R^{1,3}$ and momenta $p_\mu\in\R^{1,3}$ along straight lines
\begin{equation}\label{2.3}
x^\mu (\tau)=x^\mu + \frac{p^\mu}{m}\,\tau\und p_\mu (\tau)=p_\mu\ ,
\end{equation}
where $x^\mu = x^\mu(0)$ and $p_\mu = p_\mu(0)$ are the initial data. These initial data are parametrized by a manifold
\begin{equation}\label{2.4}
T^*H_+^3\cup T^*H_-^3: \quad \eta^{\mu\nu}p_\mu p_\nu -m^2=0\und p_\mu x^\mu=0\ ,
\end{equation}
where $(\eta_{\mu\nu})=\diag (1,-1,-1,-1)$ is the Minkowski metric and $H_\pm^3$ are the two sheets of the hyperboloid in momentum space with $q_\pm :=\,$sign$(p^0)=p^0/|p^0|=\pm 1$.

To introduce spin for a particle with $q_+=1$, the covariant phase space $T^*H_+^3$ is extended to a manifold
\begin{equation}\label{2.5}
T^*H_+^3\times \CPP_L^1\subset T^*\R^{1,3}\times \C_L^2\ ,
\end{equation}
where $\CPP_L^1$ is the projectivization of the space $\C_L^2$ \cite{Wood1}. The space $\CPP_L^1$ is introduced as follows. In the space $\C_L^2$, a three-sphere $S^3$ is defined by the equation
\begin{equation}\label{2.6}
N_{int}^L :=\frac{p^{\al\dot\al}}{m} z_\al\bar z_{\dot\al}=2s\ ,
\end{equation}
where $z_\al\in\C_L^2$, $\bar z_{\dot\al}\in\bar\C_L^2$, and $s\in\R^+$. This $p$-dependent 3-sphere is projected onto the 2-sphere $S^2$ by factorizing over the group U(1),
\begin{equation}\label{2.7}
\sU(1)\ni g=e^{-\im\omega\tilde\tau}:\quad S^3 \stackrel{\sU(1)}{\longrightarrow} \ S^2=\sSU(2)/\sU(1)\ ,
\end{equation}
where $\omega$ is a frequency parameter, and $\tilde\tau$ is the parameter on orbits $S^1$ of the group U(1) in the sphere $S^3\subset \C^2_L$. In fact, this is a symplectic reduction of the space $\C_L^2$ under the action of the group U(1), $\C_L^2\to\C_L^2//\sU(1)\cong\CPP_L^1$ \cite{MW,Wood1}. To introduce the spin of antiparticle with $q_-=p^0/|p^0|=-1$, the covariant phase space $T^*H_-^3\times\overline{\CPP}^1_L$ is used, where $\overline{\CPP}^1_L$ is the space with conjugate complex structure.

\noindent {\bf Dynamics in internal space.} Note that the reduction described above is a description of the space of initial data ${\CPP}^1_L$ of the motion of a harmonic oscillator with phase space $\R^4\cong\C_L^2$, evolution parameter $\tilde\tau$ and Hamiltonian $H_{int}^L=\omega N_{int}^L$ (see e.g. \cite{Hurt}). Namely, the particle moves in $\C_L^2$ along orbits $S^1\subset S^3\subset \C_L^2$ defined by the equations
\begin{equation}\label{2.8}
z_\al (\tilde\tau) = e^{-\im\omega\tilde\tau}z_\al\ ,
\end{equation}
where $z_\al :=z_\al (0)$, and ${\CPP}^1_L$ parametrizes these orbits. Recall that relativistic particle moves in space $T^*\R^{1,3}$ along a trajectory \eqref{2.3} with the evolution parameter $\tau$ and the space of initial data \eqref{2.4}. Using two different evolution parameters $\tau$ and $\tilde\tau$ for the motion in $T^*\R^{1,3}$ and $\C_L^2$ seems unnatural, so we will identify $\tau$ and $\tilde\tau$, making spin a dynamical variable.

Note that the motion \eqref{2.3} follows from the Hamiltonian function $H_0=\sfrac1m\,\eta^{\mu\nu}p_\mu p_\nu$ commuting with the function \eqref{2.6} with respect to the Poisson bracket on the phase space \eqref{2.5}. Therefore, $H_0$ and $H_{int}^L$ are constant on the trajectory independently of each other. Let us emphasize that the use of dynamical spin variables \eqref{2.8} with $\tilde\tau =\tau$ does not change the definition of quantum spin. In fact, the use of dynamical spin variables $z_\al (\tau)$ allows one to see the prototype of quantum spin at the classical level as a $\Z_n$-symmetry of motion of a particle of spin $s=\sfrac12 n$ in space $\C_L^2$, where $\Z_n$ is the cyclic group of order $n$.

Recall that spin is a non-relativistic concept. It is a half-integer number $s$ that parametrizes representation $\C^{2s+1}$ of the group SU(2). In the relativistic case, the representation of the Lorentz group for a massive particle must be decomposed into irreducible representations of its subgroup SO(3), and then a relativistic equation should be given that specify a projection onto the irreducible representation of the group SU(2)$\cong$SO(3) \cite{Wigner, BW}. This is why, we will begin analysis of the dynamics of spin variables \eqref{2.8} with $\tilde\tau=\tau$ in non-relativistic classical and quantum mechanics, where the phase space of a particle with spin is $T^*\R^3\times\C^2$, and equation \eqref{2.6} reduces to the equation
\begin{equation}\label{2.9}
\delta^{\al\dot\al}z_\al\bar z_{\dot\al}=2s\ ,
\end{equation}
which is independent of the particle momentum. Along with spin, we also introduce and analyze the concept of a non-relativistic antiparticle. Only after considering the dynamics of non-relativistic particles and antiparticles with spin and their quantization we will move on to describing the relativistic case.

\noindent {\bf Discrete transformations.} Recall that quantization of the covariant phase spaces $T^*H^3_+\times{\CPP}^1_L$ and $T^*H^3_-\times\overline{\CPP}^1_L$ yields fields in representations of the Lorentz group of types $(s,0)$ and $(0,s)$ \cite{Wood1}. We want to describe the first quantized fields in representations of the Lorentz group of types $(s,j)$ with arbitrary half-integer numbers $s$ and $j$. To do this, we extend the internal spin space to a space  $\C_L^2\times\C_R^2$ that transforms into itself under the action of the parity operator $P: \ \C_L^2\leftrightarrow\C_R^2$. The time reversal operator is given by the antilinear map $T:\ \C_L^2\times\C_R^2\to \overline{\C}^2_L\times\overline{\C}^2_R$ and finally we have the map $PT:\ \C_L^2\times\C_R^2\to \overline{\C}^2_R\times\overline{\C}^2_L$. As a result, we see that the space $\C_L^2\times\C_R^2$ is associated with particles, and the space $\overline{\C}^2_R\times\overline{\C}^2_L$ is associated with antiparticles, and the charge conjugation operator $C$ maps them to each other similarly to the operator $PT$. 

The above transformations together with the identity form a discrete group $\{1, P, T, PT\}=\sO(1,3)/\sSO^+(1,3)$, and we have
\begin{equation}\label{2.10}
\begin{split}
P: \ \C_L^2\times \C_R^2\to\C_R^2\times\C_L^2\ ,
\quad T:\ \C_L^2\times\C_R^2\to \overline{\C}^2_L\times\overline{\C}^2_R\ ,
\\[2pt]
PT:\ \C_L^2\times\C_R^2\to \overline{\C}^2_R\times\overline{\C}^2_L\ ,
\quad C:\ \C_L^2\times\C_R^2\to \overline{\C}^2_R\times\overline{\C}^2_L\ ,
\end{split}
\end{equation}
where in the last line we also wrote out the action of the charge conjugation operator $C$. Note that on Dirac spinors $\C^4=\C_L^2\oplus\C_R^2$ the map $P$ is given by multiplication by a matrix $\gamma^0$ that permutes the left and right spinors, $T$ is defined by complex conjugation and multiplication by the matrix $\ga^1\ga^3$, and $PT$ is defined by a combination of these mapping.

\subsection{Spinless particles}

\noindent{\bf Phase space.} A classical nonrelativistic spinless particle is defined as a  point $(x^a, p_a)$ in the phase space $T^*\R^3=\R^3\times\R^3$ with coordinates $x^a$ and momenta $p_a$, $a=1,2,3$. On the phase space of the particle we define a symplectic two-form
\begin{equation}\label{2.11}
\omega_{\R^6}^{}=\dd x^a\wedge\dd p_a = \omega_{a\,b+3}^{}\dd x^a\wedge\dd x^{b+3}=\dd(-p_a\dd x^a)=:\dd\theta_{\R^6}^{}
\end{equation}
with components
\begin{equation}\label{2.12}
\omega_{a\,b+3}^{}=\frac{1}{w^2}\delta_{ab}^{}=-\omega_{b+3\,a}^{},\quad 
\omega^{a\,b+3}=-{w^2}\delta^{ab}=-\omega^{b+3\,a}\ .
\end{equation}
Here $x^{a+3}:=w^2p^a=w^2\delta^{ab}p_b$, where $w\in\R^+$ is a length parameter. The symplectic 2-form \eqref{2.11} allows one to define the Poisson bracket for any smooth functions $f, h$ on $\R^6$,
\begin{equation}\label{2.13}
\{f, h\}=\omega^{a\,b+3}\dpar_a f \dpar_{b+3}h + \omega^{b+3\,a}\dpar_{b+3}f\dpar_ah\ ,
\end{equation}
where $\dpar_a=\dpar/\dpar x^a$ and $\dpar_{a+3}=\dpar/\dpar x^{a+3}$. For any smooth function $H$ (Hamiltonian a.k.a. the particle energy) on $\R^6$, a Hamiltonian vector field is introduced,
\begin{equation}\label{2.14}
V_H=\omega^{a\,b+3}\dpar_a H \dpar_{b+3} + \omega^{b+3\,a}\dpar_{b+3}H\dpar_a=
\frac{\dpar H}{\dpar p_a}\frac{\dpar}{\dpar x^a}-\frac{\dpar H}{\dpar x^a}\frac{\dpar}{\dpar p_a}\ ,
\end{equation}
so that $V_Hf=\{H,f\}$.

\noindent{\bf Dynamics.} The dynamics is defined as the motion of a particle along the integral curves of the Hamiltonian vector field $V_H$ defined by the flow equations,
\begin{equation}\label{2.15}
\dot x^a=V_H x^a=\frac{\dpar H}{\dpar p_a}\und \dot p_a=V_H p_a=-\frac{\dpar H}{\dpar x^a}\ ,
\end{equation}
where $\dot x=\dd x/\dd\tau$ and $\tau\in\R$ is evolution parameter. For example, a free spinless particle is defined by the Hamiltonian
\begin{equation}\label{2.16}
H_0=\frac{p^2}{2m}=\frac{1}{2m}\delta^{ab}p_a p_b\quad\Rightarrow\quad
V_{H_0}^{}=\frac{p^a}{m}\frac{\dpar}{\dpar x^a}=v^a\dpar_a\ .
\end{equation}
For this Hamiltonian integral curves are
\begin{equation}\label{2.17}
\dot x^a=V_{H_0}^{}x^a=v^a,\ \dot p_a=V_{H_0}^{}p_a=0\quad\Rightarrow\quad
x^a(\tau)=x^a + v^a\tau, \ p_a(\tau)=p_a\ ,
\end{equation}
where $x^a:=x^a(0)$ and $p_a:=p_a(0)$.

\noindent{\bf Angular momentum}. The angular momentum for a nonrelativistic particle is determined by three functions,
\begin{equation}\label{2.18}
L_a=\veps_{ab}^c x^b p_c\quad\Rightarrow\quad V_{L^a}^{}=\veps_{ab}^c\left (x^b\frac{\dpar}{\dpar x^c}+
p^b\frac{\dpar}{\dpar p^c}\right )\ ,
\end{equation}
and vector fields $V_{L^a}^{}$ are generators of the rotation group SO(3) acting on the phase space $\R^3\times\R^3$. It is easy to see that
\begin{equation}\label{2.19}
\{L_a, L_b\}=\veps^c_{ab}L_c\quad\Rightarrow\quad[V_{L^a}^{}, V_{L^b}^{}]=\veps^c_{ab}V_{L^c}^{}\ .
\end{equation}
Note that the functions $L_a$ commute with the Hamiltonian $H_0$,
\begin{equation}\label{2.20}
\{H_0, L_a\}=0\quad\Rightarrow\quad [V_{H_0}^{}, V_{L^a}^{}]=0\ ,
\end{equation}
and therefore for a free particle the functions $L_a$ are constant  on the particle's trajectory.

\subsection{Spinning particles}

\noindent{\bf Internal phase space.} As the phase space of a particle with spin, we consider the space $T^*\R^3\times\C^2$, where the group SU(2) of rotation of $\C^2$ double covers the group SO(3) of rotations of the space $T^*\R^3$ with generators \eqref{2.18}. On the internal space $\C^2$ of particle, we introduce a symplectic 2-form
\begin{equation}\label{2.21}
\omega_{int}=\im\delta^{\al\dot\al}\dd z_\al\wedge\dd\zb_{\dot\al}=\dd\left [\sfrac{\im}{2} \delta^{\al\dal}( z_\al\dd\zb_\dal-  \zb_\dal\dd z_\al )\right ]=:\dd\theta_{int}\ ,
\end{equation}
where $z_\al$ are complex coordinates on $\C^2\cong\R^4$ and $\zb_\dal$ are complex conjugate coordinates, $\al , \dal =0,1$. The components of the 2-form $\omega_{int}$ are
\begin{equation}\label{2.22}
\omega^{\al\dal}=\im\delta^{\al\dal}\und \omega_{\al\dal}=\im\delta_{\al\dal}
\end{equation}
Symplectic two-form on the entire phase space $T^*\R^3\times\C^2$ is given by the sum of two-forms \eqref{2.11} and \eqref{2.21}.

On the space $\C^2$ one can also define holomorphic and anti-holomorphic two-forms
\begin{equation}\label{2.23}
\omega^{}_{\C^2}=\veps^{\al\beta}\dd z_\al\wedge\dd z_\beta = -\veps_{\al\beta}\dd z^\al\wedge\dd z^\beta \und
\omega^{}_{\bar\C^2}=\veps^{\dal\dot\beta}\dd \zb_\dal\wedge\dd \zb_{\dot\beta} = -\veps_{\dal\dot\beta}\dd \zb^\dal\wedge\dd \zb^{\dot\beta}\ ,
\end{equation}
where
\begin{equation}\label{2.24}
\left(\veps^{\al\beta}\right)=\begin{pmatrix}0&-1\\1&0\end{pmatrix}=\left(\veps^{\dal\dot\beta}\right)\und
\left(\veps_{\al\beta}\right)=\begin{pmatrix}0&1\\-1&0\end{pmatrix}=\left(\veps_{\dal\dot\beta}\right)\ .
\end{equation}
These tensors $\veps_{\al\beta}$ are used to rise and lower indices, $z^\al = \veps^{\al\beta}z_\beta$ and $z_\al =\veps_{\al\beta}z^\beta$, etc. Note that the two-forms \eqref{2.23} are invariant under the action of the group 
SL(2, $\C$)$\,\supset\,$SU(2), so they are used in both non-relativistic and relativistic cases.

\noindent{\bf Complex structure.} Complex structure on the spin space $\R^4$ (and on any even-dimensional space $\R^{2n}$) is given as follows (see e.g. \cite{KobNom}). The complexification of the vector space $\R^4\to\C^4$ is considered and an operator $J: \C^4\to \C^4$ is introduced such that $J^2=-\unit_4$. In a complex basis it has the form
\begin{equation}\label{2.25}
J=\begin{pmatrix}\im\delta^\al_\beta&0\\ 0& -\im\delta^\dal_{\dot\beta}\end{pmatrix},\ 
J\left(\frac{\dpar}{\dpar z_\al}\right)=J^\al_\beta\frac{\dpar}{\dpar z_\beta}=\im\frac{\dpar}{\dpar z_\al}\und
J\left(\frac{\dpar}{\dpar \zb_\dal}\right)=J^\dal_{\dot\beta}\frac{\dpar}{\dpar \zb_{\dot\beta}}=-\im\frac{\dpar}{\dpar \zb_\dal}\ ,
\end{equation}
where $\dpar_{z_\al}^{}$ and $\dpar_{\zb_\dal}^{}$ are bases in the spaces $\C^2$ and $\bar\C^2$. As a result, spin spaces are $(\R^4, J)=\C^2\ni z_\al$ and  $(\R^4, -J)=\bar\C^2\ni \zb_\dal$. Having a symplectic and complex structure on $\R^4$, we can introduce a metric
\begin{equation}\label{2.26}
g_{int}=\delta^{\al\dal}\dd z_\al\dd\zb_\dal
\end{equation}
compatible with them. Note that the coordinates we used are dimensionless.

In the non-relativistic case, the metric \eqref{2.26} can be used to raise and lower indices along with the tensors \eqref{2.24}. The simultaneous use of these metrics allows us to define an antilinear mapping
\begin{equation}\label{2.27}
z_\al\ \mapsto\ \zb_\dal\ \mapsto\ \veps^{\dal\dot\beta}\zb_{\dot\beta}\ \mapsto\ \delta_{\al\dal}\veps^{\dal\dot\beta}\zb_{\dot\beta}=:\hat z_\al\ ,
\end{equation}
which is a smooth isomorphism of representations $\bold 2$ and $\bold{\bar 2}$ of the group SU(2), i.e. the defining representation of SU(2) is pseudo-real. However, mapping \eqref{2.27} is not a holomorphic isomorphism. In fact, the map \eqref{2.27} defines the charge conjugation operator $C: z_\al\mapsto\hat z_\al$ for two-component spinors $z_\al\in\C^2$. As in the relativistic case \eqref{2.10}, this operator defines a mapping of particles into antiparticles. 

\noindent{\bf Spin.} To define a dynamics of a particle in the internal space $\C^2$, we introduce the Hamiltonian 
\begin{equation}\label{2.28}
H_{int}=\omega N_{int}\with N_{int}=\delta^{\al\dal}z_\al\zb_{\dal}\ .
\end{equation}
The full Hamiltonian of a particle with spin, defining its motion in phase space $T^*\R^3\times\C^2$, is $H=H_0+H_{int}$. It is obvious that $H_0$ and $H_{int}$ commute with respect to the Poisson bracket on this phase space. Therefore, they are conserved along the particle's trajectory independently of each other,
\begin{equation}\label{2.29}
H_{0}=E_0\in\R^+\und  N_{int}=2s\in\R^+\ ,
\end{equation}
where $E_0$ and $2s$ are some constants.

The space $\C^2$ can be regarded as a cone over the unit 3-sphere $S^3$,
\begin{equation}\label{2.30}
\C^2\backslash\{0\}=C(S^3)
\end{equation}
with the metric
\begin{equation}\label{2.31}
g_{int}=\delta^{\al\dal}\dd z_\al\dd\zb_{\dal}=\dd\rho^2 + \rho^2\dd\Omega^2_{S^3}\ ,
\end{equation}
where $\rho^2=N_{int}$. This dimensionless scaling factor $\rho^2$ can always be eliminated by including it in $\omega$ from \eqref{2.28}. Therefore, we will set $2s=1$, that is, we will consider a unit 3-sphere. Next, we will show that when considering the dynamics of spin variables, this sphere corresponds to spin $s=\sfrac12$, and for higher spins the unit 3-sphere is replaced by the lens space $S^3/\Z_n$ with $n=2s>1$.

\noindent{\bf Group U(1).} The function $N_{int}$ defines a Hamiltonian vector field of the form
\begin{equation}\label{2.32}
V^{}_{N_{int}}=-\im \left(z_\al\frac{\dpar}{\dpar z_\al}-\zb_\dal\frac{\dpar}{\dpar \zb_\dal}\right)=
-\left(J_\al^\beta z_\beta\frac{\dpar}{\dpar z_\al}+ J_\dal^{\dot\beta}\zb_{\dot\beta}\frac{\dpar}{\dpar \zb_\dal}\right)=:-J\ .
\end{equation}
This vector field is the generator of the group U(1) acting on the level surface $S^3$:
\begin{equation}\label{2.33}
\C^2\supset S^3:\quad N_{int}=\delta^{\al\dal}z_\al\zb_\dal =1\ .
\end{equation}
The orbits of this group in $S^3$ are given by the formulae
\begin{equation}\label{2.34}
z_\al (\tau )=\exp({\omega\tau V^{}_{N_{int}}})z_\al =e^{-\im\omega\tau}z_\al\quad\Rightarrow\quad
\dot z_\al(\tau)=-\im\omega z_\al (\tau)\ ,
\end{equation}
where $z_\al =z_\al(0)$. The orbit space is parametrized by a 2-sphere $S^2\subset S^3$ defined by equivalence relations $z_\al (\tau)\sim z_\al(0)$.

\noindent{\bf Covariant phase space.} The spheres $S^3$ and $S^2$ are related by the Hopf projection \eqref{2.7}, where $S^2$ parametrizes the space of initial data of the particle motion along $S^1$ in $S^3\subset\C^2$. The above sphere $S^2$ is defined by three functions $S_a$ quadratic in the coordinates $z_\al , \zb_\dal$,
\begin{equation}\label{2.35}
S_a=z\sigma_a z^\+\for z=(z_0, z_1), \ z^\+=\begin{pmatrix}\zb_{\dot 0}\\\zb_{\dot 1}\end{pmatrix}
\quad\Rightarrow\quad \delta^{ab}S_aS_b=1\ ,
\end{equation}
where $\sigma_\al$ are the Pauli matrices. The Poisson brackets for the functions $S_a$ are of the form
\begin{equation}\label{2.36}
\{S_a, S_b\}=\veps^c_{ab}S_c
\end{equation}
and the corresponding Hamiltonian vector fields are
\begin{equation}\label{2.37}
V_{S^a}=\frac{1}{2\im}\,\sigma^\al_{a\beta}\left(z_\al\frac{\dpar}{\dpar z_\beta}+\hat z^\al \frac{\dpar}{\dpar \hat z_\beta}  \right)
\quad\Rightarrow\quad [V_{S^a}, V_{S^b}]=\veps^c_{ab}V_{S^c}\ ,
\end{equation}
where $\hat z_\al =\delta_{\al\dal}\veps^{\dal\dot\beta}\zb_{\dot\beta}$. These formulae are similar to formulae \eqref{2.19} for functions $L_a$ on $T^*\R^3$. Moreover, the functions $S_a$ commute with the function $N_{int}$ and therefore the vector fields \eqref{2.32} and \eqref{2.37} are generators of the group U(2) acting on the spinor space $\C^2\supset S^3\supset\CPP^1$ and on the complex conjugate space $\bar\C^2\supset S^3\supset\overline{\CPP}^1$.

\noindent{\bf Hidden topological variable.} In the non-relativistic case, $S_a$ are often considered as fundamental variables, the coordinates of the space $\rsu (2)^*=\R^3$, and are called the components of classical spin vector, with $\delta^{ab}S_aS_b= const$ (see e.g. \cite{Sni}). The particle is as if at rest on a 2-sphere $S^2$, and this is considered sufficient for quantization, where usually only the space of initial data is considered necessary. However, the space of initial data $S^2$ can be embedded in different 3-dimensional manifolds in which the particle will move, and information about this embedding can be obtained if the fundamental variables of the classical spin are chosen to be the coordinates $z_\al\in\C^2$, and $S_a$ are given by formulae \eqref{2.35}.

Next we will show that the possible motions of the particle are described  by embedding $S^2\cong\CPP^1$ into the lens spaces, $S^2\hra S^3/\Z_n$, which in turn can be embedded into the orbifold $\C^2/\Z_n$ and into the total space  of the holomorphic line bundle $\CO(-n)$ over $\CPP^1$. In other words, the level surface in $\C^2$ can be not only $S^3/\Z_1\equiv S^3$ but also the lens space $S^3/\Z_n$, in which the particle moves along the circle $S^1/\Z_n$. In this case, the projection
\begin{equation}\label{2.38}
S^3/\Z_n\ \stackrel{S^1/\Z_n}{\longrightarrow}\ S^2
\end{equation}
is given and the space of initial data will remain $S^2$. Thus, to specify the motion of a particle, it is necessary to specify a hidden discrete (topological) variable $s=\sfrac12 n=\sfrac12, 1, \sfrac32,...$. We will show that $s$ is the spin of the particle, which is discrete already at the classical level. The number $n$ is given by the first Chern class of the bundle $\CO(-n)$ as $n=-c_1(\CO(-n))$.

The motion of a particle along a circle $S^1/\Z_n$ in \eqref{2.38} corresponds to the replacements of orbits \eqref{2.34} with orbits
\begin{equation}\label{2.39}
z_\al (\tau , n)=e^{-\im\omega\tau/n}\,z_\al\ ,
\end{equation}
which cannot be seen by setting $\tau =0$. Note that in \eqref{2.39} $\vph =\omega\tau =0$ and $\vph =2\pi$ are considered to correspond to the same point for $\vph\in S^1/\Z_n$. Accordingly, the embedding of the lens space \eqref{2.38} in $\C^2$ in terms of coordinates \eqref{2.31} is given by the formulae
\begin{equation}\label{2.40}
z_0=\frac{\rho e^{-\im\vph/n}}{(1+z\zb)^{1/2}}\und z_1=\frac{\rho z e^{-\im\vph/n}}{(1+z\zb)^{1/2}}\ ,
\end{equation}
where $z:=z_1/z_0$ is a local coordinate on $\CPP^1$ and $\exp(-\im\vph/n):=(z_0/\zb_{\dot 0})^{1/2}$. Thus, the initial data should be taken as a pair $(z,n)\in (\CPP^1, \mathbb N)$, where $n$ is a hidden topological variable indicating the embedding $\CPP^1\hra S^3/\Z_n\subset\C^2/\Z_n$. We will discuss all this in detail below.

\subsection{Differential geometry and spin $s=1/2$}

\noindent{\bf Hopf fibration and spin.} Earlier we showed that on the trajectory of particle \eqref{2.34} the function $N_{int}$ from \eqref{2.28} is constant and defines the 3-sphere \eqref{2.33}. With this sphere we can associate matrices from the group SU(2),
\begin{equation}\label{2.41}
g=\begin{pmatrix}z_0&z_1\\-\zb_{\dot 1}&\zb_{\dot 0}\end{pmatrix}=\begin{pmatrix}z_0&z_1\\\zh_{0}&\zh_{1}\end{pmatrix}\ ,
\end{equation}
where the first row corresponds to the fundamental representation $\C^2\sim\bold 2$ and the second row corresponds to the representation $\hat\C^2\sim\bar\C^2\sim\bold{\bar 2}$ (charge conjugate representation) with coordinates $\hat z_\al$ from \eqref{2.27}. It is easy to see that the determinant of matrix \eqref{2.41} is equal to one.

The vector field $V_{N_{int}}$ from \eqref{2.32} defines the group U(1) acting on $z_\al$ according to formula \eqref{2.34} and this group is embedded in the group SU(2) as follows:
\begin{equation}\label{2.42}
\begin{pmatrix}z_0&z_1\\-\zb_{\dot 1}&\zb_{\dot 0}\end{pmatrix}
\mapsto 
\begin{pmatrix}e^{-\im\vph}z_0&e^{-\im\vph}z_1\\-e^{\im\vph}\zb_{\dot 1}&e^{\im\vph}\zb_{\dot 0}\end{pmatrix}=
\begin{pmatrix}e^{-\im\vph}&0\\0&e^{\im\vph}\end{pmatrix}
\begin{pmatrix}z_0&z_1\\-\zb_{\dot 1}&\zb_{\dot 0}\end{pmatrix}\ ,
\end{equation}
where $\vph =\omega\tau$. Thus the group U(1) preserves the level surface \eqref{2.33}, i.e. $\sU(1)\subset\sSU(2)\cong S^3$. The set of elements of the form $hg$, for $h\in\sU(1)$ and $g\in\sSU(2)$, is the right coset of the group SU(2) with respect to the subgroup U(1), denoted by U(1)$\setminus$SU(2). This is a homogeneous space with a right action of the group SU(2). Using factorization of the form \eqref{2.42} one can define a projection of $S^3$ on $S^2\cong\CPP^1$. In local coordinates \eqref{2.40} on SU(2) it has the form
\begin{equation}\label{2.43}
S^3\ni\begin{pmatrix}z_0&z_1\\-\zb_{\dot 1}&\zb_{\dot 0}\end{pmatrix}{=}
\begin{pmatrix}e^{-\im\vph}&0\\0&e^{\im\vph}\end{pmatrix}
\frac{1}{(1{+}z\zb)^{1/2}}
\begin{pmatrix}1&z\\-\zb&1\end{pmatrix}
 \longrightarrow
\frac{1}{(1{+}z\zb)^{1/2}}
\begin{pmatrix}1&z\\-\zb&1\end{pmatrix}\in\CPP^1.
\end{equation}
Quotienting of the sphere $S^3$ by the action \eqref{2.42} of the group U(1) yields the Hopf fibration
\begin{equation}\label{2.44}
\pi :\ S^3\ \stackrel{U(1)}{\longrightarrow}\ S^2\ ,
\end{equation}
such that $S^3$ is a non-trivial principal U(1)-bundle $S^3{=}P(S^2, \sU(1))$ over the 2-sphere $S^2$ \cite{Hopf, Seifert}.

The symplectic potential $\theta_{int}$ for the two-form $\omega_{int}$ from \eqref{2.21} induces in the bundle \eqref{2.44} the unique SU(2)-invariant connection $a_{-1}=\im\theta_{int}$ and curvature $f_{-1}=\im\omega_{int}$. In local coordinates they have the form
\begin{equation}\label{2.45}
a_{-1}=\frac{1}{2(1{+}z\zb)}(\zb\dd z -z\dd\zb)\ \Rightarrow\ f_{-1}=\dd a_{-1}=-\frac{\dd z\wedge\dd\zb}{(1{+}z\zb)^2}\ .
\end{equation}
With the principal U(1)-bundle \eqref{2.44} over $\CPP^1$ one can associate a holomorphic line bundle $\CO(-1)\to\CPP^1$ with the same connection and curvature \eqref{2.45} as in the bundle \eqref{2.44}. Here ``$-1$" in the indices means the first Chern number $c_1(\CO(-1))=-1$ of the bundle $\CO(-1)$. Recall that the bundle $\CO(-1)$ over $\CPP^1$ is called tautological and we consider it without zero section, setting $\CO(-1)\cong\C^2\setminus\{0\}$. To add the zero section, space $\C^2$ must be replaced by the blow-up $\tilde\C^2$ of this space at the origin $\{0\}\in\C^2$. We will not complicate the discussion by introducing blow-up spaces.

Note that the Hopf fibration \eqref{2.44} describes the Dirac monopole of charge one, and the fibration \eqref{2.38} describes the monopole of charge $n$ (see e.g. \cite{Tafel}). The connection $a_{-1}$ multiplied by $2s$, used to state that the classical spin can be any positive number, is incompatible with the geometry of the bundles \eqref{2.38} and $\CO(-n)$ in whose fibres the particle moves, unless $2s$ is an integer. In other words, when considering dynamical spin variables $z_\al (\tau)$, the spin $s$ becomes a topological discrete observable already at the classical level.

In case \eqref{2.44}, \eqref{2.45} we have $s=\sfrac12$, and spins $s>\sfrac12$ we will discuss later. Note that the bundle $\CO(-1)\cong\C^2\setminus\{0\}=(\R^4\setminus\{0\}, J)$ over $\CPP^1$ describes particles, and for antiparticles one should use the complex conjugate bundle $\bar\CO(-1)\cong\bar\C^2\setminus\{0\}=(\R^4\setminus\{0\}, -J)$  over $\overline{\CPP}^1$ with the Chern number $c_1(\bar\CO(-1))=1$. On the bundle $\bar\CO(-1)$ the connection has the opposite sign compared to formula \eqref{2.45}, $\bar a_{-1}=-a_{-1}$. Spin of the particle and antiparticle in the case under consideretion is equal to $s=\sfrac12 |c_1|=\sfrac12$. When passing to quantization, the bundles $\CO(-1)$ and $\bar\CO(-1)$ are replaced by dual bundles $\CO(1)\to\CPP^1$ and $\bar\CO(1)\to\overline{\CPP}^1$, respectively. Sections of these bundles are first-order polynomials in $z_\al$ and $\zb_{\dal}$,
\begin{equation}\label{2.46}
\psi_+=\psi_+^\al z_\al\und \psi_-=\psi_-^\dal \zb_\dal\ ,
\end{equation}
where $(\psi_+^\al)\in\C^2$ and $(\psi_-^\dal)\in\bar\C^2$ are the spaces of quantum spin of nonrelativistic particles and antiparticles.

\noindent{\bf Functions $S_a$.} The projection $\pi$ in the Hopf fibration \eqref{2.44} can be defined not only through the factorization \eqref{2.43}, but also as a mapping
\begin{equation}\label{2.47}
\pi (z_0, z_1) = (S_1+\im S_2, S_3)\in\R^3\ ,
\end{equation}
where the vectors $S_a$ were introduced in \eqref{2.35} as
\begin{equation}\label{2.48}
S_1=z_1\zb_{\dot 0}+z_0\zb_{\dot 1}\ ,\quad S_2=\im(z_1\zb_{\dot 0}-z_0\zb_{\dot 1})\und 
S_3=z_0\zb_{\dot 0}-z_1\zb_{\dot 1}\ .
\end{equation}
We discussed this spin vector $S_a$ in \eqref{2.35}-\eqref{2.37}. In the standard approach, its components are considered as fundamental observative (see e.g. \cite{Sni}). In our approach, which is the same for the relativistic \cite{Wood1, Wood} and non-relativistic cases, the fundamental observatives are $z_\al\in\C^2$, and $S_a$ are given as quadratic combinations
 \eqref{2.48} of these coordinates. This is analogous to the definition of the angular momentum $L_a$ in terms of the coordinates and momenta in \eqref{2.18}. In this case, $L_a$ commute with the Hamiltonian $H_0$ of free particles, $S_a$ commute with the spin Hamiltonian $H_{int}$, and they commute with each other. The generators of the total angular momentum are the vector fields
\begin{equation}\label{2.49}
V^{}_{I^a}=V^{}_{L^a}+V^{}_{S^a}\ ,
\end{equation}
corresponding to the functions $I_a=L_a+S_a$ on the phase space $T^*\R^3\times\C^2$. The action of the group SO(3) on this phase space of particles of spin $s=\sfrac12$ is given by the generators \eqref{2.49} of the diagonal subgroup of the group SO(3)$\times$SU(2).

\subsection{Higher spins}

\noindent {\bf Internal spaces.} The extension of the phase space of a particle by internal spaces of spin and charge degrees of freedom is standard, and these spaces may not be compact (see e.g. \cite{Wood1, Bona, Sni, Wood, Stern, Wein, Mont}).
Moreover, the introduction of dynamics for these additional internal coordinates allows, for example, to derive the Wong equations \cite{Wong} for the motion of charged particles in Yang-Mills fields  (see e.g. \cite{Stern, Wein, Mont}). However, usually when they talk about extra dimensions, it is stated that they must have very small size, otherwise they would be ``observable", but we do not see them in our world. Such statements contain a logical error.

In addition to the spin space $\C^2_{spin}$, a particle can have an internal space $\C_{em}$ of electric charge, an internal space $\C^2_{iso}$ of weak isospin, and an internal space $\C^3_{color}$ of color charges. These are all fibres of complex vector bundles over space-time, used in the standard model. We ``observe" the space $\C_{em}$ with a variety of devices associated with electromagnetism, we ``observe" spin with devices of a different type, and so on. It is a mistake to apply the logic of Kaluza-Klein type theories to {\it vector} spaces of internal degrees of freedom, which after quantization become finite-dimensional representation spaces of gauge groups.

\noindent {\bf Spin $s>\sfrac12$.} We have shown that the dynamics of a particle in space $\C^2$ with the symplectic 2-form  \eqref{2.21} and the Hamiltonian \eqref{2.28} defines the structure of the bundle
\begin{equation}\label{2.50}
\C^2\setminus \{0\}=C(S^3)\cong\CO(-1)\stackrel{\C^*}{\longrightarrow}\CPP^1\ ,
\end{equation}
where $\C^*=\sGL(1, \C)$ and spin is related to the first Chern number as $s=\sfrac12 |c_1(\CO(-1))|=\sfrac12$. The metric on the total space of the bundle \eqref{2.50} has the form \eqref{2.31}, and connection $a_{-1}$ and curvature $f_{-1}$ are given in \eqref{2.45}. The circle subbundle of the bundle \eqref{2.50} is the Hopf bundle \eqref{2.44} with the same connection and curvature with magnetic charge $n=1$. To obtain spin greater than $\sfrac12$, one should introduce the lens space \eqref{2.38} having a magnetic charge $n=2s$ (see e.g. \cite{Tafel}), and associate with this U(1)-bundle the holomorphic line bundle
\begin{equation}\label{2.51}
\CO(-n)=(\C^2\setminus \{0\})\times_{\C^*}\C_n \ni [z_0, z_1, \psi_{-n}]\sim
[\lambda z_0, \lambda z_1, \lambda^{-n}\psi_{-n}]
\end{equation}
having the first Chern number $c_1(\CO(-n))=-n\Rightarrow s=\sfrac12|c_1|=\sfrac12 n$. Here ``$\sim$" denotes the equivalence. Spin of the particle can be seen in the coefficients $n=2s$ that enter  into the connection and curvature of the bundle \eqref{2.38} and the bundle \eqref{2.51} associated with it,
\begin{equation}\label{2.52}
f_{-n} =\dd a_{-n} = n\dd a_{-1} = nf_{-1}=-n\frac{\dd z\wedge\dd\zb}{(1+z\zb)^2}\ .
\end{equation}
For antiparticles, the bundle \eqref{2.51} is replaced by the complex conjugate bundle $\bar\CO(-n)\to\overline{\CPP}^1$ with $f_n=\dd a_n =-f_{-n}=-\dd a_{-n}$.

\noindent {\bf $\Z_n$-invariance.} The lens space $S^3/\Z_n$ can be embedded in the orbifold $\C^2/\Z_n$, which is a cone over $S^3/\Z_n$. The metric on the orbifold $\C^2/\Z_n=C(S^3/\Z_n)$ embedded in $\C^2$ in coordinates \eqref{2.40} has the form
\begin{equation}\label{2.53}
\dd s^2_{\C^2/\Z_n} =\dd\rho^2 + \rho^2\dd s^2_{S^3/\Z_n}=\dd\rho^2 + \frac{\rho^2}{n^2}(\dd\vph -\im a_{-n})^2 + \frac{\rho^2\dd z\dd\zb}{(1+z\zb)^2}\ .
\end{equation}
From \eqref{2.53} we see that locally $\C^2/\Z_n$ has the form $\C/\Z_n\times\CPP^1$. Recall that the motion of a particle in 
$\C^2/\Z_n$
of the type \eqref{2.39}, \eqref{2.40} arises from the imposition of $\Z_n$-invariance condition, that is, $\vph\sim\vph + 2\pi\ell$ is equivalent to $z_\al\sim\zeta z_\al$, where $\zeta$ is an element in the group $\Z_n$ of $n$-th roots of unity, $\zeta =\exp(2\pi\im\ell/n)$, $\ell =0,...,n-1$.

In terms of the bundle $\CO(-n)$, $\Z_n$-invariance follows from the definition \eqref{2.51} of $\CO(-n)$. Namely, its total space can be considered as the set of points
\begin{equation}\label{2.54}
\left([z_0: z_1], (z_0^n, z_1^n)\right)\in\CPP^1\times\C^2\ ,
\end{equation}
where $[z_0: z_1]\in\CPP^1$ are homogeneous coordinates on $\CPP^1$ and $(z_0^n, z_1^n)\in\C^2$ is a vector representing the point in $\CPP^1$. This vector defines a straight line in $\C^2$ parametrized by the complex coordinate $\psi_{-n}$ on the fibres of the bundle $\CO(-n)$. Obviously, \eqref{2.54} is invariant under mappings $z_\al\mapsto\zeta z_\al$ for $\zeta\in\Z_n$. From formulae \eqref{2.50}-\eqref{2.54} we see that spin $s=\sfrac12 n$ is related to the topological characteristics of the bundles $S^3/\Z_n\to S^2$ and $\CO(-n)\to\CPP^1$ and not to the radii of the spheres $S^3$ and $S^2$, and spin is discrete already at the classical level, and not only after the introduction of the quantum bundle $\CO(n)\to\CPP^1$ dual to the bundle \eqref{2.51}.

Thus, the bundle \eqref{2.51} associated with the Seifert fibration \eqref{2.38} is related to classical particles of spin $s=\sfrac12 n$ and $\Z_n$-invariance of the particle motion in the space $\C^2$. Antiparticles are given by complex conjugate bundles $\bar\CO(-n)$ over $\overline{\CPP}^1$. To define quantum particles, one introduces bundles $\CO(n)$ and $\bar\CO(n)$, dual to $\CO(-n)$ and $\bar\CO(-n)$, with global sections of the form
\begin{equation}\label{2.55}
\psi_+(n)=\psi_+^{\al_1...\al_n}z_{\al_1}...z_{\al_n}\und \psi_-(n)=\psi_-^{\dal_1...\dal_n}\zb_{\dal_1}...\zb_{\dal_n}\ ,
\end{equation}
where symmetric in indices coefficients parametrize the representations of the group SU(2) on the spaces $\C^{n+1}$ and
$\bar\C^{n+1}$. The functions \eqref{2.55} are obviously invariant under transformations $z_\al\mapsto\zeta z_\al$ of the group $\Z_n$.

\noindent  {\bf Orbifold geometry}. Let us explain the concept of orbifold and $\Z_m$-invariance using the example of the cone $\C/\Z_m=C(S^1/\Z_m)$ over $S^1/\Z_m$. Consider function $\psi_m(z)=z^m$ for $z\in\C$. This function for $m\ge 2$ defines the map $\psi_m:  \C\to \C/\Z_m$, which is the branched covering of degree $m$, where $z=0$ is the branch point. 
On $\C\setminus\{0\}$ this mapping is a regular covering. The function $\psi_m(z)$ is an ordinary function and its inverse is a multivalued function $z=\psi_m^{1/m}$ corresponding to the map $\psi_m^{-1}: \C/\Z_m\to \C$, embedding $\C/\Z_m$ with coordinate $\psi_m$ in the space $\C$. The group  $\Z_m$ acts on $\R^2\cong\C$ by a rotation through the angle $2\pi/m$ about the origin and quotient is a cone with the cone  angle $2\pi/m$:
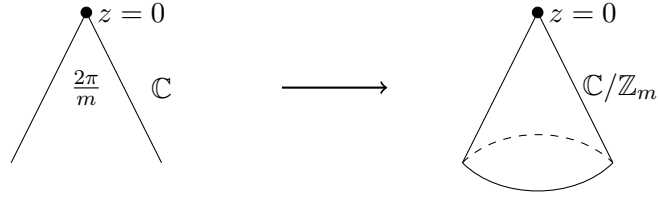
\begin{center}
\begin{tikzpicture}
\draw (0,0) -- (-1,-2);
\draw (0,0) -- (1,-2);
\filldraw[black] (0,0) circle (2pt);
\node at  (0.6,0) {$z=0$};
\node at  (0,-1) {$\frac{2\pi}{m}$};
\node at  (1,-1) {$\mathbb{C}$};
\draw[->, thick] (2.6,-1) -- (4,-1); 
\draw (6,0) -- (5,-2);
\draw (6,0) -- (7,-2);
\filldraw[black] (6,0) circle (2pt);
\node at  (6.6,0) {$z=0$};
\node at  (7.1,-1) {$\mathbb{C}/\mathbb{Z}_m$};
\draw (5,-2) .. controls (5.5,-2.5) and (6.5,-2.5) .. (7,-2);
\draw[dashed](5,-2) .. controls (5.5,-1.5) and (6.5,-1.5) .. (7,-2);
\end{tikzpicture}
\end{center}
The interior of angle $2\pi/m$ on the left is the preimage of cone $\C/\Z_m$ in $\C$. The angle $0\le\theta <2\pi/m$ corresponds to the full circle $0\le\vph <2\pi$ in  $\C/\Z_m$ and therefore $\theta =\vph/m$.

Note that on $\C{\setminus}\{0\}$ in mapping $\psi_m:\C\to\C/\Z_m$ there are $m$ different points $z_\ell=\zeta^\ell z\in\C$, $\ell =0,...,m-1$, mapped to the same point $\psi_m=z^m$ on $\C/\Z_m$, where  $\zeta =\exp (2\pi\im/m)$. The action of group $\Z_m\subset \sU(1)$ on $\C$ defines an equivalence relation and the part of the plane $\C$, which is cut out by rays with angle $\frac{2\pi}{m}$, is a representative of the cone $\C/\Z_m$.
Metric on $\C/\Z_m$ is induced from the metric on $\C$. We have
\begin{equation}\label{2.56}
\dd s^2_{\C/\Z_m} = \dd z\dd\zb|_{\C/\Z_m}^{} = \frac{1}{m^2}(\psi_m^*\psi_m)^{\frac{1-m}m}\dd\psi_m\dd\psi_m^* =\dd\rho^2 + \frac{\rho^2}{m^2}\dd\vph^2\ ,
\end{equation}
where $\psi_m=\rho^m\exp(-\im\vph)$ and $0\le\vph<2\pi$ is an angular variable on the cone. 
The  Riemann curvature of the metric \eqref{2.56} is
\begin{equation}\label{2.57}
\CR=\frac{2\pi(m-1)}{m}\,\delta (\rho)\,\dd\psi_m\wedge\dd\psi_m^*\ ,
\end{equation}
where the delta-function $\delta (\rho)$ indicates the singularity of curvature at the tip of the cone.

\noindent{\bf $\Z_{n-m}\times\Z_m$ symmetry}. So far we have considered an isotropic system, when the motion of a particle in the internal space $\C^2$ is determined by the SU(2)-invariant Hamiltonian function \eqref{2.28} and $\Z_n$-symmetry. However, we have the right to violate isotropy by putting $z_0\zb_{\dot 0}=\rho_0^2$ and  $z_1\zb_{\dot 1}=1-\rho_0^2$ instead of the equation \eqref{2.33} and imposing the symmetry $\Z_{n-m}\times\Z_m$ on the particle's motion,
\begin{equation}\label{2.58}
\Z_{n-m}\times\Z_m :\quad (z_0, z_1)\sim (\zeta_0z_0, \zeta_1z_1)\for \zeta_0^{n-m}=1, \zeta_1^m=1\ .
\end{equation}
In this case, the level surface is defined not by the lens space \eqref{2.38} but by the torus $S^1/\Z_{n-m}\times S^1/\Z_m$ embedded in the orbifold $\C/\Z_{n-m}\times \C/\Z_m$. Here we assume $\C/\Z_k=\{point\}$ if $k=0$. The particle moves along a circle in the above torus. After quantization this particle motion will correspond to the spin $s=\sfrac12n$ and the projection of the spin vector onto the 3rd axis equal to $s_3=s-m$.

\section{Quantum mechanics and differential geometry}

\subsection{Quantum mechanics as gauge theory}

\noindent{\bf Classical spin.} We introduced the phase space of a nonrelativistic particle with spin $s=\sfrac12n$ as a space $T^*\R^3\times\C^2$ with a symplectic 2-form $\omega_{\R^6}^{} + \omega_{int}^{}$ and a $\Z_n$-symmetry reducing $\C^2$ to $\C^2/\Z_n$. We introduced the Hamiltonian $H_0+H_{int}$ that defines the motion of a particle in space $\R^6\times\C^2$. The Hamiltonian $H_{int}$ coincides with the Hamiltonian of the oscillator, so a particle with spin is not free. We also introduced a charge $q_\pm^{}=\pm 1$ associated with particles and antiparticles at the classical level, relating it to the orientation $\pm\tau$ on the particle trajectories. The map $\tau\mapsto -\tau$ is antilinear, so it maps the complex structure on spin space to the conjugate: $\C^2=(\R^4, J)\to\bar\C^2=(\R^4, -J)$. The Hamiltonians $H_0$ and $H_{int}$ commute, so we get free motion of the particle in $T^*\R^3$ and rotation in spin space $\C^2$. The fundamental classical observables are the coordinates $(x^a, p_a, z_\al , \zb_\dal)$ on the phase space $\R^3\times\R^3\times\C^2$ and all other observables are functions of them.

\noindent{\bf Vacuum gauge field.} Quantization of a nonrelativistic system is a transition from a phase manifold $X$ with a symplectic 2-form $\omega_X\ (=\dd\theta_X$ locally) to a principal $\sU(1)_{\sf v}$-bundle $P(X, \sU(1)_{\sf v})$ over $X$ with a connection $\Av :=\im\theta_X$ and curvature $\Fv =\dd\Av =\im\omega_X$ \cite{Sour}. Both $\Av$ and $\Fv$ take values in the Lie algebra $\ru(1)_{\sf v}=\,$Lie$\sU(1)_{\sf v}$, where $\sU(1)_{\sf v}$ is the structure group of the above bundle. The abbreviation ``$\sf v$" and ``$\sf vac$"  here mean ``vacuum" since $\theta_X$ and $\omega_X$ (and, therefore, $\Av$ and $\Fv$) have no sources and define a symplectic structure on $X$. Thus we have a principal bundle $P(X, \sU(1)_{\sf v})$ over $X$ and a background connection on it. 

The next step is to introduce a complex line bundle $\Lv$ associated with $P(X, \sU(1)_{\sf v})$ and to impose on sections of this bundle the condition of constancy along integrable Lagrangian subbundle $\CT$ of the complexified tangent bundle $T^{\C}X$ of $X$ \cite{Sour}-\cite{Wood}.  In the symplest cases, this is the condition of independence either from momenta, or from coordinates, or holomorphicity, or antiholomorphicity of sections (wave functions). Thus quantum mechanics is a special kind of Abelian gauge theory on phase space $X$ described by the set $(\Lv, \Av, \CT)$, where the connection $\Av =\im\theta_X$ is not dynamical.

\noindent{\bf Quantum observables.} Let us emphasize that the group $ \sU(1)_{\sf v}$ is not related to the group SU(3)$\times$SU(2)$\times$U(1) of the standard model, and background connection $\Av$ is not related to the gauge potentials of the standard model. This is a new Abelian gauge field that defines the covariant derivative $\nabla_{\sf vac}$ in the bundle $\Lv$ over the phase space $X$. In fact, components of covariant derivative $\nabla_{\sf vac}$ are ``quantum" coordinates  for the phase space $X$ and their commutators define the canonical commutation relations (CCR) via the curvature $\Fv$. For example, for $X=T^*\R^3$ we have $\hat x^a=\im\nabla_{p_a}^{}$ and $\hat p_a=-\im\nabla_{x^a}^{}$.
The polarization $\CT$ defines the Hilbert space of sections of $\Lv$ on which CCRs are irreducibly realized. Any quantum Hamiltonian is given by a combination of covariant derivatives $\nabla_{\sf vac}$ (e.g. covariant Laplacian) in the bundle $\Lv\to X$.

\noindent{\bf Antiparticles.} In this paper we will consider $X=T^*\R^3\times\C^2$ and introduce $P(X, \sU(1)_{\sf v})$, $L_\C^+:=\Lv$ and $\Av$. We choose the independence of sections of the bundle $L_\C^+$ from $x^a$ and holomorphicity along $\C^2$. We accentuate the view of quantum mechanics as a gauge theory with an Abelian connection $\Av$ on $L_\C^+$. Recall that the mapping of particles into antiparticles is given by the mapping $\tau\mapsto -\tau$. This map is antilinear, so it maps the spin phase space $\C^2=(\R^4, J)$ into the space $\bar\C^2=(\R^4, -J)$. Similarly, it maps the complex line bundle $L_\C^+$ into the complex conjugate line bundle $L_\C^-:=\bar\Lv\equiv L^*_{\sf v}$. Sections of the bundle $L_\C^-$ define antiparticles. As polarized sections of this bundle we choose sections that are independent of the coordinates $x^a$ and antiholomorphic along the spin space $\C^2$. Let us emphasize that we will look at quantum particles with spin from a new angle and show that this leads to new results and new understanding.

\subsection{Quantum bundles}

\noindent{\bf Principal bundle $P(\R^4, \sU(1)_{\sf v})$}. As a first step of quantization we should introduce the principal $\sU(1)_{\sf v}$-bundle over the phase space $T^*\R^3\times\C^2$. On this space we have a symplectic 2-form
$\omega_X=\omega_{\R^6}^{}+\omega_{int}^{}=\dd\theta_{\R^6}^{}+\dd\theta_{int}^{}=\dd\theta_X$, where the symplectic potentials $\theta_{\R^6}^{}$ and $\theta_{int}^{}$ are given in \eqref{2.11} and \eqref{2.21}. Since the potentials $\theta_{\R^6}^{}$ and $\theta_{int}^{}$ are independent, the description of the bundle $P(X,  \sU(1)_{\sf v})$ reduces to the description of the bundles $P(T^*\R^3,  \sU(1)_{\sf v})$ and $P(\R^4,  \sU(1)_{\sf v})$. The description of these bundles is of educational nature, so we will illustrate it only for the internal space of spin $\R^4\cong \C^2$.

Principal bundle $P$ over $\R^4$ is a direct product,
\begin{equation}\label{3.1}
P(\R^4, \sU(1)_{\sf v})=\R^4\times S^1\cong \C^2\times S^1\with S^1\cong\sU(1)_{\sf v}\ .
\end{equation}
Let us introduce the following basis of vector fields on this five-dimensional manifold:
\begin{equation}\label{3.2}
\nabla^\al =\frac{\dpar}{\dpar z_\al}+A^\theta_{z_\al}\frac{\dpar}{\dpar\theta}\ ,\quad 
\nabla^\dal =\frac{\dpar}{\dpar \zb_\dal}+A^\theta_{\zb_\dal}\frac{\dpar}{\dpar\theta}
\und
\nabla_\theta=\dpar_\theta=\frac{\dpar}{\dpar\theta}\ .
\end{equation}
Here $A^\theta_{z_\al}$ and $A^\theta_{\zb_\dal}$ are arbitrary functions of the coordinates $z_\al , \zb_\dal$ that do not depend on $\theta$. We choose these vector fields as frame vector fields on the manifold \eqref{3.1} and introduce dual one-forms
\begin{equation}\label{3.3}
\Theta_\al=\dd z_\al\ ,\quad \bar\Theta_\dal=\dd \zb_\dal\und \Theta^\theta =\dd\theta -A^\theta\ \for A^\theta =A_{z_\al}^\theta\dd z_\al + A_{\zb_\dal}^\theta\dd \zb_\dal \ ,
\end{equation}
which are co-frame fields. Accordingly, the metric on manifold \eqref{3.1} has the form
\begin{equation}\label{3.4}
\dd s^2_P=\delta^{\al\dal}\Theta_\al\bar\Theta_\dal + \Theta^\theta\bar\Theta^\theta\ .
\end{equation}
Obviously, when $A^\theta$ is not equal to zero, this metric is not flat, that is, fibred manifolds are a special type of curved manifolds. 

\noindent{\bf Connection and curvature.} The one-form $A^\theta\dpar_\theta$ on $\R^4$ with values in the Lie algebra $\ru(1)_{\sf v}$ with generator $\dpar_\theta$ defines a connection (Abelian gauge potential) on the principal bundle  \eqref{3.1}. The deviation of this manifold from flat manifold is characterized by the curvature tensor with components
\begin{equation}\label{3.5}
F^{\al\dal}=[\nabla^\al , \nabla^\dal ]=\left (\dpar_{z_\al }A_{\zb_\dal}^\theta -  \dpar_{\zb_\dal} A_{z_\al}^\theta   \right)\dpar_\theta\ .
\end{equation}
We emphasize once again that the connection \eqref{3.2} and its curvature (field strength) have nothing to do with the Abelian  fields of standard model. The transition from the phase space $(\R^4, \omega_{int}, \theta_{int}$) of a classical particle to the bundle \eqref{3.1} over this phase space is quantization if we equate the components of the connection $A^\theta$ in \eqref{3.2}-\eqref{3.5} to the components of the potential $\theta_{int}$ defining the symplectic 2-form $\omega_{int}$,
\begin{equation}\label{3.6}
A_{z_\al}^\theta  =\theta_{z_\al}=-\sfrac{\im}{2}\delta^{\al\dal}\zb_\dal\und 
A_{\zb_\dal}^\theta  =\theta_{\zb_\dal}=\sfrac{\im}{2}\delta^{\dal\al}z_\al\ .
\end{equation}
Accordingly, the curvature components of this connection coincide with the components of the symplectic 2-forms $\omega_{int}$,
\begin{equation}\label{3.7}
F^{\al\dal}=\im\delta^{\al\dal}\Jv\quad\Rightarrow\quad 
F^{\al\dal\theta}=\im\delta^{\al\dal}=\omega_{int}^{\al\dal}\ ,
\end{equation}
where $\Jv :=\dpar_\theta$ is the generator of the group SO(2)$_{\sf v}\cong\sU(1)_{\sf v}$.

Note that the background field \eqref{3.6}, \eqref{3.7} satisfy Maxwell's equations on the spin space $\R^4\cong\C^2$ and have no sources, so we consider them to be vacuum. If we will consider the fields $A^\theta$ as dynamical and introduce sources through sections of associated bundles, then this will mean the influence of matter on the vacuum, which is in principle possible. For background gauge fields \eqref{3.6}, \eqref{3.7} we have an effect of the vacuum on the matter fields, but we have no reverse effect of matter on the vacuum.

Quantization of the space $\C^2$ consists of introducing creation and annihilation operators, which, as we will show, coincide with the covariant derivatives $\nabla^\dal$ and $\nabla^\al$, and the commutator (curvature, holonomy) of these covariant derivatives define the canonical commutation relations of quantum mechanics. It is this view of QM that is persistently advanced in this paper.

\noindent{\bf Associated bundles $L_\C^\pm$.} Passing from the phase space to the principal $\sU(1)_{\sf v}$-bundle over it, we introduce ``quantum" coordinates and ``quantum" momenta (or a complex combination of them) as covariant derivatives, but we have no wave function. The point is that wave functions must be complex and add up like vectors, which means they must be described by complex vector bundles of rank one \cite{Sour}-\cite{Wood}. To introduce such a bundle, we should embed the circle $S^1$ from \eqref{3.1} into the vector space $\R^2$, and also replace the generator $\dpar_\theta$ of SO(2)-rotations with the matrix 
\begin{equation}\label{3.8}
\Jv =\begin{pmatrix}0&-1\\1&0\end{pmatrix}\ ,\quad \Jv^2=-\unit_2\ ,
\end{equation}
because it is more convenient to deal with matrices than with vector fields. After introducing $\R^2$ and $\Jv$, we consider a chain of spaces
\begin{equation}\label{3.9}
\R^4\times S^1\hra\R^4\times\R^2\to \R^4\times\C^2=\R^4\times V^+\oplus \R^4\times V^-=:
L_\C^+\oplus L_\C^-\ ,
\end{equation}
where $L_\C^+$ and $L_\C^-$ are two conjugate complex line bundles over $\R^4$ associated with the principal bundle $P(\R^4, \sU(1)_{\sf v})$ from \eqref{3.1}-\eqref{3.7}. 

The meaning of the chain \eqref{3.9} is as follows. On the vector space $\R^2$ there acts the group SO(2)$_{\sf v}$ with generator \eqref{3.8} which has no real eigenvectors. Therefore, we consider the complexification $\C^2$ of the space $\R^2$ on which $\Jv$ has eigenvalues $\pm\im$. In the vector space $\C^2$ we introduce a basis of eigenvectors of the matrix $\Jv$,
\begin{equation}\label{3.10}
\Jv v_\pm =\pm\im v_\pm\quad\Rightarrow\quad v_\pm =\frac{1}{\sqrt 2}
\begin{pmatrix}1\\\mp\im\end{pmatrix},\quad v_-=v_+^*,\ v_\pm^\+ v_\pm =1\ ,
\end{equation}
where ``*" means complex conjugation. These vectors $v_\pm$ are basis vectors in the complex one-dimensional subspaces $V^\pm\cong\R^2$ of $\C^2\cong\R^4$, i.e. $\C^2=V^+\oplus V^- = \C\oplus\bar\C$, where bar also means complex conjugation. Any vector $\Psi$ from $\C^2$ can be expanded in $V^\pm$-parts:
\begin{equation}\label{3.11}
\Psi=\begin{pmatrix}\psi^1\\\psi^2\end{pmatrix}=\psi_+v_+ + \psi_-v_-=\Psi_+ + \Psi_-\in V^+\oplus V^-\with\psi_\pm=\frac{1}{\sqrt 2}(\psi^1\pm\im\psi^2)\ .
\end{equation}
These $\psi_\pm$ are complex coordinates on fibres $V^\pm$ of the bundles $L_\C^\pm$ in \eqref{3.9},
\begin{equation}\label{3.12}
L_\C^\pm =\R^4\times V^\pm\ .
\end{equation}
Note that $\psi^1, \psi^2$ in \eqref{3.11} are complex, therefore in general case $\psi_-$ is not complex conjugate to $\psi_+$ despite the fact that $v_-=v_+^*=\bar v_+$. Thus, with the principal bundle \eqref{3.1} there are always associated two complex line bundles,
\begin{equation}\label{3.13}
L_\C^\pm = P\times_{\sU(1)_{\sf v}}V^\pm=\left\{P\times V^\pm\ni (p, \psi_\pm)\sim (pg_\pm^{-1}, g_\pm \psi_\pm )\in P\times V^\pm\right\}\ ,
\end{equation}
where $g_\pm =\exp(\pm\im\theta )$. The sign ``$\sim$" means equivalence under the action of the group $\sU(1)_{\sf v}$ on the direct product $P\times V^\pm$ of spaces $P$ and $V^\pm$.

\noindent{\bf Quantum charge.} Historically, it so happened  that in quantum mechanics only the bundle $L_\C^+$ was implicitly introduced, the sections of which are wave functions $\psi_+$ describing particles. This is not surprising, since complex vector bundles were introduced in mathematics only in the second half of the thirties of the twentieth century, and the Chern classes distinguishing holomorphic line bundles $\CO(n)$ and $\CO(-n)$, discussed in Section 2 of this paper, were introduced \cite{Chern} in 1946. But Hermitian complex vector bundles are always given in pairs, and if $L_\C^+$ describes particles, then $L_\C^-$ must describe antiparticles. These bundles are characterized by a charge $\qv =q_\pm=\pm 1$ defined as an eigenvalue of the operator $\Qv$,
\begin{equation}\label{3.14}
\Qv:=-\im\Jv\ ,\quad \Qv v_\pm =\qv v_\pm =\pm v_\pm .
\end{equation}
The background field $\Av$ acts precisely on this charge $\qv =\pm 1$, and if the wave functions $\Psi_\pm$ did not have this charge (i.e. were real), then the covariant derivatives would be reduced to partial derivatives with zero commutators and there would be no quantization. In other words, the transition to quantum mechanics is the introduction of interaction with the vacuum field $\Av$.

Let us emphasize that the bundles $L_\C^+$ and $L_\C^-$ are not isomorphic and their sections must be summed as vectors according to formulae \eqref{3.11}. When generalizing quantum mechanics to the relativistic case, it was not taken into account that particles and antiparticles take values in different bundles $L_\C^+$ and $L_\C^-$ with basis $v_+$ and $v_-$, which led to negative energies and negative probabilities. In \cite{Popov2, Popov3}, using the example of relativistic Klein-Gordon and Dirac oscillators, it was shown that using definition \eqref{3.10}-\eqref{3.13} and taking into account the charge \eqref{3.14} of fields $\Psi_\pm$ changes the definition of inner products and currents and ultimately eliminates problems with non-physical states in relativistic quantum mechanics.

\noindent{\bf Holomorphic structures.} The covariant derivatives \eqref{3.2} on the bundle $P(\R^4, \sU(1)_{\sf v})$ with a fixed connection \eqref{3.6} after replacing $\dpar_\theta$ with $\Jv$ from \eqref{3.8} become covariant derivatives in the bundle
\begin{equation}\label{3.15}
L_{\C^2}^{}=L_\C^+\oplus L_\C^-\ .
\end{equation}
Note that $\Jv=\pm\im$ on the subbundles $L_\C^\pm$, which reduces the covariant derivatives given on $L_{\C^2}^{}$ to covariant derivatives on $L_\C^\pm$, which are of the form
\begin{equation}\label{3.16}
\begin{split}
L_\C^+ :&\quad 
\nabla_+^\al =\dpar_{z_\al}^{}+\sfrac12\delta^{\al\dal}\zb_\dal\und
\nabla_+^\dal =\dpar_{\zb_\dal}^{}-\sfrac12\delta^{\dal\al}z_\al\ ,
\\[2pt]
L_\C^- :&\quad
\nabla_-^\al =\dpar_{z_\al}^{}-\sfrac12\delta^{\al\dal}\zb_\dal\und
\nabla_-^\dal =\dpar_{\zb_\dal}^{}+\sfrac12\delta^{\dal\al}z_\al\ .
\end{split}
\end{equation}
The difference in signs in front of the connection components in  \eqref{3.16} reflects the opposite signs of the charge  \eqref{3.14} of sections of these bundles.

As already noted, for each of the bundles $L_\C^\pm$, one must specify polarizations $\CT_\pm$ that leaves sections $\Psi_\pm$ of these bundles dependent only on part of coordinates. We will use the complex Segal-Bargmann representation \cite{Segal, Bar}, in which functions $\Psi_\pm$ can be either holomorphic or antiholomorphic. The ground state (vacuum) in this representation is given by the function
\begin{equation}\label{3.17}
\psi_0=\exp(-\sfrac12|z|^2)\ \for |z|^2:=\delta^{\al\dal}z_\al\zb_\dal\ .
\end{equation}
From formulae \eqref{3.16} we see that this ground state is annihilated by the operators $\nabla_+^\al$ and $\nabla_-^\dal$ and therefore they must be chosen as the annihilation operators on the spaces of sections of the bundles $L_\C^+$ and $L_\C^-$. Accordingly, the polarized sections of these bundles have the form
\begin{equation}\label{3.18}
\Psi_+=\psi_+(z,\tau )\,\psi_0v_+\und \Psi_-=\psi_-(\zb,\tau )\,\psi_0v_-\ ,
\end{equation}
where $\psi_+$ is a holomorphic function of $z_\al\in\C^2$, and $\psi_-$ is an antiholomorphic function of these coordinates. It is easy to see that the covariant derivarives \eqref{3.16} act on sections \eqref{3.18} as follows:
\begin{equation}\label{3.19}
\begin{split}
\nabla_+^\al \Psi_+&=(\dpar_{z_\al}^{}\psi_+)\, \psi_0v_+\und
\nabla_+^\dal \Psi_+=-\delta^{\dal\al}z_\al\,\psi_+\,\psi_0v_+\ ,
\\[2pt]
\nabla_-^\al \Psi_-&=-\delta^{\al\dal}\zb_\dal\,\psi_-\,\psi_0v_-\und
\nabla_-^\dal \Psi_-=(\dpar_{\zb_\dal}^{}\psi_-)\, \psi_0v_-\ .
\end{split}
\end{equation}
These formulae show that the covariant derivatives  $\nabla_+^\dal$ and $\nabla_-^\al$ act as creation operators.

\noindent{\bf Coordinates $z_\al^\pm\in\C_\pm^2$.} It is convenient to make formulae \eqref{3.16}-\eqref{3.19} uniform by introducing the notation 
\begin{equation}\label{3.20}
z_\al^+:=z_\al\ ,\quad\zb_\dal^+=\zb_\dal\ ,\quad z_\al^-:=\zb_\dal\und\zb_\dal^-=z_\al\ .
\end{equation}
In these coordinate, formulae \eqref{3.19} have the form
\begin{equation}\label{3.21}
\nabla_{z_{\al}^{\pm}}^{}\Psi_\pm=\left(\dpar_{z_{\al}^{\pm}}^{}\psi_\pm\right) \psi_0v_\pm\und
\nabla_{\zb_{\dal}^{\pm}}^{}\Psi_\pm=-\delta^{\dal\al}z_\al^\pm\psi_\pm\,\psi_0v_\pm\ .
\end{equation}
In other words, functions $\psi_\pm$ are holomorphic functions of coordinates $z^\pm_\al$ . In this notations, the operators of annihilation and creation on the spaces of sections of bundles $L_\C^\pm$ have the form:
\begin{equation}\label{3.22}
a_\pm^\al =\nabla_{z_{\al}^{\pm}}^{}\ ,\quad a_{\pm\beta}^\+=-\delta_{\beta\dot\beta}\nabla_{\zb_{\dot\beta}^{\pm}}^{}\ \Rightarrow\ [a_\pm^\al , a_{\pm\beta}^\+]=\delta_\beta^\al\ .
\end{equation}
Thus, particles are described by holomorphic sections $\Psi_+$ of the bundle $L_\C^+$ over $\C_+^2:=\C^2\ni z_\al$, and antiparticles are described by holomorphic section $\Psi_-$ of the bundle $L_\C^-$ over $\C_-^2:=\bar\C^2$. Their inner products are given by formulae
\begin{equation}\label{3.23}
\Psi_\pm^\+\Psi_\pm =\psi_\pm^*\psi_\pm\psi_0^2 =\psi_\pm^*\psi_\pm e^{-|z|^2}\ .
\end{equation}
These functions can be integrated over the spaces $\C_\pm^2\cong\R^4$. Accordingly, one can introduce two Hilbert spaces $\CH^\pm$ of square integrable holomorphic sections of bundles $L_\C^\pm$.

\subsection{Schr\"odinger equations}

\noindent{\bf Bundles over $T^*\R^3$.} We have described in detail the introduction of ``quantum coordinates" $\hat z_\al$ and $\hat\zb_\dal$ as covariant derivatives \eqref{3.21} in bundles \eqref{3.13} over spin space $\C^2$. For the phase space $T^*\R^3$, everything is completely analogous and we will simply write out the formulae. For the covariant derivatives in the bundle $L_{\C^2}^{}$ over $T^*\R^3$ we have
\begin{equation}\label{3.24}
\nabla_{x^a}^{}=\frac{\dpar}{\dpar x^a}- p_a\Jv\ ,\quad \nabla_{p_a}^{}=\frac{\dpar}{\dpar p_a}\quad\Rightarrow\quad
\hat x^a=\Jv\nabla_{p_a}^{}\und\hat p_a=\im\nabla_{x^a}^{}\ .
\end{equation}
On the phase space $T^*\R^3$ we will use the momentum representation, that is, use functions depending on $p_a\in\R^3$. Accordingly, operators  \eqref{3.24} on such functions from $L_\C^\pm$ are reduced to expressions
\begin{equation}\label{3.25}
\nabla_{x^a}^{\pm}=\mp\im p_a\ ,\quad \nabla_{p_a}^{\pm}=\frac{\dpar}{\dpar p_a}\quad\Rightarrow\quad
\hat x^a_\pm=\pm\im\frac{\dpar}{\dpar p_a}\und 
\hat p_a^\pm =\pm p_a\ ,
\end{equation}
which are used when specifying Hamiltonians in quantum mechanics. These operators act on functions that depend on $p_a$.

\noindent{\bf Quantum Hamiltonians.} We have introduced covariant derivatives \eqref{3.21}-\eqref{3.25} in bundles $L_\C^\pm$ over the extended phase space $T^*\R^3\times\C^2$. These covariant derivatives, playing the role of ``quantum coordinates" on $T^*\R^3\times\C^2$ act on polarized sections $\Psi_\pm$ of bundles $L_\C^\pm$ of the form
\begin{equation}\label{3.26}
\Psi_\pm =\psi_\pm (p, z^\pm, \tau)\psi_0\,v_\pm^{}\quad \for\ \psi_0=e^{-\sfrac12|z|^2}\ .
\end{equation}
They belong to the Hilbert spaces of square-integrable functions on $\R^3\times\C^2\ni (p_a, z^\pm)$ with inner product
\eqref{3.23} and integration over $p_a$ and $z_\al^\pm$.

To introduce the Schr\"odinger equation for $\Psi=\Psi_+ +\Psi_-\in L_{\C^2}$, it is necessary to specify quantum Hamiltonian $\hat H=\hat H_0 +\hat H_{int}$. For function $H_0$ from \eqref{2.16} we obtain $\hat H_0=H_0$,  and for function $H_{int}$ from \eqref{2.28} we introduce $\hat H_{int}$ as covariant Laplacian on the bundle $L_{\C^2}$,
\begin{equation}\label{3.27}
\hat H_{int}=\omega\hat N_{int}\ ,\quad \hat N_{int}=-\Delta_2=-\sfrac12\delta_{\al\dal}(\nabla_{z_\al}\nabla_{\zb_\dal}+ \nabla_{\zb_\dal}\nabla_{z_\al})\ ,
\end{equation}
acting on $\Psi =\Psi_+ +\Psi_-$. For holomorphic sections \eqref{3.26} of bundles $L_\C^\pm$ we obtain operators
\begin{equation}\label{3.28}
\Delta_2^\pm=\sfrac12\delta_{\al\dal}(\nabla_{z_\al^\pm}\nabla_{\zb_\dal^\pm}+ \nabla_{\zb_\dal^\pm}\nabla_{z_\al^\pm})\ ,
\end{equation}
acting on $\Psi_\pm$ such that
\begin{equation}\label{3.29}
-\Delta_2^\pm\Psi_\pm=\left[(z_\al^\pm\frac{\dpar}{\dpar z_\al^\pm}+1)\psi_\pm\right]\psi_0\,v_\pm\ .
\end{equation}
In this case, the Schr\"odinger equation for $\Psi\in L_{\C^2}$ has the form
\begin{equation}\label{3.30}
\Jv\dpar_\tau\Psi =(\hat H_0 + \hat H_{int})\Psi\quad\Rightarrow
\quad\pm\im\dpar_\tau\Psi_\pm =(\frac{p^2}{2m} -\omega\Delta_2^\pm)\Psi_\pm\ ,
\end{equation}
where the matrix $\Jv$ plays the role of the imaginary unit.

It is easy to show that from \eqref{3.30} follow two continuity equations,
\begin{equation}\label{3.31}
\dpar_\tau\rho_\pm +\nabla_{z_\al^\pm}j^\pm_\al + \nabla_{\zb_\dal^\pm}j^\pm_\dal =0\ ,
\end{equation}
where
\begin{equation}\label{3.32}
\rho_\pm =\pm\Psi_\pm^\+\Psi_\pm\ , \quad j_\al^\pm =\im\omega\delta_{\al\dal}\left(\Psi_\pm^\+\nabla_{\zb_\dal^\pm}\Psi_\pm - (\nabla_{\zb_\dal^\pm}\Psi)^\+\Psi_\pm\right)=-(j_\dal^\pm)^\+\ .
\end{equation}
Here $\rho_\pm$ are densities of quantum charges $\qv=\pm 1$ of sections $\Psi_\pm$ of bundles $L_\C^\pm$. The probability densities for $\Psi_\pm$ are given by the inner products \eqref{3.23} and are the moduli of the charge densities $\rho_\pm$.

\noindent{\bf Solutions.} Hamiltonians $\hat H_0$ and $\hat H_{int}$ commute, so solutions can always be represented in the form
\begin{equation}\label{3.33}
\Psi_\pm =\Psi_\pm^A(p,\tau)\Phi_A^\pm (z^\pm, \tau)\for \Phi_A^\pm =\phi_A^\pm (z^\pm , \tau)\,\psi_0\, v_\pm\ ,
\end{equation}
where ``$A$" is some generalized index over which the summation is performed. After substituting \eqref{3.33} into the 
Schr\"odinger equations \eqref{3.30}, we obtain
\begin{equation}\label{3.34}
\pm\im\dpar_{\tau}\Psi_\pm^A=\hat H_0\Psi_\pm^A\quad\Rightarrow\quad
\pm\im\dpar_{\tau}\Psi_\pm^A=\frac{p^2}{2m}\Psi_\pm^A\ ,\quad\quad\quad\quad
\end{equation}
\begin{equation}\label{3.35}
\pm\im\dpar_{\tau}\Phi^\pm_A=\hat H_{int}^\pm\Phi^\pm_A\quad\Rightarrow\quad
\pm\im\dpar_{\tau}\phi^\pm_A=\omega (z^\pm_\al\dpar_{z^\pm_\al}+1)\phi^\pm_A\ ,
\end{equation}
where we used formulae \eqref{3.27}-\eqref{3.29}.

The solutions of equations \eqref{3.35} are
\begin{equation}\label{3.36}
\Phi^\pm_{\al_1...\al_n}=\ga_n z_{\al_1}^\pm(\tau)...z_{\al_n}^\pm(\tau)\,v_\pm^c(z, \zb, \tau)\ ,
\end{equation}
where
\begin{equation}\label{3.37}
z_{\al}^\pm(\tau)=e^{\mp\im\omega\tau}z_\al\und
v_\pm^c(z, \zb, \tau)=e^{-|z|^2/2}e^{\mp\im\omega\tau}v_\pm\ .
\end{equation}
We emphasize that $z_{\al}^\pm(\tau)$ are exactly the solutions of the classical system, and $v_\pm^c(\tau)$ are $\tau$-dependent bases in the fibres of the bundles $L_\C^\pm$. It is the rotation of the bases $v_\pm^c(\tau)$ gives the ``vacuum energy" independent of $n=2s=0,1,2,...$. The constant $\ga_n$ in the formula \eqref{3.36} is a normalization factor. In the eigenfunctions \eqref{3.36} we select the part
\begin{equation}\label{3.38}
Z_{\al_1...\al_n}^\pm = \ga_n z_{\al_1}^\pm...z_{\al_n}^\pm
\end{equation}
that does not depend on $\tau$, so that
\begin{equation}\label{3.39}
\Phi_{\al_1...\al_n}^\pm =e^{\mp\im\omega(n+1)\tau}\, Z_{\al_1...\al_n}^\pm \psi_0\,v_\pm\ ,\quad
E_n=\omega (n+1)\ ,
\end{equation}
where $E_n$ is the eigenvalue of the operator $\hat H_{int}$ on the eigenfunctions \eqref{3.39} and the eigenfunctions \eqref{3.38} defines the basis of the space $\C^{n+1}$ of the quantum spin $s=\sfrac12 n$.

The solutions of equations \eqref{3.34} are functions
\begin{equation}\label{3.40}
\Psi^{\al_1...\al_n}_\pm =e^{\mp\im E_{kin}\tau}\psi^{\al_1...\al_n}_\pm (p)\ ,\quad E_{kin}=\frac{p^2}{2m}\ .
\end{equation}
Finally, we obtain solutions to the Schr\"odinger equations  \eqref{3.30} of the form
\begin{equation}\label{3.41}
\Psi_\pm  (n, p, z^\pm, \tau)=e^{\mp\im (E_{kin}+E_n)\tau}
\psi^{\al_1...\al_n}_\pm (p)\, Z_{\al_1...\al_n}^\pm \psi_0 \,  v_\pm\ .
\end{equation}
Note that the energies of particles and antiparticles are positive and equal,
\begin{equation}\label{3.42}
\CE_n =E_{kin}+E_n=\frac{p^2}{2m}+\omega (n+1)\ .
\end{equation}
Formulae  \eqref{3.36}-\eqref{3.42}  suggest that spin-zero fields should be considered as the $n=0$ case of these formulae.

\noindent{\bf Lorentz forces.} Classical particles and antiparticles differ in charge $q_\pm =\pm 1$ corresponding to the orientation on their trajectories of motion. Their motion in the spin space $\C^2$ passes along circles defined by the equations
\begin{equation}\label{3.43}
\zd_\al^\pm =-\im q_\pm\omega z_\al^\pm\ .
\end{equation}
We found out that after quantization, the charge $q_\pm$ coincides with the charge $\qv$ associated with the bundles $L_\C^\pm$ and entered in the covariant derivatives \eqref{3.16}. This is not accidental, it is this charge that is responsible for the forces leading to equations \eqref{3.43} similar to the Lorentz forces for electrically charged particles.

The total spaces of the bundles \eqref{3.1} and $L_\C^\pm$ are curved manifold, as can be seen from the metric \eqref{3.4}
and the curvature \eqref{3.7}, the components of which in the bundles $L_\C^\pm$ have the form
\begin{equation}\label{3.44}
F_\pm^{\al\dal}=-q_\pm\CF^{\al\dal}=-q_\pm\delta^{\al\dal}\ .
\end{equation}
In the case of Abelian connections, the motion of charged particles along geodesics in the total spaces of bundles is described by the Lorentz force equations (see e.g. \cite{Stern, Wein, Mont}). For bundles $L_\C^\pm$, these equations have the form
 \begin{equation}\label{3.45}
\zd_\al^\pm =\im q_\pm\omega\CF^{\beta\dal}\delta_{\dal\al} z_\beta^\pm
\end{equation}
and substituting \eqref{3.44} into them, we obtain \eqref{3.43}.

The motion of particles with charges $q_\pm =\pm 1$ along circles \eqref{3.43} in $\C^2$ is similar to the motion of electrically charged particles in a constant magnetic field. In the case under consideration \eqref{3.43}-\eqref{3.45}, the constant Abelian field $\Fv =\im\omega_{int}$ is the vacuum field, which is introduced for the transition to quantum mechanics.

\subsection{Wave function collapse}

\noindent{\bf Three wave functions.} In the previous sections we described in detail the introduction of spin of particles at the classical and quantum levels. In particular, we introduced wave functions \eqref{3.41} parametrized by $n+1$ components \eqref{3.40} in basis \eqref{3.39}. In this section we use this description of classical and quantum spin to discuss the question of wave function reduction, i.e. a transition of a quantum system from a superposition of multiple states to a single definite state upon measurement. For this purpose, we focus on solutions of equations \eqref{3.35} in the space $\C^2$.

We will rewrite equations \eqref{3.35} in the form
 \begin{equation}\label{3.46}
(\hat N_{int}-1)\psi = 2s\,\psi\quad\Leftrightarrow\quad z_\al\frac{\dpar}{\dpar z_\al}\psi (z) =2s\,\psi (z)\ ,
\end{equation}
omitting sign ``$\pm$".  Eigenvalue $2s=n=0,1,...$ corresponds to $n+1$ eigenfunctions
 \begin{equation}\label{3.47}
\psi_{nm}(z)=\ga_{nm}z_0^{n-m}z_1^m\ ,\quad m=0,...,n\ ,
\end{equation}
where $\ga_{nm}$ is a normalization factor chosen so that the integral of $|\psi_{nm}|^2\exp(-|z|^2)$ over the space $\C^2$ is equal to unity. Thus, the state with spin $s=\sfrac12 n$ is degenerate and is given by the function
 \begin{equation}\label{3.48}
\psi_n(z) =\sum_{m=0}^n b_m\,\psi_{nm}(z)\ ,
\end{equation}
which can be normalized to unify similarly to functions \eqref{3.47}.

The spin operators act on functions \eqref{3.47} and \eqref{3.48} as vector fields 
 \begin{equation}\label{3.49}
\hat S_a=\sfrac12{\sigma_a}^\al_\beta\, z_\al \frac{\dpar}{\dpar z_\beta}\quad\Rightarrow\quad
\hat S_3\psi_{nm}=(s-m)\psi_{nm}\ .
\end{equation}
Therefore, function \eqref{3.48} defines the space $\C^{n+1}$ of fixed spin $s=\sfrac12 n$, and functions \eqref{3.47} define  one-dimensional subspaces in $\C^{n+1}$, where not only the spin $s$ is fixed, but also the value of the projection $s_3=s-m$ of the spin vector onto the 3rd axis. All states $\psi_n$ with $n=0,1,2,...$ are admissible and their superposition
 \begin{equation}\label{3.50}
\psi (z) =\sum_{n=0}^\infty c_n\psi_n(z)
\end{equation}
belongs to the Segal-Bargmann space of square-integrable holomorphic functions on $\C^2$.

\noindent{\bf Reductions of state vectors.}  Spin is observable usually characterized by the numbers $s=\sfrac12 n$ and $m$ from \eqref{3.49}. According to the standard Copenhagen-type interpretation, if we have a wave function \eqref{3.50}, then we do not know the value of the particle's spin and can only calculate the probability that when measured the spin will be a fixed number $s=\sfrac12 n$. In this case, according to the standard interpretation, the measurement will result in a collapse of the wave function: $\psi\to\psi_n$. On the other hand, the spin of particle $s$ is always considered fixed and the above mentioned collapse of the wave function is not discussed. Let me emphasize that what is being discussed here is the mathematical concept of spin, and not the spin of a particular elementary particle.

Note that the spin operators \eqref{3.49} are defined on the whole Hilbert space of functions of the form \eqref{3.50}, and they are irreducible on the subspaces of functions \eqref{3.48} for any $n\in\Nbb$. It turns out that if the function $\psi_n(z)$ in \eqref{3.48} is interpreted as an eigenfunction of an isotropic oscillator for the eigenvalue $E_n=\hbar\omega (n+1)$, then the values $|c_n|^2$ from the wave function \eqref{3.50} set the probability of detecting energy $E_n$ during the measurement and there is a collapse $\psi\to\psi_n$ of the wave function. On the other hand, if in the same mathematical model the function $\psi_n$ is interpreted as the wave function of a quantum particle of spin $s=\sfrac12 n$, then there is no collapse and the number $s$ is initially fixed, which looks contradictory.

Let us now assume that we have somehow fixed the spin $s=\sfrac12 n$ and are considering the space $\C^{n+1}$ of functions of the form \eqref{3.48}. It is claimed, and generally accepted, that the value of $s_3$ in the projection of the spin vector onto the 3rd axis is intrinsically indeterministic, with probability $|b_m|^2$ of finding $s_3=s-m$ upon measurement. Accordingly, upon measurement, a collapse of the wave function $\psi_n\to\psi_{nm}$ occurs.

\noindent{\bf Incompleteness of information.} Note that the wave function \eqref{3.47} is invariant under the transformations of the group $\Z_{n-m}\times\Z_m$ given in \eqref{2.58},
\begin{equation}\label{3.51}
\psi_{nm} (\zeta_0z_0, \zeta_1z_1) =\psi_{nm} (z_0, z_1) \for \zeta_0\in\Z_{n-m}\und\zeta_1\in\Z_m\ .
\end{equation}
This means that the function $\psi_{nm}(z)$ is reduced from the space $\C^2$ to the orbifold $\C/\Z_{n-m}\times\C/\Z_m\subset\C^2$ discussed at the end of Section 2.  It is in this orbifold that a particle with fixed numbers $(s, s_3)$ moves.
Note also that all functions \eqref{3.47} are invariant under the action of the group $\Z_n$ for any $m=0,1,...,n$. Therefore, their superposition \eqref{3.48} is also invariant with respect to the action of the group $\Z_n$, 
\begin{equation}\label{3.52}
\psi_{n} (\zeta z_0, \zeta z_1) =\psi_{n} (z_0, z_1) \for \zeta\in\Z_{n}\ ,
\end{equation}
and consequently the function $\psi_n(z)$ is defined on the space $\C^2/\Z_n\subset\C^2$.

When we consider solutions of evolutionary differential equations, we must fix not only initial data but also the geometry of the manifold on which the equations are given. In the case under consideration, this is equivalent to fixing the symmetry $\Z_{n-m}\times\Z_m$ or $\Z_n$ that the solution must satisfy. This can also be formally attributed to the choice of initial data. Comparing the description of classical spin in Section 2 and quantum spin  in Section 3, we can make the following statements.

\begin{itemize}
\item
A classical particle moving in phase space  $\C/\Z_{n-m}\times \C/\Z_m$ has fixed parameters $(s, s_3)=(\sfrac12 n, s-m)$. Quantization of this phase space defines a function $\psi_{nm}(z)$ on it and this is a quantum particle with fixed parameters $(s, s_3)$ of spin and its projection.
\item
A classical particle moving in phase space  $\C^2/\Z_{n}$ has a fixed spin $s=\sfrac12 n$ and an unfixed parameter $s_3$. Quantization of this phase space defines a function $\psi_{n}(z)$ on it and this is a quantum particle with a fixed spin $s$  and an unfixed projection $s_3$ of the spin.
\item
A classical particle moving in phase space  $\C^2$ has an unfixed spin $s$ and an unfixed parameter $s_3$. Quantization of this phase space defines a function $\psi(z)$ on it and this is a quantum particle with an unfixed spin $s$  and unfixed $s_3$.
\end{itemize}
Thus, the probability of detecting certain parameters $(s, s_3)$ in an experiment arises in the case where we do not know completely the initial conditions of the particle's motion. This is equivalent to how probabilities arise when throwing dice.

\subsection{From  Schr\"odinger to Pauli equation}

\noindent{\bf Pauli equation.} It is generally accepted that the Pauli equation generalizes the Schr\"odinger equation to particles with spin $s=\sfrac12$. The spin itself is considered as an exclusively quantum property of particles. In Section 2 we argue the opposite, introducing clasical spin as the genuine rotation in the internal space $\C^2$ of particles. We also argue that this internal angular momentum is related to the $\Z_n$-symmetry of the particle's motion in space $\C^2$.

In this section we considered the quantization of the extended phase space $T^*\R^3\times\C^2$, that is, the transition to quantum particles with arbitrary spin. To do this, we introduced line bundles $L_\C^\pm$ over the space $T^*\R^3\times\C^2$ and introduced the Schr\"odinger equations on sections $\Psi_\pm$ of these bundles. For free particles we obtained equations  \eqref{3.34}, \eqref{3.35} for non-relativistic fields of any spin and wrote out their solutions  \eqref{3.36}-\eqref{3.42}.

Of greatest interest are two-component spinors of spin $s=\sfrac12$. The equations for them are obtained from fields $\Psi_\pm$ of first order in $z_\al$,
\begin{equation}\label{3.53}
\Psi_+=\psi_+^\al z^+_\al v_+\und \Psi_-=\psi_-^\al z_\al^- v_-\ ,
\end{equation}
where we use holomorphic bases in the bundles $L_\C^\pm$. The action of vector fields \eqref{3.49}, which are quantum spin operators, is defined on functions \eqref{3.53}. Using this action, one can go from the vector fields  \eqref{3.49} to the matrix operators $\hat S_a=\sfrac12\sigma_a$ according to the obvious rule,
\begin{equation}\label{3.54}
\hat S_a(\psi_+^\al z_\al )=(\sfrac12 {\sigma_a}^\al_\beta\psi_+^\beta)z_\al
\quad\Rightarrow\quad
(\hat S_a\psi_+)^\al =\sfrac12 {\sigma_a}^\al_\beta\psi_+^\beta\ ,
\end{equation}
where matrices $\hat S_a$ act on columns
\begin{equation}\label{3.55}
\psi_+ =\begin{pmatrix}\psi_+^0\\\psi_+^1\end{pmatrix}\in\C^2\und
\psi_- =\begin{pmatrix}\psi_-^0\\\psi_-^1\end{pmatrix}\in\bar\C^2\ .
\end{equation}
From formulae \eqref{3.34}-\eqref{3.42} we obtain that these spinors satisfy the equations
\begin{equation}\label{3.56}
\left(E_{kin} -\frac{p^2}{2m}\right)\Phi_\pm =0\ ,\quad \Phi_\pm =\psi_\pm(p)\,v_\pm\ ,
\end{equation}
where the columns of $\Phi_\pm$ contain the basis vectors $v_\pm$ in the fibres of the bundle $L_\C^\pm$. The transition to matrices and columns is carried out by integrating over the internal space $\C^2$, after which we obtain, for example, expressions
\begin{equation}\label{3.57}
\langle\Psi_\pm , \Psi_\pm\rangle =\Phi_\pm^\+\Phi_\pm =\delta_{\al\dal} \psi_\pm^\al\bar\psi_\pm^\dal\und
\rho_\pm=\pm\Phi_\pm^\+\Phi_\pm 
\end{equation}
for the inner products of spinors and for the densities $\rho_\pm$ of quantum charges of these fields.

It is generally believed that in the 3-dimensional space $\R^3$ of non-relativistic quantum mechanics one two-component spinor is sufficient. This is not entirely true. The point is that the Pauli matrices $\sigma_a$ are generators of the Clifford algebra Cl(3,0) of the space $\R^3$ only over the field of real numbers, Cl(3,0)$\cong$ Mat(2, $\C$). 
This means that matrix $\sigma_1\sigma_2=\im\sigma_3$ is considered to be different from matrix $\sigma_3$, i.e. matrices and spinors can only be multiplied by real numbers. Therefore, it is more correct to consider the complexified Clifford algebra Cl$^\C$(3)$\cong$Mat(2, $\C)\oplus$Mat(2, $\C$) with generators
\begin{equation}\label{3.58}
\tilde\ga^a=\begin{pmatrix}\sigma^a&0\\0&-\sigma^a\end{pmatrix}\quad\Rightarrow\quad
\tilde\ga^1\tilde\ga^2=\begin{pmatrix}\im\sigma^3&0\\0&\im\sigma^3\end{pmatrix}\ne\im\tilde\ga^3
\end{equation}
acting on the columns $\Phi =\Phi_++\Phi_-$ introduced in formulae \eqref{3.53}-\eqref{3.56}.

\noindent{\bf From Dirac to Pauli equations.} The Pauli equation for $\psi_+$ from \eqref{3.55}-\eqref{3.57} is usually considered as a non-relativistic limit of the Dirac equation for positive frequency bispinors $\Psi_+$. The Dirac equation for negative frequency bispinors $\Psi_-$ is considered unphysical and its reduction to the Pauli and Schr\"odinger equations is not considered. We will show that the Dirac equation for both $\Psi_+$ and $\Psi_-$ reduces to the Pauli equations \eqref{3.56} if we take into account the charges $\qv =\pm 1$ of the spinors $\Psi_\pm$.

Let us take the standard Dirac $\ga$-matrices, $\ga^0=\sigma^3\otimes\unit_2$ and $\ga^a=\im\sigma^2\otimes\sigma^a$, acting on bispinors $\Psi_+$. In the momentum representation, the Dirac equations for them have the form
\begin{equation}\label{3.59}
\begin{pmatrix}q_\pm\tilde E-m&p_a\sigma^a\\-p_a\sigma^a&-(q_\pm\tilde E+m)\end{pmatrix}
\begin{pmatrix}\phi_\pm\\\chi_\pm\end{pmatrix}=0\ ,\quad
\Psi_\pm =\begin{pmatrix}\phi_\pm\\\chi_\pm\end{pmatrix}\in\C^4\ ,
\end{equation}
where $p^0=q_\pm\tilde E$ for $q_\pm =p^0/|p^0|=\pm 1$ and $\tilde E=|p^0|>0$. In the non-relativistic limit we have the following reductions of the energy-momentum relation:
\begin{align}
\label{3.60}
p_0^2&=p^2 + m^2\quad\Rightarrow\quad (p^0-m)(p^0+m)=p^2<<m^2\ ,\\
\label{3.61}
E:&= p_0 - m\ ,\quad p^0 + m\cong 2m \for p^0>0\quad\Rightarrow\quad E=\frac{p^2}{2m}\ ,\\
\label{3.62}
E:&= -p_0 - m\ ,\quad -p^0 + m\cong 2m \for p^0<0\quad\Rightarrow\quad E=\frac{p^2}{2m}\ .
\end{align}
In case \eqref{3.61} with $\qv =q_+=1$ we see that $\chi_+\in\C^2$ is very small compared to $\phi_+$ and equation \eqref{3.59} reduces to the equations
\begin{equation}\label{3.63}
\left(E-\frac{p^2}{2m}\right)\phi_+(p)=0\und\chi_+=\frac{p_a\sigma^a}{2m}\phi_+\ .
\end{equation}
This reduction is well known, in contrast to the case \eqref{3.62} for which \eqref{3.59} reduces to the conjugate version of Pauli equation with $\tau\mapsto -\tau$. From \eqref{3.62} with $\qv =q_-=-1$ and the Dirac equation \eqref{3.59} for $\Psi_-$ we see that $\phi_-\in\C^2$ is very small compared to $\chi_-$ so that \eqref{3.59} reduces to the equations
\begin{equation}\label{3.64}
\left(E-\frac{p^2}{2m}\right)\chi_-=0\und\phi_-=\frac{p_a\sigma^a}{2m}\chi_-\ .
\end{equation}
Thus, positive frequency Dirac spinors $\Psi_+$ are reduced to two-component spinors $\psi_+$ from  \eqref{3.53}-\eqref{3.56}, and negative frequency Dirac spinors $\Psi_-$ are reduced to two-component spinors $\psi_-$ from  \eqref{3.53}-\eqref{3.56}.

\noindent{\bf Inner products.} Note that nonrelativistic spinors  \eqref{3.53}, \eqref{3.56} contain basis vectors $v_\pm$ of the bundles $L_\C^\pm$. This reflects the fact that they are two-component spinors with values in different complex line subbundles $L_\C^\pm$ of the bundle $L_{\C^2}=L_\C^+\oplus L_\C^-$ and therefore add up according to formulae \eqref{3.11} as spinors with values in $\C^2$. Particles and antiparticles take values in conjugate complex line bundles $L_\C^+$ and $L_\C^-$. Therefore, when moving from the non-relativistic Pauli equation to the relativistic Dirac equation, one should not only replace the 2-component spinors $\psi_\pm$ from \eqref{3.53}-\eqref{3.56} with the 4-component Dirac spinors $\Psi_\pm$, but also pair them with bases $v_\pm$ of complex line bundles $L_\C^\pm$, obtaining spinor field with values in $L_{\C^2}=L_\C^+\oplus L_\C^-$.

The density $\rho_\pm$ of quantum charge $q_\pm$ of Dirac spinors in the non-relativistic limit must go over to expression \eqref{3.57}, which leads to the definition 
\begin{equation}\label{3.65}
\rho_\pm = q_\pm\bar\Psi_\pm\ga^0\Psi_\pm =q_\pm\Psi_\pm^\+\Psi_\pm\with \bar\Psi_\pm =\Psi_\pm\ga^0\ ,
\end{equation}
where the negative sign of $\rho_-$ is given by the charge $q_-$. This leads to a modification of the definition of the inner product:
\begin{equation}\label{3.66}
\begin{split}
&\bar\Psi^\qv :=\Psi^\+\ga^0\Qv=(\Psi_+^\+v_+^\++\Psi_-^\+v_-^\+)\ga^0\Qv= \bar\Psi_+v_+^\+-\bar\Psi_-^\+v_-^\+\ ,
\\[2pt]
\Rightarrow\ &\bar\Psi^\qv \Psi =(\bar\Psi_+v_+^\+-\bar\Psi_-v_-^\+)(\Psi_+v_++\Psi_-v_-)=\bar\Psi_+\Psi_+-\bar\Psi_-\Psi_-\ .
\end{split}
\end{equation}
From \eqref{3.66} it follows that $\bar\Psi^\qv\Psi$ is positive-definite and Lorentz invariant (it is a scalar that does not change sign under $PT$-transformation due to $C$-transformation), and the quantum charge density has the form
\begin{equation}\label{3.67}
\rho=\bar\Psi^\qv\ga^0\Psi =\Psi^\+\Qv\Psi = \Psi_+^\+\Psi_+- \Psi_-^\+\Psi_-=\rho_++\rho_-\ ,
\end{equation}
that is, it is positive for particle $\Psi_+$ and negative for antiparticle $\Psi_-$, and the charge conjugate operator $C$ changes these signs to opposite.

Recall that fields $\Psi_+(p)$ and $\Psi_-(p)$ are defined on different sheets $H_+^3$ and $H_-^3$ of a two-sheeted hyperboloid in momentum space. The line bundles $L_\C^+$ and $L_\C^-$ are defined on $T^*H_+^3$ and $T^*H_-^3$, respectively, and their sections have quantum charges $q_+=1$ and $q_-=-1$. Taking these charges into account requires changes \eqref{3.65}-\eqref{3.67} to the definitions of the inner products and current for Dirac spinors, which leads to elimination of negative energies and probabilities from the first quantized theory. Note that the positive definite inner product \eqref{3.66} was introduced by Woodhouse \cite{Wood2} by changing the sign of the complex structure on the negative frequency spinors. This is equivalent to introducing the matrix $\Qv$ into definition \eqref{3.66}. In the next section we introduce classical Dirac particles and show that formulae \eqref{3.65}-\eqref{3.67} arise as a consequence of the definition of non-negative energies for antiparticles at classical level.

\section{Relativistic Hamiltonian mechanics}

\subsection{Klein-Gordon particles}

\noindent {\bf Preliminary remarks.} A free classical nonrelativistic spinless particle with phase space $T^*\R^3$ was described in Sect.2.2. It is given by the Hamiltonian \eqref{2.16} identified with  the particle energy, and both the Hamiltonian and the evolution parameter $\tau$ are scalars from the point of view of the transformation groups of the phase space $T^*\R^3$. Note that not only for the Hamiltonian $H_0$ of a free particle, but also for any Hamiltonian $H(x^a, p_b)$, the space of initial data ($\equiv$ covariant phase space) determining the trajectory of motion coincides with the space $T^*\R^3$. There are no restrictions on the initial data of spinless particles in nonrelativistic mechanics. In other words, the constancy of energy does not reduce the phase space $T^*\R^3$ to a submanifold, since the numerical value of energy is arbitrary.

In this paper we develop the approach \cite{Popov2, Popov3} to relativistic mechanics with a Hamiltonian function $H(x^\mu, p_\nu)$ on the phase space $T^*\R^{1,3}$, where $\R^{1,3}$ is Minkowski space. This function $H$ must be invariant under the Lorentz group transformations and hence it is not the energy of the particle. In the relativistic case, the energy $E$ is proportional to the component $T^{00}$ of the stress-energy tensor, that is, it is not a scalar. The scalar function $H$ fixes not the energy, but the mass of the particle and for free particle specifies the energy-momentum relation in the form $H_0:=\sfrac1m\eta^{\mu\nu}p_\mu p_\nu =m$. The evolution parameter is not the coordinate time $x^0$, but the parameter $\tau$ on the particle trajectory, which is a scalar for the Lorentz group. As a result, we have standard Hamiltonian mechanics, but on a non-Euclidean phase space $T^*\R^{1,3}\cong\R^{1,3}\times\R^{1,3}\cong\R^{2,6}$. Any Lorentz invariant Hamiltonian function $H$ defines a 7-dimensional hypersurface in $\R^{2,6}$, and the dynamics are parametrized by a 6-dimensional symplectic submanifold in this 7-dimensional hypersurface.

In papers \cite{Popov2, Popov3}  effectiveness of the approach described above was illustrated using the example of a relativistic oscillator and its supersymmetric version. Here we will apply this approach to particles with spin. Particles with spin and additional charges (electric, isotopic, color, etc.) are defined by extended phase manifolds. In this case, the constant values of the functions $H(x,p,$ extra coordinates) fix not energy, but such observables as mass, spin, and various charge characteristics of particles. Accordingly, these observables enters in the dependence of functions on the evolution parameter $\tau$ of the form $\exp(\mp\im m\tau)$, etc. The operator $\im\dpar/\dpar\tau$ defines a natural polarization of the space of functions depending on $\tau$, i.e. a splitting into positive and negative eigenspaces of this operator. In other words, positive frequency (particles) and negative frequency (antiparticles) should be defined relative to $\im\dpar/\dpar\tau$, and not relative to $\im\dpar/\dpar x^0$.

\noindent {\bf Relativistic Hamiltonian equations.}  The symplectic structure on $T^*\R^{1,3}$ is
\begin{equation}\label{4.1}
\Omega_0=\dd p_\mu\wedge\dd x^\mu =\Omega_{\mu\,\nu+4}\dd x^\mu\wedge\dd x^{\nu +4}\ ,
\end{equation}
where 
\begin{equation}\label{4.2}
x^{\mu +4}:=-w^2p^\mu = -w^2\eta^{\mu \nu}p_\nu\ ,
\end{equation}
\begin{equation}\label{4.3}
\Omega_{\mu\,\nu+4}=\frac{1}{w^2}\eta_{\mu\nu}=-\Omega_{\nu+4\, \mu}\ ,\quad
\Omega^{\mu\,\nu+4}=-{w^2}\eta^{\mu\nu}=-\Omega^{\nu+4\, \mu}\ ,
\end{equation}
where the parameter $w\in\R^+$ was introduced in \eqref{2.12} and $(\eta_{\mu\nu})=\diag(1, -1, -1,- 1)$ is the Minkowski metric.

Classical spinless particle of mass $m$ is a point in $T^*\R^{1,3}$ moving along a trajectory defined by a Hamiltonian vector field
\begin{equation}\label{4.4}
V_H^{}=\Omega^{\mu+4\,\nu}\dpar_{\mu+4}H\dpar_\nu + \Omega^{\mu\,\nu+4}\dpar_\mu H\dpar_{\nu +4}\with
\dpar_\mu :=\frac{\dpar}{\dpar x^\mu},\ \dpar_{\mu +4}:=\frac{\dpar}{\dpar x^{\mu +4}}\ ,
\end{equation}
where $H$ is a Hamiltonian function. The Hamiltonian flow equation are
\begin{equation}\label{4.5}
\dot x^\mu =V_H^{}x^\mu\und \dot x^{\mu +4}=V_H^{}x^{\mu+4}\ ,
\end{equation}
where $\dot x=\dpar x/\dpar\tau$.

\noindent {\bf Hamiltonian function $H_0$.}  For a free relativistic spinless particle the Hamiltonian function is
\begin{equation}\label{4.6}
H_0=\frac{1}{m}\,\eta^{\mu\nu}p_\mu p_\nu \quad\Rightarrow\quad
V_{H_0}^{} = \frac{p^\mu}{m}\,\frac{\dpar}{\dpar x^\mu} =v^\mu\dpar_\mu\
\end{equation}
and the solution of equations \eqref{4.5} for $H_0$ is written out in \eqref{2.3}. The function $H_0$ is constant on the trajectories and define a hypersurface $X_7$ (level surface) in $T^*\R^{1,3}$,
\begin{equation}\label{4.7}
X_7=H^3\times\R^{1,3}=\bigl\{x,p\in T^*\R^{1,3}\mid \eta^{\mu\nu}\,p_\mu p_\nu =m^2\bigr\}\ .
\end{equation}
Here $H^3 = H^3_+\cup H^3_-$ is the two-sheeted hyperboloid in the momentum space,
\begin{equation}\label{4.8}
H^3_\pm\cong\sSL(2,\C)/\sSU(2):\ \  p^0=\pm E=\pm\sqrt{p^2+ m^2}\ ,
\end{equation}
where $p^2=\delta^{ab}p_ap_b$.

\noindent {\bf Orbit space $X_6$.} On the manifold \eqref{4.7} there acts a one-parameter group
\begin{equation}\label{4.9}
\sGL(1, \R)=\R^*=\bigl\{g=\exp (\tau V_{H_0})=\exp (\tau v^\mu\dpar_\mu )\bigr\}\ .
\end{equation}
The orbits of this group are defined as
\begin{equation}\label{4.10}
g\cdot x^\mu= x^\mu (\tau)=x^\mu +v^\mu\tau\ , \ \ g\cdot p_\mu = p_\mu\quad \Rightarrow\quad \dot x^\mu (\tau ) =\dot gx^\mu =\frac{p^\mu}{m}\ ,
\end{equation}
and they are straight lines written in \eqref{2.3}. For $m\ne 0$, we can always choose $x^\mu =x^\mu (0)$ to belong to a 3-dimensional plane orthogonal to $p_\mu\in H^3_\pm$ by putting
\begin{equation}\label{4.11}
x^\mu =\epsilon_a^\mu s^a\with \epsilon_a^\mu p_\mu =0\ .
\end{equation}
Here $\epsilon_a=(\epsilon_a^\mu)$ are the basis vectors in the 3-dimensional plane orthogonal to $p_\mu$, and the parameters $s_a$ are real numbers, $a=1,2,3$.

Quotienting by the action of the dynamical group \eqref{4.9} is a covariant phase space \eqref{2.4},
\begin{equation}\label{4.12}
X_6=X_7\slash\R^* = T^*H^3_+\cup T^*H^3_-\ ,
\end{equation}
that parametrizes the orbits of the group   \eqref{4.9}. We have a principal bundle
\begin{equation}\label{4.13}
H^3\times\R^{1,3}\ \stackrel{\sGL(1,\R)}{\longrightarrow}\ T^*H^3
\end{equation}
with projection onto the orbit space $T^*H^3$. According to the standard description, cotangent bundles $T^*H^3_+$ and $T^*H^3_-$ in  \eqref{4.12} correspond to particles $(q_+=p^0/|p^0|=1)$ and antiparticles $(q_-=p^0/|p^0|=-1)$. Note that the energy of particles and antiparticles is defined by the energy-momentum relation  \eqref{4.7},
\begin{equation}\label{4.14}
E^2:=p_0^2=p^2 + m^2\ .
\end{equation}
Energy $E$ is always positive and equal $E=\qv p^0$, $\qv =q_\pm =\pm 1$. Thus, the covariant phase space of a free particle is six-dimensional as in nonrelativistic case.

\noindent {\bf Comparison of $T^*H^3$ and $T^*\R^3$.} What is the difference between the description of a free relativistic and non-relativistic particles? For their Hamiltonian function we have
\begin{equation}\label{4.15}
\mbox{non{-}relativistic}\quad H_0=\frac{p^2}{2m}=E\ ,
\end{equation}
\begin{equation}\label{4.16}
\quad{\rm relativistic}\quad H_0=\frac{1}{m}(p_0^2-p^2)=mc^2\ ,
\end{equation}
where for clarity we have temporarily restored speed of light. The energy of the particle on the right side of \eqref{4.15} is constant but arbitrary. Therefore, there are no restrictions on the momentum $p_a\in\R^3$ and the space of initial data of the motion of a nonrelativistic particle is $T^*\R^3$. At the same time, the right hand side in \eqref{4.16} is fixed at the rest energy of the particle, so the Lorentz invariant space of initial data of a free relativistic particle is the manifold $T^*H^3$ and not $T^*\R^{1,3}$. From \eqref{4.16} it can be seen that the relativistic Hamiltonian function specify not the energy, but some parameters of the particle, in the case \eqref{4.16} this is the mass of the particle. In more general cases, this will also be  spin of the particle and various charges it possesses.

\noindent {\bf Generic Hamiltonian function $H$.} The above-described {\it symplectic reduction} of the phase space  $T^*\R^{1,3}$ to a six-dimensional submanifold of $T^*\R^{1,3}$ is general and can be defined for any Lorentz invariant function $H(x,p)$ with $x^\mu, p_\nu\in T^*\R^{1,3}$. A constant value of the function $H(x,p)=m$ defines a 7-dimensional hypersurface $Y_7\subset T^*\R^{1,3}$ in the phase space $T^*\R^{1,3}$. This function also defines a Hamiltonian vector field \eqref{4.4} generating a one-parameter group $\CG_H$ with elements $\exp(\tau V_H)\in \CG_H$ acting on the manifold $Y_7$. Here $\tau$ is a parameter on the orbits in $Y_7$ along which the particle moves. The covariant phase space $Y_6$ is obtained by quotienting $Y_7$ by the action of the group $\CG_H$,
\begin{equation}\label{4.17}
Y_7\ \stackrel{\CG_H}{\longrightarrow}\ Y_6=Y_7/\CG_H\ .
\end{equation}
The geometry of the manifold $Y_6$ depends on the Hamiltonian function $H$ and dictates the choice of the canonical commutation relation. Note that in the relativistic case there is no analogue of the Stone-von Neumann theorem.

\noindent {\bf Klein-Gordon oscillator.} A good illustration of the symplectic reduction scheme formulated above is Klein-Gordon oscillator \cite{Dirac} and its supersymmetric version, described  in detail in \cite{Popov2, Popov3}. This relativistic oscillator is given by the Hamiltonian function 
\begin{equation}\label{4.18}
H_\omega^{}=\frac{1}{m}\,(\eta^{\mu\nu}p_\mu p_\nu + m^2\omega^2\eta_{\mu\nu}x^\mu x^\nu )
\quad\Rightarrow\quad
V_{H_\omega}^{}=\frac{p^\mu}{m}\,\frac{\dpar}{\dpar x^\mu} - m\omega^2x^\mu\frac{\dpar}{\dpar p^\mu} \ .
\end{equation}
The dynamics is given by the Hamiltonian vector field $V_{H_\omega}^{}$, which is the generator of the group $\sU(1)\ni g=\exp(\tau V_{H_\omega}^{})$ acting on the level surface
\begin{equation}\label{4.19}
H_\omega =m\ \Leftrightarrow\ \Ad S_7:\quad 2\omega^2 \eta_{\mu\bar\nu}^{}\, z^\mu \zb^{\bar\nu} =1\ \for \ 
z^\mu =\frac{1}{\sqrt 2}(x^\mu - \frac{\im}{m\omega}p^\mu )\ .
\end{equation}
The covariant phase space is a K\"ahler-Einstein manifold $Z_6$ obtained by quotienting $Z_7:=\Ad S_7$ by the action of the dynamical group U(1),
\begin{equation}\label{4.20}
\Ad S_7\ \stackrel{\sU(1)}{\longrightarrow}\ \sU(1,3)/\sU(1)\times \sU(3)=: Z_6\ ,
\end{equation}
which can be compared with the case  \eqref{4.12} of free particle. In the limit $\omega\to 0$, the manifold $Z_6$ turns into the manifold $T^*H^3$. Note that the coordinate time $x^0$ cannot coinside with the evolution parameter $\tau$ for $H_\omega$ in \eqref{4.18}. For a free particle this is possible but only in the rest frame.

The level surface \eqref{4.19} is given by the equation 
\begin{equation}\label{4.21}
E^2:= p_0^2 + m^2\omega^2x_0^2 = p^2+m^2\omega^2\delta_{ab}x^ax^b + m^2\ ,
\end{equation}
which generalizes the energy-momentum relations \eqref{4.14} and transforms into it when $\omega\to 0$. From \eqref{4.21} it follows that the energy $E$ coincides with the {\it radius} of the circle $S^1$ in the $(x^0, p_0)$-plane and therefore it cannot be negative. It is also obvious that in the limit $\omega\to 0$ we will get two points $p^0=E$ and $p^0=-E$ with $E>0$, and not the equality $E=\pm p^0$ with $p^0>0$. Energy is the length of segments $[0, p^0]$ and $[-p^0, 0]$, not points $p^0$ and $-p^0$.  This can be written as $p^0=q_\pm E$, where $q_\pm =p^0/|p^0|=\pm 1$ correspond to two directions on the $\tau$-axis (orientation).

Quantization of the model \eqref{4.18} is well defined in the complex Segal-Bargmann representation \cite{Popov2, Popov3}. It was shown that the general solution of this model is given by functions from the weighted Bergman space of square-integrable holomorphic (for particles) and antiholomorphic (for antiparticles) functions on the K\"ahler-Einstein manifold $Z_6$ from \eqref{4.20}. This relativistic model is Lorentz covariant, unitary and does not contain non-physical states.

In this paper we will show that in the theory of spin $s=\sfrac12$ particles there are also no non-physical states. To do this,
 we generalize the approach described above to relativistic Hamiltonian mechanics of particles with spin. For this we extend the phase space $T^*\R^{1,3}$ with an additional space of spin degrees of freedom $\C_L^2\times\C_R^2$. On this space, we introduce positive-definite  Lorentz invariant Hamiltonian functions that take into account the charges $q_\pm =\pm 1$ of particles and antiparticles. In the next section we will introduce a relativistic analogue of the Schr\"odinger equation for the evolution in $\tau$ of $\C$-valued wave functions on the extended phase space of particles with spin. We will show the Klein-Gordon, Dirac, Proca and other equations follow from this equation after expanion in spin variables from $\C_L^2\times\C_R^2$. The positivity of energies and probabilities at the first quantized level follows from the use of positive-definite Hamiltonian functions at the classical level.

\subsection{Spin in relativistic Hamiltonian mechanics}

\noindent {\bf Preliminary remarks.} In Sect.2.1 we discussed the currently accepted procedure \eqref{2.5}-\eqref{2.8} for introducing spin. We noted that by quantizing the phase spaces $T^*H_+^3\times\CPP_L^1$ and $T^*H_-^3\times\overline{\CPP}_L^1$  one can obtain representations of the Lorentz group of type $(s,0)$ and $(0,s)$, but not of type $(s, j)$. This is because description \eqref{2.5}-\eqref{2.8} is not preserved under parity transformation $P\in\sO(1,3)$. For this reason, we consider a relativistic parametrization of the spin degrees of freedom by the spaces $\C_L^2\times\C_R^2$ (particles) and $\bar\C_R^2\times\bar\C_L^2$ (antiparticles), such that they are mapped into themselves or each other under discrete transformations from the Lorentz group O(1,3). Note, however, that spin is a non-relativistic concept and is described by representations of the group $\sSU(2)\cong\sSO(3)\subset\sO(1,3)$ and not the Lorentz group. Therefore, it is necessary to define a Lorentz covariant reduction of the space $\C_L^2\times\C_R^2$ to the space $\C^2$ of non-relativistic mechanics, which we will discuss using the example of classical Dirac particles. It is precisely this reduction task that relativistic equations perform at the first quantized level \cite{Wigner, BW}. In the relativistic case we will not strive for a maximum generality and will concentrate our attention on the representations of the Lorentz group of the type $(\sfrac12, 0)\oplus (0,\sfrac12)$ (Dirac particles) and $(\sfrac12,\sfrac12)$ (Proca particles a.k.a. massive vector bosons).

\noindent {\bf Bispinors.} The above representations do not require the introduction of orbifolds, which we discussed in Section 2. As a phase space for describing relativistic particles with spin, we take space
\begin{equation}\label{4.22}
T^*\R^{1,3}\times\C^4\ ,
\end{equation}
where the internal space
\begin{equation}\label{4.23}
\C^4 =\C_L^2\oplus\C_R^2\cong \C_L^2\times\C_R^2\supset\CPP_L^1\times\CPP_R^1
\end{equation}
is the space of the spinor representation of the Lorentz group. Note that the direct sum and direct product in  \eqref{4.23} are canonically isomorphic. Dirac spinors from the space  \eqref{4.23} transform according to the group Spin(1,3) doubly covering the group SO(1,3)$\subset$O(1,3). Discrete transformations act on them according to formulae  \eqref{2.10}.

\noindent {\bf Spinor indices.} We introduce matrices 
\begin{equation}\label{4.24}
\sigma^\mu =(\unit_2, \sigma^a)\und \bar\sigma^\mu =(\unit_2, -\sigma^a)\ ,
\end{equation}
where $\sigma^a$ are the Pauli matrices. Using these matrices, any complex (co-)vector $p_\mu\in\C^4$ can be assigned matrices
\begin{equation}\label{4.25}
p^{\al\dal}=p_\mu\sigma^{\mu\al\dal}\und p_{\dal\al}=p_\mu\bar\sigma^{\mu}_{\dal\al}\ .
\end{equation}
These matrices are Hermitian for real $p_\mu\in\R^{1,3}$. In fact, the maps
\begin{equation}\label{4.26}
\sigma:\ p_\mu\mapsto p^{\al\dal}\und \bar\sigma :\ p_\mu\mapsto p_{\dal\al}
\end{equation}
provides two homomorphisms between proper orthochronous Lorentz group SO$^+$(1,3) and the group SL(2, $\C )/\Z_2$. Indices $\al$ and $\dal$ are the indices of the representations of type $(\sfrac12, 0)$ and $(0, \sfrac12)$ of group SL(2, $\C$). On the vectors \eqref{4.25} a representation of the complexified special Lorentz group SO(4, $\C$)=SL(2, $\C$)$\times$SL(2, $\C)/\Z_2$ is defined,
\begin{equation}\label{4.27}
p^{\al\dal}\mapsto L^\al_\beta\,  p^{\beta\dot\beta}\,R^\dal_{\dot\beta}\ ,
\end{equation}
where $(L^\al_\beta)$, $(R^\dal_{\dot\beta})\in\sSL(2, \C )$. This action preserves the Minkowski metric,
\begin{equation}\label{4.28}
\eta^{\mu\nu}p_\mu p_\nu =\veps_{\al\beta}\veps_{\dal\dot\beta}p^{\al\dal}p^{\beta\dot\beta}\ ,\quad
p_\mu\in\C^4\ ,
\end{equation}
where the $\veps$-symbols were introduced in \eqref{2.24}, $\veps_{01}=1=\veps_{\dot 0\dot 1}, \veps^{01}=-1=\veps^{\dot 0\dot 1}$. These tensors are used to define an SL(2, $\C$)-invariant inner products on $\C_L^2$ and $\C_R^2$:
\begin{equation}\label{4.29}
\begin{split}
\C_L^2\ni z_\al& :\quad   z^\al\tilde z_\al =-z_\al\tilde z^\al=-\veps^{\al\beta}z_\al\tilde z_\beta\ ,
\\[2pt]
\C_R^2\ni y^\dal& :\quad  y^\dal\tilde y_\dal =-y_\dal\tilde y^\dal =\veps_{\dal\dot\beta}y^\dal\tilde y^{\dot\beta}  \ .
\end{split}
\end{equation}
For real $p_\mu\in\R^{1,3}$, the matrices $(R_{\dot\beta}^\dal)$ in \eqref{4.27} must be complex conjugates of the matrices $(L_\beta^\al )$. In complex conjugation, the index $\al$ changes to index $\dal$ and vice versa: 
$\overline{\mu^\al}=\bar\mu^\dal$ and  $\overline{\lambda_{\dal}}=\bar\lambda_\al$.

\noindent {\bf Gamma matrices.} Using matrices \eqref{4.24}, we introduce the generators of the complexified Clifford algebra Cl$^\C$(4) of the Minkowski space in the form
\begin{equation}\label{4.30}
\ga^\mu =\begin{pmatrix}0&\sigma^\mu\\\bar\sigma^\mu&0\end{pmatrix}\ \Rightarrow\ \ga^\mu\ga^\nu + \ga^\nu\ga^\mu =2\eta^{\mu\nu}\cdot\unit_4\ ,\quad \ga^5:=\im\ga^0\ga^1\ga^2\ga^3=
\begin{pmatrix}-\unit_2&0\\0&\unit_2\end{pmatrix}\ .
\end{equation}
We use a chiral representation of $\ga^\mu$ in which the Dirac spinors are a direct sum of left and right Weyl spinors:
\begin{equation}\label{4.31}
\Psi =(\Psi^i)=\begin{pmatrix}\psi_L\\\psi_R\end{pmatrix}=\begin{pmatrix}\mu^\al\\\lambda_\dal\end{pmatrix},\ 
\Pi_L\Psi=\begin{pmatrix}\psi_L\\0\end{pmatrix}, \ \Pi_R\Psi=\begin{pmatrix}0\\\psi_R\end{pmatrix},     \
\Pi_{L,R}=\sfrac12(\unit_4\mp\ga^5)\ .
\end{equation}
In the relativistic Hamiltonian mechanics that we are developing, we use variables of the form
\begin{equation}\label{4.32}
Z=(Z_i):=(z_\al , y^\dal)\und\bar Z=(\bar Z^i):=\ga^0Z^\+=\begin{pmatrix}\bar y^\al\\\zb_\dal\end{pmatrix}\ ,\quad i=1,...,4\ ,
\end{equation}
that are dual to $\Psi$ and $\bar\Psi=\Psi^\+\ga^0$, so that functions $\Phi$ of $Z$ have the form $\Phi=\Psi_0+\Psi^iZ_i+\Psi^{\bar\imath}Z_{\bar\imath}+...\ $. The Klein-Gordon, Dirac and Proca equations will be obtained from such expansions of $\C$-valued wave functions.

\noindent {\bf Charge conjugation and self-duality.} Particles and antiparticles are related by a charge conjugation map, which is always antilinear, that is, it contains complex conjugation. In the case under consideration, the internal space of particles is the space $\C^4\ni Z$, and charge conjugation has the form:
\begin{equation}\label{4.33}
C:\quad Z\mapsto Z^*\mapsto Z^*(\im\ga^2)=:Z^c\quad\Leftrightarrow\quad (z_\al , y^\dal)\mapsto (\bar y_\al , \zb^\dal)\ .
\end{equation}
This action of the operator $C$ on the space $\C_L^2\times\C_R^2$ was indicated earlier in \eqref{2.10}.

Spaces $\C_L^2$ and $\C_R^2$ are also related to the concepts of self-duality and anti-self-duality. Namely, the generators of the group Spin(1,3) are matrices
\begin{equation}\label{4.34}
\Sigma^{\mu\nu}=\frac{\im}2 [\ga^\mu , \ga^\nu]=
\begin{pmatrix}\sfrac{\im}2 [\sigma^\mu , \bar\sigma^\nu]&0\\ 0& \sfrac{\im}2 [\bar\sigma^\mu , \sigma^\nu]  \end{pmatrix}=:
\begin{pmatrix}\sigma^{\mu\nu}&0\\0&\bar\sigma^{\mu\nu}\end{pmatrix}\ .
\end{equation}
Matrices $\sigma^{\mu\nu}$ and $\bar\sigma^{\mu\nu}$ in \eqref{4.34} satisfy the duality equations,
\begin{equation}\label{4.35}
\begin{split}
\sfrac12\veps^{\mu\nu\lambda\varrho}\sigma_{\lambda\varrho}=&\im\sigma^{\mu\nu}\quad\mbox{(self-duality)}\ ,\\
\sfrac12\veps^{\mu\nu\lambda\varrho}\bar\sigma_{\lambda\varrho}=&-\im\bar\sigma^{\mu\nu}\quad\mbox{(anti-self-duality)}\ ,
\end{split}
\end{equation}
where $\veps^{\mu\nu\lambda\varrho}$ is the completely antisymmetric tensor in Minkowski space.

\noindent {\bf Momentum Hilbert space.} In \eqref{4.6}-\eqref{4.14} we described the dynamics of a particle in the relativistic phase space $T^*\R^{1,3}$ of coordinates and momenta and introduced the covariant phase space \eqref{4.12}, i.e. the space of initial data of this motion. In the following, when quantizing, we will use the momentum representation, in which all functions will depend on momenta from the space $H_+^3\cup H^3_-$. The space of all of such functions is splitted into a direct sum of Hilbert spaces of $L^2$-functions,
\begin{equation}\label{4.36}
\CH =(\CH_p\oplus\CH_p^P)\oplus(\CH_p^T\oplus\CH_p^{PT})\ ,
\end{equation}
where $\CH_p$ is a space of unitary representation of the group SO$^+$(1,3), $\CH_p^P=P(\CH_p)$, $\CH_p^T=T(\CH_p)$ and $\CH_p^{PT}=PT(\CH_p)$, and the first parenthesis in \eqref{4.36} correspond to functions on $H_+^3$, and the second parenthesis in \eqref{4.36} correspond to functions on $H_-^3$. 

\noindent {\bf Internal phase space.} In this paper we want to understand the reasons for the emergence of negative energies in antiparticle solutions of the Dirac equations. To do this, we describe classical Dirac particles and antiparticles as points moving in phase space \eqref{4.22} and define the Hamiltonian function so that it is Lorentz invariant and non-negative. When reducing the phase space $T^*\R^{1,3}$ to $T^*H_+^3\cup T^*H_-^3$, leading to a direct sum of Hilbert spaces \eqref{4.36}, it is necessary to indicate from the very beginning which of these spaces are associated with Dirac particles, and which with antiparticles. Namely, over the hyperboloids $H_+^3$ and $H_-^3$ we have spinor bundles with fibres of the form
\begin{equation}\label{4.37}
\begin{split}
H_+^3: \ \C_+^4=&  \C_L^2\times\C_R^2\ni Z^+=(z_\al^+, y_+^\dal ):=(z_\al , y^\dal )\und v_+^{\al\dal}=v^{\al\dal},
\\[1pt]
H_-^3: \ \C_-^4=&  \bar\C_R^2\times\bar\C_L^2\ni Z^-=(z_\al^-, y_-^\dal ):=(\bar y_\al , \bar z^\dal )\und v_-^{\al\dal}=-v^{\al\dal},
\end{split}
\end{equation}
where $v^{\al\dal}=p^{\al\dal}/m$ is the velocity vector $v^\mu$ with $v^0>0$. Note that the internal phase space $\C_+^4$ corresponds to particles, and the internal phase space $\C_-^4$ corresponds to antiparticles, defined by complex conjugate coordinates. In what follows we will use coordinates $(z_\al^\pm, y_\pm^\dal )$ on charge-conjugate spaces $\C_+^4$ and $\C_-^4$ $(Z^-=CZ^+=Z^+_c$) to control the difference between particles and antiparticles.

\subsection{Dynamics of classical spin variables}

\noindent {\bf Symplectic 2-form on $\C^4$.} In formula \eqref{4.1} we introduced a symplectic 2-form $\Omega_0$ on a subspace $T^*\R^{1,3}$ in the extended phase space $T^*\R^{1,3}\times\C^4$. On the internal space $\C^4$ we introduce a two-form
\begin{equation}\label{4.38}
\Omega_{int}^{}=\im q_\pm g^{i\bar\jmath}\dd Z_i^\pm\wedge\dd \bar Z_{\bar\jmath}^\pm =\im (v^{\al\dal}_\pm\dd z_\al^\pm\wedge\dd\zb_\dal^\pm + v_{\dal\al}^\pm\dd y^\dal_\pm\wedge\dd\bar y^\al_\pm )\ ,
\end{equation}
with components
\begin{equation}\label{4.39}
\Omega_{\pm}^{i\bar\jmath}=\im q_\pm g^{i\bar\jmath}=-\Omega_{\pm}^{\bar\jmath i}\ ,\quad
\Omega^\pm_{\bar\jmath i} =\im q_\pm g_{\bar\jmath i}=-\Omega^\pm_{i \bar\jmath}
\ ,
\end{equation}
where
\begin{equation}\label{4.40}
(g^{i\bar\jmath})=\diag (v^{\al\dal}, v_{\dal\al})\ ,\quad (g_{\bar\jmath i})=\diag (v_{\dal\al}, v^{\al\dal})\ ,
\end{equation}
and $v^{\al\dal}$ are introduced in \eqref{4.37}. Note that 2-form \eqref{4.38} has the same form in the coordinates $Z_i^+=(z_\al^+, y_+^\dal)=(z_\al , y^\dal)$ and $Z_{i}^-=(z_\al^-, y_-^\dal)=(\bar y_\al , \zb^\dal)$ due to the charge $q_\pm =\pm 1$ distinguishing the spaces $\C_+^4$ and $\C_-^4$ from \eqref{4.37}. The K\"ahler metric on $\C^4$ in coordinates $Z_i^\pm$ has the form
\begin{equation}\label{4.41}
g_{int}=v_\mu\dd Z^\pm\ga^\mu\dd \bar Z^\pm =g^{i\bar\jmath}\dd Z^\pm_i\dd \bar Z^\pm_{\bar\jmath}=
v^{\al\dal}\dd z^\pm_\al\dd \bar z^\pm_{\dal}+ v_{\dal\al}\dd y_\pm^\dal\dd \bar y_\pm^{\al}\ .
\end{equation}
This momentum-dependent metric is positive definite. The two-form \eqref{4.38} and the metric \eqref{4.41} are Lorentz invariant since the transformation $PT: v_\mu^+\ga^\mu\mapsto  v_\mu^-\ga^\mu$  is accompanied by the mapping $\C_+^4\to \C_-^4$, i.e. change of complex structure $J_+\to J_- =-J_+$.

\noindent{\bf Hamiltonian function $N_{int}$}. We use the metric \eqref{4.41} on the internal space $\C^4$ to define a positive definite Hamiltonian function of the form:
\begin{equation}\label{4.42}
\begin{split}
H_{int}=&\omega N_{int}\ ,\quad N_{int}=g^{i\bar\jmath}Z_i^\pm \bar Z^\pm_{\bar\jmath} =N_{int}^L + N_{int}^R\ ,
\\[1pt]
N_{int}^L=&v^{\al\dal}z_\al\zb_\dal\und N_{int}^R=v_{\dal\al}y^\dal\bar y^\al\ .
\end{split}
\end{equation}
We wrote $N_{int}^{+L}=N_{int}^L$ and $N_{int}^{+R}=N_{int}^R$ for the space $\C_+^4$, and for the space $\C_-^4$ we have $N_{int}^{-L}=N_{int}^{+R}$ and $N_{int}^{-R}=N_{int}^{+L}$, but the sum $N_{int}$ does not change. Note that the scalar Hamiltonian function $H_{int}$ is not energy. Recall that in the non-relativistic case we introduced
\begin{equation}\label{4.43}
S_L^0:=zz^\+ =\delta^{\al\dal}z_\al\zb_\dal\und S^a_L=z\sigma^a z^\+\ ,
\end{equation}
so that 
\begin{equation}\label{4.44}
v_\mu S_L^\mu =v_\mu z\sigma^\mu z^\+=v^{\al\dal}z_\al\zb_\dal =N_{int}^L\ .
\end{equation}
Similarly, one can introduce
\begin{equation}\label{4.45}
S_R^\mu =y\bar\sigma^\mu y^\+\und 
N_{int}^{R}=v_\mu S_R^\mu =v_\mu y\bar\sigma^\mu y^\+=v_{\dal\al}y^\dal\bar y^\al\ .
\end{equation}
As energy one can consider the function $m(S_L^0 + S_R^0)$.

The reason for the appearance of negative energies in solutions of the Dirac equations can be seen when considering  the Hamiltonian function \eqref{4.42} of classical Dirac particles moving not only in Minkowski space but also in the internal space  $\C^4$. We know from \eqref{4.37} that under the $PT$-transformation $v_+^{\al\dal}\mapsto v_-^{\al\dal}$ the spinor $Z^+$ is mapped into the charge conjugated spinor $Z^-=(Z^+)^c$, and the symplectic 2-form \eqref{4.38} must be invariant under the transformations of the Lorentz group O(1,3). It is invariant if its components have the form \eqref{4.39} with the multiplier $q_\pm =\pm1$ on $T^*H^3_\pm\times\C_\pm^4$. And since under this mapping  the complex structure  also changes sign, $J_+\to J_-=-J_+$, the metric on $\C^4$ must have the form \eqref{4.41} with $q_+v_+^{\al\dal}=v^{\al\dal}$ on $\C_+^4$ and  $q_-v_-^{\al\dal}=v^{\al\dal}$ on $\C_-^4$. If we remove $q_\pm$ from the 2-form \eqref{4.38}, then $q_\pm$ appears in the metric \eqref{4.41}, and the Hamiltonian function \eqref{4.42} ceases to be Lorentz invariant and positive definite.

\noindent{\bf Covariant phase space.} The Hamiltonian functions $H_0$ from \eqref{4.6} and $H_{int}$ from \eqref{4.42} commute with respect to the Poisson bracket given by the 2-form $\Omega_0+\Omega_{int}$ on the phase space $T^*\R^{1,3}\times\C^4$. Therefore, these functions take constant values independently of each other. For the function $H_0$ we obtain the level surface \eqref{4.7}, and the commuting functions $H_{int}^L$ and $H_{int}^R$ define 3-spheres,
\begin{equation}\label{4.46}
\C_L^2\supset S_L^3: \quad N_{int}^L=1\und \C_R^2\supset S_R^3: \quad N_{int}^R=1\ ,
\end{equation}
with metrics depending on points $p_\mu$ in the momentum space. Spheres in space $\bar\C_R^2\times\bar\C_L^2$ are defined similarly.

The function $N_{int}$ from \eqref{4.42} defines a vector field
\begin{equation}\label{4.47}
V_{N_{int}}^{} =-\im\left(Z_i\frac{\dpar}{\dpar Z_i} - Z_{\bar\imath}^*\frac{\dpar}{\dpar Z_{\bar\imath}^*}\right)=
-\im\left(z_\al\frac{\dpar}{\dpar z_\al} - \zb_\dal\frac{\dpar}{\dpar \zb_\dal}+y^\dal\frac{\dpar}{\dpar y^\dal}-
\bar y^\al\frac{\dpar}{\dpar \bar y^\al}\right)\ ,
\end{equation}
which is the generator of the group U(1) acting on the level surface $S_L^3\times S^3_R\subset \C_L^2\times\C_R^2$. The orbits of this group are given by formulae
\begin{equation}\label{4.48}
z_\al(\vph)=\exp(\vph V_{N_{int}}^{})z_\al =\exp(-\im\vph)z_\al \ ,\quad
y^\dal(\vph)=\exp(\vph V_{N_{int}}^{})y^\dal =\exp(-\im\vph)y^\dal \ ,
\end{equation}
where $\vph =\omega\tau$. Orbits \eqref{4.48} are parametrized by the product of Riemann spheres,
\begin{equation}\label{4.49}
\CPP_L^1\times\CPP_R^1\subset S_L^3\times S_R^3\subset \C_L^2\times\C_R^2\ ,
\end{equation}
with homogeneous coordinates $[z_0:z_1]$ and $[y^{\dot 0}:y^{\dot 1}]$.

Thus, we arrive at the covariant phase space of particles with spin of the form
\begin{equation}\label{4.50}
T^*H_+^3\times\CPP_L^1\times\CPP_R^1\ \cup\  T^*H_-^3\times{\overline \CPP}_R^1\times{\overline \CPP}_L^1\subset
T^*\R^{1,3}\times\R^8\ ,
\end{equation}
where the connected component with $q_+=p_0/|p_0|=1$ corresponds to particles, and the connected component with  $q_-=p_0/|p_0|=-1$ corresponds to antiparticles. Recall that the discrete transformations $T$ and $PT$ map $\C_+^4$ and $\C_-^4$ to each other, which is also what the charge conjugation operator $C$ does. Therefore, Lorentz covariant theories cannot consider only particles without antiparticles. However, at a more fundamental level, the map of particles into antiparticles is associated with a mapping of the evolution parameter $\tau\mapsto -\tau$ that induce a mapping of all complex structures on extended phase spaces, and on complex vector bundles over them, into conjugated complex structures.

\noindent{\bf Bargmann-Wigner particles.} In this paper we do not strive for maximum generality; our goal is to consider particles with spin $s\le 1$. That is why, we restrict ourselves to particles moving in phase space \eqref{4.22} with $\Z_n$ symmetry in the internal space $\C^4$, which is equivalent to moving in phase spaces
\begin{equation}\label{4.51}
T_+^*\R^{1,3}\times \C_+^4/\Z_n\und T_-^*\R^{1,3}\times \C_-^4/\Z_n
\end{equation}
for particles ($q_+=1$) and antiparticles ($q_-=-1$). Here $T_\pm^*\R^{1,3}$ denotes subspaces of $T^*\R^{1,3}$ with $q_\pm=p_0/|p_0|=\pm 1$.

We will call the particles described above Bargmann-Wigner particles. By quantizing their phase spaces \eqref{4.51}, we obtain Bargmann-Wigner multispinor fields of the form
\begin{equation}\label{4.52}
\Psi_{n\,\pm}^{BW}=\Psi_\pm^{i_1...i_n}(p)Z_{i_1}^\pm...Z_{i_n}^\pm
\end{equation}
with symmetric components, $i_k=1,...,4$. The components $\Psi_\pm^{i_1...i_n}$ in \eqref{4.52} parametrize the symmetrization of the $n$-fold tensor product
\begin{equation}\label{4.53}
D_n^{BW}=[(\sfrac12, 0)\oplus (0, \sfrac12)]^{\otimes n} =\mathop\oplus^n_{m=0}(\sfrac12 n-\sfrac12 m, \sfrac12 m)\ ,
\end{equation}
of the spinor representation $(\sfrac12, 0)\oplus (0, \sfrac12)$ of the Lorentz group. In this paper we will consider particle with $n=0$ (Klein-Gordon), $n=1$ (Dirac) and $n=2$ (Proca).

Fields with spin $s=0, \sfrac12$ and 1 have 1, 2 and 3 complex components, respectively. To find the spin content of the representations \eqref{4.53} of the Lorentz group, we must restrict it to the subgroup SU(2) and perform the Clebsch-Gordan decomposition. In particular, for $n=1$ and $n=2$ we have
\begin{equation}\label{4.54}
\begin{split}
n=1:&\quad (\sfrac12,0)\sim\C^2\und (0,\sfrac12)\sim\C^2\ ,
\\[1pt]
n=2:&\quad (1,0)\sim\C^3, \ (0,1)\sim\C^3\und (\sfrac12,\sfrac12)\sim\C^3\oplus\C\ ,
\end{split}
\end{equation}
where $s=\sfrac12n$. Thus, fields \eqref{4.52} have four components instead of two at $n=1$, and fields with $n=2$ have 10 components intead of 3, and we must define the projection of the spaces $\C^4$ and $\C^{10}$ from  \eqref{4.54} onto the spaces $\C^2$ and $\C^3$, respectively. These projections are given by relativistic equations. The spaces of their solutions are these spaces of irreducible representations of the group SU(2) given by Lorentz covariant way \cite{Wigner, BW}. In fact, these projections can also be defined at the level of relativistic Hamiltonian mechanics. In the next section we will show this using the example of Dirac and Proca particles.

\subsection{Dirac and Proca particles}

\noindent{\bf Projection $\C_\pm^4{\to}\C_\pm^2$.} Dirac spinors $\Psi_\pm =(\Psi_\pm^i)$ arise from the linear in $Z_i^\pm$ terms of the form  \eqref{4.52} with $n=1$ in the expansion of $\C$-valued wave functions in $Z_i^\pm$ coordinates on the extended phase spaces $T^*_\pm\R^{1,3}\times\C_\pm^4$. Relativistic particles with spin move not only in phase space $T^*\R^{1,3}$ but also in internal spaces $\C_\pm^4$, where $\C^4_+=\C^4$ and $\C^4_-=\bar\C^4$. At the same time, we know that in the non-relativistic case, particles with spin move in the internal spaces $\C_\pm^2$. The Wigner approach requires the specification of a Lorentz covariant projection
\begin{equation}\label{4.55}
\C_\pm^4\ \longrightarrow\ \C_\pm^2\ ,
\end{equation}
which can be realized by specifying Lorentz covariant equations expressing the coordinates of the space $\C_R^2$ through the coordinates on $\C_L^2$.

Recall that the extension of $\C_L^2$ to $\C_L^2\times\C_R^2$ was necessary to preserve $P$-invariance. Note that in the rest frame of the particle the $P$-invariance condition is preserved if we reduce the space $\C_+^4=\C_L^2\times\C_R^2$ to the diagonal subspace $\C_+^2=\diag(\C_L^2\times\C_R^2)$ and reduce the space $\C_-^4=\bar\C_R^2\times\bar\C_L^2$ to the antidiagonal subspace $\C_-^2\hra\bar\C_R^2\times\bar\C_L^2$. Passing by boosts to the general momentum $p_\mu\in\sO(1,3)/\sO(3)$, we obtain from the above diagonal/antidiagonal constraints the following Lorentz covariant equations
\begin{equation}\label{4.56}
\begin{split}
Z^\pm (v_\mu^\pm\ga^\mu -1)=0\ &\Rightarrow\ Z^\pm(p_\mu\ga^\mu\mp m)=0
\\[1pt]
\Rightarrow\ y_\pm^\dal = v^{\al\dal}_\pm z_\al^\pm\ \for\  
v^{\al\dal}_\pm &=\pm v^{\al\dal}=\pm v_\mu\sigma^{\mu\al\dal}\ ,\quad v^0=\frac{p^0}{m}>0\ .
\end{split}
\end{equation}
These equations define the coordinates $y^\dal$ of the space $\C_R^2(p)$ as linear functions of the coordinates  $z_\al$ of space $\C_L^2$ and the momenta $p_\mu\in H_+^3$, and similarly for $\bar\C_R^2$ and $\bar\C_L^2$. Thus, for each fixed value of $p_\mu$, these equations define a Lorentz covariant projection \eqref{4.55} of $\C_\pm^4$ onto $\C_\pm^2(p)$. Equations \eqref{4.56} also reduce the space of initial data $\CPP_L^1\times\CPP_R^1$ of the particle's motion in $\C_L^2\times\C_R^2$ to the space $\CPP_+^1$ and the space $\overline{\C P}_R^1\times\overline{\C P}_L^1$ to the space $\CPP_-^1$. 

Note that equations \eqref{4.56}, as well as functions \eqref{4.52}, are invariant with respect to the action of the group $\Z_n$ of the form $Z_i^\pm\mapsto\zeta Z_i^\pm$, $\zeta\in\Z_n$. Therefore, equations \eqref{4.56} are also satisfied for coordinates  on the space $\C_\pm^4/\Z_n$ for which they define the projection $\C_\pm^4/\Z_n\to \C_\pm^2/\Z_n$. Recall that $\Z_1\equiv 1$ and Dirac particles correspond to $n=1$, while the Proca particles are given by $n=2$. They are special cases of Bargmann-Wigner particles.

\noindent{\bf Bargmann-Wigner equations.} So we have shown that the geometric meaning of equations \eqref{4.56} is that the Bargmann-Wigner $\Z_n$-particles move not in space $\C_\pm^4/\Z_n$, but in space $\C_\pm^2(p)/\Z_n$, where $n=1,2,...$. Looking ahead, we note that the Bargmann-Wigner equations for symmetric multispinors automatically follow from equations \eqref{4.56}. Namely, let us rewrite equations \eqref{4.56} as
\begin{equation}\label{4.57}
Z_j^\pm =(v_\mu^\pm\ga^\mu )^i_j Z_i^\pm
\end{equation}
and replace one $Z_i^\pm$ in \eqref{4.52} with the right-hand side of this equality:
\begin{equation}\label{4.58}
\begin{split}
\Psi^{BW}_{n\,\pm}\mid_{\C^2_\pm/\Z_n}^{}&=\Psi_\pm^{j\,i_2...i_n}
Z^\pm_jZ^\pm_{i_2}...Z^\pm_{i_n}=(v_\mu^\pm\ga^\mu )^{i_1}_j\Psi_\pm^{j\,i_2...i_n}Z^\pm_{i_1}...Z^\pm_{i_n}\ ,
\\[2pt]
\Rightarrow\quad & (p_\mu\ga^\mu)_j^{(i_1}\Psi_\pm^{i_2)j...i_n}(p) -
q_\pm m \Psi_\pm^{i_1...i_n}(p)=0\ , 
\end{split}
\end{equation}
where $q_\pm =p_0/|p_0|=\pm 1$. We have obtained the Bargmann-Wigner equations, which for $n=1$ is the Dirac equation and for $n=2$ they are equivalent to the Proca equations (we will show this later) for massive vector bosons. Thus, the meaning of equations  \eqref{4.58}, and the Dirac equations in particular, is that the fields \eqref{4.52} are in fact not fields on the space $\C_\pm^4/\Z_n$, but free fields on the space $\C_\pm^2/\Z_n$, defined in a Lorentz covariant way.

\noindent{\bf Remarks.} It is generally accepted  that the Pauli equation is a generalization of the Schr\"odinger equation to spin degree of freedom, and the Dirac equation is a relativistic generalization of both the Pauli equation and the Schr\"odinger equation. However, in Section 3 we showed that the Pauli equation follows from the Schr\"odinger equation on the phase space $T^*\R^3\times\C^2$ for the  terms linear in coordinates $z_\al$ on $\C^2$ in the expansion of the wave function in these coordinates. In the next section we show that the Bargmann-Wigner equations \eqref{4.58} follow from the relativistic Schr\"odinger equation for the evolution over $\tau$ of the wave function on the phase space $T^*\R^{1,3}\times\C^4$. Considering that equations \eqref{4.58} are the Dirac equation for $n=1$, the Proca equations for $n=2$ and the Maxwell equations in the limit of zero mass of the vector bosons of the Proca equations, then it would be correct to consider the Schr\"odinger equation as the fundamental equation, and not the Klein-Gordon, Dirac, Proca and Maxwell equations derived from it. Besides, it is strange to consider  equations derived from the Schr\"odinger equation as equations for classical fields.

\noindent{\bf Dirac metric.} We have shown that constraint equations \eqref{4.57} define Lorentz covariant subspaces $\C_\pm^2/\Z_n$ in the spaces $\C_\pm^4/\Z_n$ in which the Bargmann-Wigner particles must move. We also know that these particles move in spheres \eqref{4.46} or their orbifold subspaces $S_L^3/\Z_n\times S_R^3/\Z_n$. We should establish a correspondence between this notion in $S_L^3\times S_R^3$ and the constraint equation \eqref{4.57}. To understand this, we note that in addition to the momentum-dependent symplectic 2-form \eqref{4.38} on the spinor space $\C^4$, one can introduce a momentum-independent 2-form,
\begin{equation}\label{4.59}
\Omega_D =\im\,\dd Z^+\wedge\dd\bar Z^+=\im\,\dd Z^-\wedge\dd\bar Z^-\ ,
\end{equation}
where $Z^+=(z_\al , y^\dal )$, $Z^-=CZ^+=(\bar y_\al , \zb^\dal )$ and $\bar Z^\pm =\ga^0(Z^\pm)^\+$. This 2-form is associated with a pseudo-Hermitian metric
\begin{equation}\label{4.60}
g_D =-\im\,\dd Z^\pm J\dd\bar Z^\pm=\dd Z^+\dd\bar Z^+=-\dd Z^-\dd\bar Z^-=
\dd z_\al\dd\bar y^\al + \dd y^\dal\dd\zb_\dal\ ,
\end{equation}
where $Z^\pm J=\pm\im Z^\pm$ for the operator of complex structure $J$. This metric on $\C^4$ is independent of momenta, has signature $(++--)$ and does not change sign under charge conjugation. Using the metric \eqref{4.60} we define on $\C^4$ a function
\begin{equation}\label{4.61}
N_D=q_+Z^+\bar Z^+ = q_-Z^-\bar Z^-=z_\al\bar y^\al + y^\dal\zb_\dal\ ,
\end{equation}
where the multipliers $q_\pm =p_0/|p_0|=\pm 1$ correspond to the components of the metric in the coordinates $Z^+$ and $Z^-$. This function is positive definite on spaces $\C_\pm^4$ despite the signature $(++--)$.  This is due to the change in the sign of the complex structure on the charge conjugate space $\C_-^4$ associated with the space $H_-^3$ (negative frequency, antiparticles). The necessity of such a change in the sign of the complex structure was noted by Woodhouse \cite{Wood2}.

\noindent{\bf Positive energy.} The function $N_D$ introduced above commutes with both Hamiltonian functions $H_0$ and $H_{int}$, so it is a constant of motion. The numerical value of the function $N_D$ can be chosen equal to the numerical value of function $N_{int}$, obtaining equations
\begin{equation}\label{4.62}
N_D=2=N_{int}\quad\Rightarrow\quad z_\al^\pm = y_\pm^\dal v^\pm_{\dal\al}
\end{equation}
that coincide with equations \eqref{4.56}. In fact, the equality of functions in \eqref{4.62} means that the particles move in the intersection of two hypersurfaces with initial data parametrized by Riemannian spheres $\CPP_\pm^1\subset\C_\pm^2$.

Note that on the subspaces $\C_\pm^2$ the Hamiltonian function $N_{int}$ coincides with the function $N_D$ and it is this equivalence that defines the Bargmann-Wigner equations \eqref{4.58} and, in particular, the Dirac equation. If we remove $q_\pm$ from definition \eqref{4.61}, then in equality \eqref{4.62} it will be necessary to replace the positive definite function $N_{int}$ with the function $q_\pm N_{int}$, which will give a negative energy of the antiparticles. This corresponds to a change of sign in the metric \eqref{4.40}, that is, to considering the antiparticles as spinors from the same space $\C^4_+=\C_L^2\times\C_R^2$ as the particles. However, the antiparticles belong to the charge conjugated space $\C^4_-=\bar\C_R^2\times\bar\C_L^2$ and not to the space $\C_+^4$.

\noindent{\bf Comparison with QFT.} The Dirac equation \eqref{4.58} with $n=1$ has two solutions $\psi_+^\al$ for particles and two solutions $\psi_-^\al$ for antiparticles. In the standard approach, the charges $q_+$ and $q_-$ are ignored and $\psi_+^\al$ and $\psi_-^\al$ are assumed to belong to the same complex space $\C^4_+=\C^4$, which leads to negative energies for the antiparticles $\psi_-^\al$. In quantum field theory (QFT), the following trick is used to correct this. One considers operator-valued Grassmann variables $a_\al$ and charge conjugated variables $b_\al$. After this, the operator-valued solution of the Dirac equation is given in the form
\begin{equation}\label{4.63}
\psi = a_\al\psi_+^\al + b_\al^\+\psi_-^\al\ ,
\end{equation}
which is the column $\C^4$ with operator-valued components. Next, the anticommutation relation of operators $a_\al$ and  $b_\al$ is used to represent the negative energy of antiparticles as a finite positive energy minus the infinite ``vacuum" energy.

The described trick make sense only in QFT, because if $a_\al$ and  $b_\al$ are not operators but simply Grassmann variables, then \eqref{4.63} is not a Dirac spinor but something meaningless. That is why it is declared that the Dirac equation describes not first quantized wave functions but ``classical fields". This contradicts to all the arguments and facts presented in this paper.

\section{Relativistic quantum mechanics}

\subsection{Antiparticles and positive energy}

\noindent{\bf Quantization.} Let us sum up the preliminary results. In non-relativistic Hamiltonian mechanics we introduced spin degrees of freedom of particles as the space $\C^2$ of the fundamental representation of the group SU(2). The particles we are considering move in straight lines in $\R^3$ with constant velocity and in circles in the inner space $\C^2$ with constant angular velocity.

Quantization from the point of view of differential geometry is a transition from phase manifold $X$ to two  conjugate complex line bundles $L_\C^+$ and $L_\C^-$ over $X$. Coordinates on the manifold $X$ are mapped to ``quantum coordinates", which are covariant derivatives in bundles $L_\C^\pm$ along the coordinate directions. The connections in these bundles  are given by $A_{\sf{vac}}^\pm =\pm\im\theta_X$, where $\theta_X$ is the potential of a symplectic 2-form $\omega_X=\dd\theta_X$ on $X$. The difference in signs of the fields $A_{\sf{vac}}^\pm$ is related to the opposite signs of the charges $\qv =\pm 1$ of sections $\Psi_\pm$ of these conjugate bundles $L_\C^\pm$. Particles are sections of the bundle $L_\C^+$, and antiparticles are sections of the bundle $L_\C^-$. At the level of classical mechanics, these charges correspond  to charges $q_\pm =\pm 1$ associated  with the orientation on the particle's trajectory, parametrized by $\tau$, so that the mapping $\tau\mapsto -\tau$ corresponds to the mapping $q_\pm^{}\mapsto -q_\pm^{}$.

\noindent{\bf Antiparticles.} The map $\tau\mapsto -\tau$ is an antilinear transformation mapping all complex structures on the extended phase space and complex vector bundles over it to the conjugate complex structures. In particular, the operator $\Qv =-\im\Jv$ defined on fibres of the quantum bundle $L_{\C^2}^{}=L_\C^{+}\oplus L_\C^-$ and having eigenvalues $\qv$ is mapped at $\tau\mapsto -\tau$ into the operator $-\Qv$, which corresponds to the mapping $L_\C^\pm\to L_\C^\mp$ with $\qv =q_\pm\mapsto -\qv$. This is the mapping of particles $\Psi_+\in L_\C^{+}$ into antiparticles $\Psi_-\in L_\C^{-}$ and vice versa. Time reversal is not a Galilean transformation, therefore the introduction of antiparticles in non-relativistic quantum mechanics is not mandatory, but it seems reasonable \cite{Popov1}.

In relativistic theory, consideration of antiparticles is mandatory, since the antilinear mappings $T: x^0\mapsto -x^0$ and $PT: x^\mu\mapsto -x^\mu$ are elements of the Lorentz group O(1,3). In relativistic theory we have quantum bundles $L_\C^\pm$ over phase space $T^*\R^{1,3}$ with sections $\Psi_\pm$ having charges $\qv =\pm 1$. For free particles we identify these charges with $q_\pm =p_0/|p_0|=\pm 1$, defining the bundle $L_\C^+$ over a region $T^*_+\R^{1,3}$ with $p_0\ge m$ and the bundle $L_\C^-$ over a region $T^*_-\R^{1,3}$ with $p_0\le -m$. Let us recall that the operators $T$ and $PT$ from the Lorentz group are antilinear, they change the signs of complex  structures on the internal spaces of free particles and, as a consequence, change the signs of charges associated with these spaces. Therefore, we identify charges $\qv$ and $q_\pm$. However, in the general case, particles and antiparticles are positive  and negative  frequency functions with respect to the evolution parameter $\tau$, and not with respect to the coordinate time $x^0$.

\noindent{\bf Positive energy.} Coordinate time $x^0$ cannot be identified with the evolution parameter $\tau$, for example, in the relativistic oscillator discussed in Sect.4, eqs.\eqref{4.18}-\eqref{4.21}. Its covariant phase space is a homogeneous K\"ahler-Einstein manifold $Z_6$=PU(1,3)/U(3) introduced in \eqref{4.20}. In fact, $Z_6$ can be identified with  the unit complex 3-ball in $\C^3$ with coordinates $y^a$,
\begin{equation}\label{5.1}
Z_6\cong B_\C^3=\left\{y^a:=\frac{z^a}{z^0}, \ z^\mu =\frac{1}{\sqrt 2}(x^\mu -\frac{\im}{m\omega}
p^\mu)\in\C^{3,1}\mid \delta_{a\bar b}y^a\bar y^{\bar b}<1\right\}\ .
\end{equation}
The particle moves along the circle $S^1\sim e^{-\im\omega\tau}$ in the fibres of the bundle \eqref{4.20} parametrized by $\tau$, and the antiparticle moves in the opposite direction as $e^{\im\omega\tau}$ on this circle. The map $\tau\mapsto -\tau$ defines a map from a complex structure on $Z_6$ to conjugate, and hence a map from holomorphic functions (particles) to antiholomorphic functions (antiparticles), but the map $x^0\mapsto -x^0$ does not. The reason for this is obvious: the symmetry of the relativistic oscillator is the group U(1,3), not the Lorentz group O(1,3). Note that in the simply connected manifold \eqref{5.1} it is impossible to introduce either a position or a momentum representation  -- this illustrates the fact that in the relativistic case there is no analogue of the Stone-von Neumann theorem. 

We note again that $p^0$ is not the energy of the particle even in the free case. The fallacy of this identification was discussed using the example of a classical relativistic oscillator after formula \eqref{4.21}. This positivity of energy is preserved at the quantum level, where the energy operator $\hat E$ has eigenvalues \cite{Popov2}
\begin{equation}\label{5.2}
E(n_1, n_2, n_3) =mc^2\sqrt{1{+}\frac{2\hbar\omega}{mc^2}(n_1{+}n_2{+}n_3{+}\sfrac32)}\cong mc^2 + \hbar\omega(n_1{+}n_2{+}n_3{+}\sfrac32)\ \ \mbox{for}\ c^2\to\infty ,
\end{equation}
with $n_1, n_2, n_3=0,1,...$. Eigenfunctions of this operator $\hat E$ form a basis in the weighted Bergman space which is a reproducing kernel Hilbert space. The positiveness of energy is also preserved for particles with spin, in particular for Dirac particles and antiparticles. In the previous section we discussed this for classical particles, and in this section we will discuss the positivity of energy for first quantized particles.

\subsection{Relativistic analogue of Schr\"odinger equation}

\noindent {\bf From classical to quantum.} In considering non-relativistic particles with spin, we examined in detail the transition from their phase space $T^*\R^3\times\C^2$ to complex line bundles $L_\C^\pm$ over $T^*\R^3\times\C^2$ and the Schr\"odinger equations on sections $\Psi_\pm$ of these bundles. In the previous Sect.4 we introduced and examined a new approach to relativistic mechanics, based not on energy, but on scalar Lorentz invariant Hamiltonian functions $H$ and a scalar evolution parameter $\tau$. Functions $H_0$, $H_{int}$ and others included in the function $H$ defined on the extended  phase space of the particle have a clear mathematical and physical meaning. The mathematical meaning of the functions $H_0$, $H_{int}^L$ and $H_{int}^R$ is that they define a symplectic reduction of the spaces $T^*\R^{1,3}$, $\C_L^2$ and $\C_R^2$ by the actions of one-parameter groups generated by the Hamiltonian vector fields $V_{H_0}^{}$, $V_{H_{int}^L}^{}$ and $V_{H_{int}^R}^{}$ on these spaces. The physical meaning of these functions is that they define the internal characteristics of particles. In the case under consideration, this is the mass of the particle and its spin, or the type of representation of the Lorentz group in the more general case $s>1$. This relativistic Hamiltonian mechanics differs from non-relativistic mechanics only in that it is initially defined on non-Euclidean phase space. However, the covariant phase space of particles and antiparticles has a positive definite metric, which we showed in the previous section using the example of the Dirac particle and the relativistic oscillator.

In the approach we consider to relativistic Hamiltonian mechanics, the quantization of a relativistic system is not different from the quantization of a non-relativistic system. Therefore, our description of the relativistic case will be more brief. To move to quantum mechanics of relativistic particles with spin, we must define the bundle $L_{\C^2}^{}=L_\C^+\oplus L_\C^-$ over $T^*\R^{1,3}\times\C^4$, introduce a quantum Hamiltonian $\hat H$, and define the evolution over $\tau$ of sections $\Psi=\Psi_++\Psi_-$ of the bundle $L_{\C^2}^{}$ by the equation
\begin{equation}\label{5.3}
\Jv\dpar_\tau\Psi =\hat H\Psi\quad\Leftrightarrow\quad\pm\im\dpar_\tau\Psi_\pm =
(\hat H_0+\hat H_{int}^\pm)\Psi_\pm\ ,
\end{equation}
similar to the Schr\"odinger equation  \eqref{3.30}. All Bargmann-Wigner fields \eqref{4.52} satisfy the Klein-Gordon equations when expanding the functions $\Psi_\pm$ in the coordinates $Z_i^\pm$ on $\C_\pm^4$. The Dirac, Proca and Bargmann-Wigner equations \eqref{4.58} follow from \eqref{5.3} when restricting the Hamiltonian functions $\hat H_{int}^\pm$ and the functions $\Psi_\pm$ to the subspaces $\C_\pm^2$, $\C_\pm^2/\Z_2$ and $\C_\pm^2/\Z_n$ in $\C_\pm^4$, $\C_\pm^4/\Z_2$ and $\C_\pm^4/\Z_n$. The operators $\hat H_0$ and $\hat H_{int}$ are introduced as covariant Laplacians in the bundle $L_{\C^2}^{}= L_\C^+\oplus L_\C^-$, defined in terms of covariant derivatives in $L_{\C^2}^{}$. Equations \eqref{5.3} are invariant under the Lorentz group transformations.

Let us emphasize that $\Psi$ in \eqref{5.3} is a $\C^2$-vector and the index ``$\pm$"  of the functions $\Psi_\pm$ is a vector index denoting the components of $\Psi$ in the expansion in terms of the basis vectors $v_\pm$ in the fibres of the bundle $L_{\C^2}^{}$. All indices of the Lorentz group representations arise from the expansion of  functions  $\Psi_\pm$ into series in terms of the spin variables $Z_i^\pm\in\C_\pm^4$.  Therefore, functions $\Psi_\pm$ contain an infinite number of representations, most of which are reducible. We have already noted earlier that in this paper we do not aim to describe all possible representations; we limit ourselves to representations \eqref{4.53} covering fields \eqref{4.54} of spin $s\le 1$. We will focus our attention on the fields \eqref{4.52} with $n=0, 1$ and 2.

\noindent {\bf Bundles $L_\C^\pm$}. We should define the bundle $L_{\C^2}^{}=L_\C^+\oplus L_\C^-$ over the relativistic phase space $T^*\R^{1,3}\times\C^4$ with  a symplectic 2-form $\Omega_0$ from \eqref{4.1} on $T^*\R^{1,3}$ and the symplectic 2-form $\Omega_{int}$ from \eqref{4.38} on $\C^4$. On the fibres of the bundle $L_{\C^2}^{}$ there are bases $v_\pm$ from \eqref{3.10} and operators $\Jv$ and $\Qv$ from \eqref{3.8} and \eqref{3.14}. The covariant derivatives on the bundles $L_\C^\pm$ have the form
\begin{equation}\label{5.4}
\begin{split}
L_\C^\pm&\ \to\ T^*\R^{1,3}:\quad \nabla^{}_{x^\mu}=\frac{\dpar}{\dpar x_{\mu}}\pm\im p_\mu\ ,\quad \nabla^{}_{p_\mu}=\frac{\dpar}{\dpar p_\mu}\ ,
\\[2pt]
L_\C^\pm&\ \to\ \C_\pm^4:\quad 
\begin{array}{l} 
\nabla_{z_\al^\pm}^{} = \dpar_{z_\al^\pm}^{}+\sfrac12v^{\al\dal}\zb_\dal^\pm\ , \  
\nabla_{\zb_\dal^\pm}^{} =\dpar_{\zb_\dal^\pm}^{}-\sfrac12v^{\al\dal}z_\al^\pm\ ,\\[1pt]
\nabla_{y^\dal_\pm}^{} = \dpar_{y^\dal_\pm}^{}+\sfrac12v_{\dal\al}\bar y^\al_\pm\ , \  
\nabla_{\bar y^\al_\pm}^{} =\dpar_{\bar y^\dal_\pm}^{}-\sfrac12v_{\dal\al}y^\dal_\pm\ ,
\end{array}
\end{split}
\end{equation}
where $(x^\mu , p_\mu)\in T^*\R^{1,3}$ and $Z_i^\pm = (z_\al^\pm , y_\pm^\dal )\in\C_\pm^4$ are given in \eqref{4.37}. Recall that the components of the connections $A^\pm_{\sf vac}$ in \eqref{5.4} in the bundles $L_\C^\pm$ are obtained from the components of the potential for the symplectic 2-form $\Omega_0+\Omega_{int}$.

For the $(x^\mu , p_\mu)$-part of \eqref{5.4} we will use the momentum representation, in which the functions $\Psi_\pm$ do not depend on the coordinates $x^\mu$. When acting on such functions, the quantum Hamiltonian $\hat H_0$ coincides
with the function $H_0$ from \eqref{4.6}. For bundles \eqref{5.4} over internal spaces $\C_\pm^4$ we use the Segal-Bargmann representation \cite{Segal, Bar}, in which the ground state (vacuum) is given by the function
\begin{equation}\label{5.5}
\psi_0=\exp(-\sfrac12 N_{int})\ ,\quad N_{int}=v^{\al\dal}z_\al\zb_\dal + v_{\dal\al}y^\dal\bar y^\al\ .
\end{equation}
It is easy to see that the state \eqref{5.5} is annihilated by the covariant derivatives $\nabla_{\zb_\dal^\pm}$ and $\nabla_{\bar y^\al_\pm}$, that is, they are Dolbeault operators defining holomorphic structures in the bundles $L_\C^\pm\to \C_\pm^4$. Therefore, we choose polarized sections of the bundles $L_\C^+$ and $L_\C^-$ in the form
\begin{equation}\label{5.6}
\Psi_+=\psi_+(p, Z^+, \tau)\,\psi_0v_+\und \Psi_-=\psi_-(p, Z^-, \tau)\,\psi_0v_-\ ,
\end{equation}
where $\psi_+$ and $\psi_-$ are holomorphic functions of the coordinates $Z_i^+$ and $Z_i^-$, respectively.

\noindent {\bf Hamiltonian $\hat H_{int}$.} The covariant derivatives \eqref{5.4} on functions \eqref{5.6} take the form similar to \eqref{3.19},
\begin{equation}\label{5.7}
\nabla_{Z_i^\pm}^{}\Psi_\pm =\left(\frac{\dpar}{\dpar Z_i^\pm}\psi_\pm\right)\psi_0 v_\pm\und 
\nabla_{\bar Z_{\bar\imath}^\pm}^{}\Psi_\pm =-g^{\bar\imath j}Z_j^\pm\psi_\pm\,\psi_0v_\pm\ ,
\end{equation}
where $Z_i^\pm=(z_\al^\pm , y_\pm^\dal), \bar Z_{\bar\imath}^\pm =(\zb_\dal^\pm, \bar y_\pm^\al )$ and $(g^{i\bar\jmath})=(v^{\al\dal}, v_{\dal\al})$. Recall that we use 
\begin{equation}\label{5.8}
v^{\al\dal}=v_\mu\sigma^{\mu\al\dal}\ ,
\end{equation}
where $v^\mu =p^\mu/m$ is the velocity of particle with $v^0>0$.

The quantum Hamiltonian $\hat H_{int}=\omega\hat N_{int}$ on the bundles $L_\C^\pm\to\C_\pm^4$ is defined in terms of covariant Laplacians similarly to the non-relativistic case \eqref{3.27}-\eqref{3.29},
\begin{equation}\label{5.9}
\begin{split}
\hat N_{int}^\pm=&-\sfrac12 g^{i\bar\jmath}(\nabla_{Z_i^\pm}^{}\nabla_{\bar Z_{\bar\jmath}^\pm}^{}+\nabla_{\bar Z_{\bar\jmath}^\pm}^{}\nabla_{Z_i^\pm}^{})=-\sfrac12 v_{\al\dal}(\nabla_{z_\al^\pm}\nabla_{\zb_\dal^\pm}+
\nabla_{\zb_\dal^\pm}\nabla_{z_\al^\pm})
\\[2pt]
&-\sfrac12 v^{\dal\al}(\nabla_{y_\pm^\dal}^{}\nabla_{\bar y_\pm^\al}^{}+\nabla_{\bar y_\pm^\al}^{}\nabla_{y_\pm^\dal}^{})\ .
\end{split}
\end{equation}
On holomorphic sections \eqref{5.6}, the operators \eqref{5.9} take the form
\begin{equation}\label{5.10}
\hat N_{int}^\pm=Z_i^\pm\frac{\dpar}{\dpar Z_i^\pm} + 2= z_\al^\pm\frac{\dpar}{\dpar z_\al^\pm}+y^\dal_\pm\frac{\dpar}{\dpar y^\dal_\pm}+2\ .
\end{equation}
It is easy to see that the eigenfunctions and eigenvalues of the operator $\hat H_{int}=\omega\hat N_{int}$ are
\begin{equation}\label{5.11}
\Psi_{i_1...i_n}^\pm (n, Z^\pm) = Z_{i_1}^\pm...Z_{i_n}^\pm\psi_0 v_\pm\ ,\quad
E_{int}(n)=\omega (n+2)\ ,
\end{equation}
where $\psi_0$ is the ground state \eqref{5.5}.

\noindent {\bf Solutions.} The eigenfunctions \eqref{5.11} correspond to the following solutions of equations \eqref{5.3}:
\begin{equation}\label{5.12}
\Psi_\pm(n)=e^{\mp\im m\tau\mp\im\omega (n+2)\tau}\psi_\pm^{i_1...i_n}(p)Z_{i_1}^\pm...Z_{i_n}^\pm\psi_0v_\pm\ ,
\end{equation}
where  functions $\psi_\pm^{i_1...i_n}(p)$ satisfy the Klein-Gordon equation
\begin{equation}\label{5.13}
(\eta^{\mu\nu}p_\mu p_\nu -m^2)\psi_\pm^{i_1...i_n}(p)=0\for q_\pm = p_0/|p_0|=\pm 1\ ,
\end{equation}
with $\psi_+^{i_1...i_n}$ defined on $H_+^3$ and $\psi_-^{i_1...i_n}$ defined on $H_-^3$. Solutions of the form \eqref{5.12} with $n=0, 1,...$  exhaust all solutions of equations \eqref{5.3}. Solution \eqref{5.12} can be rewritten as
\begin{equation}\label{5.14}
\Psi_\pm(n)=e^{-\im m\tau}\psi_\pm^{i_1...i_n}(p)Z_{i_1}^\pm(\tau)...Z_{i_n}^\pm (\tau)\,v_\pm^c(\tau)\ ,
\end{equation}
where $Z_i^\pm(\tau)=\exp(\mp\im\omega\tau)Z_i^\pm(0)$ corresponds to the motion of a classical particle in the space $\C_\pm^4$ and 
\begin{equation}\label{5.15}
v_\pm^c(\tau)=e^{\mp 2\im\omega\tau}\psi_0 v_\pm
\end{equation}
are rotating bases in the fibres of the bundles $L_\C^\pm$. This rotation \eqref{5.15} does not depend on the number $n$, therefore fields with spin $s=\sfrac12n=0$ should also be described by formulae \eqref{5.12}-\eqref{5.15}. Solutions \eqref{5.12} are exactly the Bargmann-Wigner (BW) fields we discussed in \eqref{4.52}-\eqref{4.58}.

\subsection{Klein-Gordon, Dirac and Proca equations}

\noindent {\bf BW fields with $n=0,1,2$.}  In what follows, we will be interested in solutions of the Schr\"odinger equations \eqref{5.3} with $n=0, 1$ and 2,
\begin{equation}\label{5.16}
\begin{split}
\Psi_{KG}&=\psi_+(p)v_++\psi_-(p)v_-\ ,
\\[1pt]
\Psi_{D}&=\bigl(\psi_+^i(p)\bigr)v_++\bigl(\psi_-^i(p)\bigr)v_-=\begin{pmatrix}\mu_+^\al\\\lambda_\dal^+\end{pmatrix}v_+ +\begin{pmatrix}\mu_-^\al\\\lambda_\dal^-\end{pmatrix}v_-\ ,
\\[1pt]
\Psi_{Pr}&=
\begin{pmatrix}\psi_+^{\al\beta}&\psi^{\ \al}_{+\dot\beta}\\\psi^{\ \beta}_{+\dal}&\psi_{+\dal\dot\beta}\end{pmatrix}v_+ +
\begin{pmatrix}\psi_-^{\al\beta}&\psi^{\ \al}_{-\dot\beta}\\\psi^{\ \beta}_{-\dal}&\psi_{-\dal\dot\beta}\end{pmatrix}v_-\ .
\end{split}
\end{equation}
They correspond to representations of the Lorentz group of type (0,0), ($\sfrac12$,0)$\oplus$(0,$\sfrac12$) and (1,0)$\oplus$($\sfrac12$, $\sfrac12$)$\oplus$(0,1) describing the Klein-Gordon, Dirac and Proca fields.

All fields \eqref{5.16} satisfy the Klein-Gordon equation \eqref{5.13}, which in Wigner's approach is the condition of mass irreducibility \cite{Wigner, BW, BR}. For Dirac and Proca fields (massive vector fields) this condition is not sufficient. It is also necessary to specify conditions for the irreducibility of the spin since fields \eqref{5.16} with $n=1$ and $n=2$ contain more components \eqref{4.54} than are necessary to describe spin $s=\sfrac12$ and $s=1$. We have already discussed these conditions in Section 4.4. We have shown that it is necessary to specify a Lorentz covariant projection
\begin{equation}\label{5.17}
\C_\pm^4/\Z_n\longrightarrow\C_\pm^2(p)/\Z_n\ ,
\end{equation}
which is equivalent to imposing the constraints \eqref{4.57} on the coordinates $Z_i^\pm$. In this case, substituting these constraints -- the restriction from $\C_\pm^4/\Z_n$ to $\C_\pm^2/\Z_n$ -- gives the Bargmann-Wigner equations \eqref{4.58}.

Recall that constraints \eqref{4.57} arise from the equality \eqref{4.62} of the functions $N_{int}$ and $N_D$. The explicit form of solutions \eqref{5.12} arises from the explicit form of eigenfunctions \eqref{5.11} of the quantum operators $\hat N_{int}^\pm$. Accordingly, equality \eqref{4.62} goes over to the fact that these functions \eqref{5.11} will also be eigenfunctions of the quantum operators $\hat N_{D}^\pm$ with the same eigenvalues. In what follows, we will introduce these operators, which, as expected, will be the Dirac operators.

\noindent {\bf  Klein-Gordon fields.} Let us consider the Schr\"odinger equations \eqref{5.3} for fields of zero spin, depending on the coordinates $Z_i^\pm$ only through the function $\psi_0$ as in \eqref{5.12}, and return to the position representation. In this case, equations \eqref{5.3} have the form
\begin{equation}\label{5.18}
\pm\im\dpar_\tau\psi_\pm =\sfrac1m\eta^{\mu\nu}\dpar_\mu\dpar_\nu\psi_\pm\ .
\end{equation}
It is easy to show that from \eqref{5.18} follow two continuity equations,
\begin{equation}\label{5.19}
\dpar_\tau\rho_\pm +\dpar_\mu j^\mu_\pm=0\ ,
\end{equation}
where
\begin{equation}\label{5.20}
\rho_\pm =\pm\psi_\pm^*\psi_\pm\und    j^\mu_\pm=\frac{1}{\im m}(\psi_\pm^*\dpar^\mu\psi_\pm - \psi_\pm \dpar^\mu\psi_\pm^*)   \ .
\end{equation}
Here $\rho_\pm$ are densities of quantum charges $\qv =\pm 1$ associated with the bundles $L_\C^\pm$. From \eqref{5.18}-\eqref{5.20} it follows that for free particles we have $\dpar_\mu j_\pm^\mu =0$ and $\dpar_\tau\rho_\pm =0$. Moreover, $j_\pm^0$ is proportional to $\rho_\pm$ and therefore also specify the density of quantum charge $\qv$ associated with quantum particles $(\qv=1)$ and antiparticles $(\qv=-1)$.

Thus, the approach we are developing to relativistic Hamiltonian mechanics and its quantization shows that $j_\pm^0$ is the quantum charge density, not the probability density. Note that this interpretation of $j_\pm^0$ as the quantum charge density  was given by Bjorken and Drell in their book \cite{BDrell}, where they write after eq.(12.63): ``More generally, we identify the quanta with positive eigenvalues of $Q$ as the particles and those with negative eigenvalues as the antiparticles...".  It is obvious that probability densities are given by functions $|\rho_\pm|=\psi_\pm^*\psi_\pm$ which, like $\rho_\pm$, do not depend on $\tau$.

\noindent {\bf  Dirac fields.} Previously we discussed in detail the geometric meaning of Dirac constraints \eqref{4.57}, \eqref{4.58} and \eqref{4.62}. Once again, a non-relativistic spin-$\sfrac12$ particle moves in Pauli phase spaces
\begin{equation}\label{5.21}
E_P^\pm =\C_\pm^2\times T^*\R^3\ \stackrel{\C_\pm^2}{\longrightarrow}\ T^*\R^3\ ,
\end{equation}
and a relativistic spin-$\sfrac12$ particle moves in Dirac phase manifolds
\begin{equation}\label{5.22}
E_D^\pm\ \stackrel{\C_\pm^2}{\longrightarrow}\ T^*H_\pm^3\ .
\end{equation}
Here $E_P^\pm$ are trivial complex vector bundles over $T^*\R^3$ and $E_D^\pm$ are complex vector bundles over $T^*H_\pm^3$, where fibres $\C_\pm^2(p)$ depend on the momentum $p_\mu\in H_\pm^3$. The Dirac constraints project the direct product $\C_\pm^4\times T^*H_\pm^3$ onto the bundle $E_D^\pm$. We see that the dimensions of the phase manifolds $E_P^\pm$ and $E_D^\pm$ are the same, the difference being that geometry \eqref{5.22} is mapped onto itself under the action of Lorentz group. 

After these clarifications, it only remains to write out the explicit form of the Dirac operator $\hat N_D$ that carries out the projection in the space of functions on $T_\pm^*\R^{1,3}\times\C_\pm^4$. According to general rules, this operator has the form
\begin{equation}\label{5.23}
\begin{split}
\hat N_D^\pm&=q_\pm(\nabla_{y_\pm^\dal}^{}\nabla_{\zb^\pm_\dal}^{}+\nabla_{\zb^\pm_\dal}^{}\nabla_{y_\pm^\dal}^{})=\{{\rm when\ acting\ on}\ \Psi_\pm\ {\rm from\ (5.6)}\}
\\[1pt]
&= v_\pm^{\al\dal}z^\pm_\al\frac{\dpar}{\dpar y_\pm^\dal} +  v^\pm_{\dal\al}y_\pm^\dal\frac{\dpar}{\dpar z^\pm_\al}=Z^\pm v_\mu^\pm\ga^\mu\frac{\dpar}{\dpar Z^\pm}\ ,
\end{split}
\end{equation}
where $Z^\pm =(Z_i^\pm)=(z_\al^\pm , y_\pm^\dal)$. Using \eqref{5.23} we obtain the constraint equations
\begin{equation}\label{5.24}
\hat N_D^\pm(\psi_\pm^i Z_i^\pm)=\hat N_{int}^\pm(\psi_\pm^i Z_i^\pm)=\psi_\pm^iZ_i^\pm\quad
\Rightarrow\ p_\mu\ga^\mu\psi_\pm\mp m\psi_\pm =0,\quad \psi_\pm = (\psi_\pm^i),\ p^0>0\ ,
\end{equation}
where $\psi_\pm$ is given in \eqref{5.16}. Thus, the Dirac equations \eqref{5.24} follow from the Schr\"odinger equation \eqref{5.3} after restricting the operator $\hat N_{int}$ to the subspaces $\C_\pm^2(p)$ in $\C^4_\pm$.

Let us now consider the quantum Dirac particles  \eqref{5.24} and their connection with quantum Pauli particles  
\eqref{3.54}-\eqref{3.56} from a geometric point of view. For the geometric introduction of quantum particles of spin $s=\sfrac12$, the spaces $\C_\pm^2$ in the fibres of bundles  \eqref{5.21} and \eqref{5.22} should be replaced by their projectivizations  $\CPP_\pm^1$, obtaining covariant phase manifolds of relativistic and non-relativistic particles of spin $s=\sfrac12$ of the form
\begin{equation}\label{5.25}
P(E_D^\pm)\ \stackrel{\CPP_\pm^1}{\longrightarrow}\ T^*H_\pm^3\quad\stackrel{c\to\infty}{\Longrightarrow}\quad
P(E_P^\pm)\ \stackrel{\CPP_\pm^1}{\longrightarrow}\ T^*\R^3\ .
\end{equation}
Here we emphasize that in the non-relativistic limit of infinite speed of light $c$ the manifolds $P(E_D^\pm)$ becomes $P(E_P^\pm)$. The transition to quantum particles is accomplished by introducing complex line bundles $\CL_\C^\pm(1)$ over $P(E_D^\pm)$ and $L_\C^\pm(1)$ over $P(E_P^\pm)$ such that their restrictions to the fibres $\CPP_\pm^1$ of bundles \eqref{5.25} are hyperplane bundles $\CO_\pm(1)$,
\begin{equation}\label{5.26}
\CL_\C^\pm(1)\ \longrightarrow\ P(E_D^\pm)\quad\stackrel{c\to\infty}{\Longrightarrow}\quad
L_\C^\pm(1)\ \longrightarrow\ P(E_P^\pm)\ .
\end{equation}
In the limit $c\to\infty$, the bundles  $\CL_\C^\pm(1)$ become the bundles $L_\C^\pm(1)$ and they can be denoted by the same letter. Quantum particles with $\qv=\pm 1$ are holomorphic sections of the bundles $\CL_\C^\pm(1)$ (relativistic) and 
$L_\C^\pm(1)$ (non-relativistic). By construction, they satisfy equations  \eqref{5.24} and  \eqref{3.63}, \eqref{3.64}, respectively.

Note that the positive definite Hermitian inner product on fibres of the bundles  \eqref{5.21} is given by the non-relativistic Hamiltonian function $H_{int}$ from  \eqref{2.28}. Similarly, the positive definite relativistic Hamiltonian function $H_{int}$ from  \eqref{4.42} gives a positive definite metric on the fibres of the bundles  \eqref{5.22}. This leads to the positive definiteness of the function $N_D$ from  \eqref{4.61} for both particles and antiparticles. The positive definiteness of the relativistic case is dictated by the correspondence principle given by the limit $c\to\infty$ in  \eqref{5.25}. This positive definitness holds at the quantum level and leads to the change in the definitions of the inner product  \eqref{3.66} and the currents  \eqref{3.67} discussed in Section 3.5. With these new definitions, charge conjugation does not change the sign of the energy, but it does change the signs of the charges, as it should.

\noindent {\bf  Massive vector  fields.} Recall that quantum nonrelativistic particles of spin $s=1$ have three complex components and are described by homogeneous polynomials of degree two on the space $\C_+^2=\C^2$, which is equivalent to the fact that they are given by sections of the bundle $\CO(2)$ over $\CPP_+^1\hra\C_+^2$. For antiparticles everything is similar but with $\CPP^1_-\hra\C^2_-=\bar\C^2$ instead of $\CPP^1_+\hra\C^2_+$. Accordingly their classical prototypes are points moving in $\C^2_+/\Z_2$ and $\C^2_-/\Z_2$.

In the relativistic case, particles with $q_\pm=\pm 1$ and $n=2$ move in spaces $\C^4_\pm/\Z_2$ and are therefore parametrized by the 10 components written out in formulae \eqref{4.54} and \eqref{5.16}. We have shown earlier that the fields $\Psi_{Pr}$ in \eqref{5.16} satisfy the Bargmann-Wigner equations \eqref{4.58} with $n=2$. It is well known that these equations are equivalent to the Proca equations, which we will now show.

For the fields $\Psi_{Pr}$ from \eqref{5.16}, the Bargmann-Wigner equations have the form
\begin{equation}\label{5.27}
\begin{pmatrix}0&p_\pm^{\al\dal}\\p^\pm_{\dal\al}&0\end{pmatrix}
\begin{pmatrix}\psi_\pm^{\al\beta}&{\psi_\pm}^{\al}_{\dot\beta}\\{\psi_\pm}^{\beta}_\dal&\psi^\pm_{\dal\dot\beta}\end{pmatrix} -m\begin{pmatrix}\psi_\pm^{\al\beta}&{\psi_\pm}^{\al}_{\dot\beta}\\{\psi_\pm}^{\beta}_\dal&\psi^\pm_{\dal\dot\beta}\end{pmatrix}=0\ ,
\end{equation}
plus we must symmetrize them as in \eqref{4.58}. It is easy to show that they are equivalent to the Proca equations
\begin{equation}\label{5.28}
(\eta^{\lambda\sigma}p_\lambda p_\sigma -m^2)B_\mu^\pm =0\ ,\quad \eta^{\mu\nu}p_\mu B_\nu^\pm =0\ ,
\end{equation}
where
\begin{equation}\label{5.29}
B^\pm =B^\pm_\mu\ga^\mu =\begin{pmatrix}0&B_\pm^{\al\dal}\\B^\pm_{\dal\al}&0\end{pmatrix}=
\begin{pmatrix}0&\sfrac1m\psi_\pm^{\al\dal}\\\sfrac1m\psi^\pm_{\dal\al}&0\end{pmatrix}
\end{equation}
and 
\begin{equation}\label{5.30}
\psi_\pm^{\al\beta}=q_\pm p^{\al\dal}B_{\dal\ga}^\pm\veps^{\beta\ga}\ ,\quad
\psi^\pm_{\dal\dot\beta}=q_\pm p_{\dal\al}B^{\al\dot\ga}_\pm\veps_{\dot\beta\dot\ga}\ .
\end{equation}
The Lorenz conditions in \eqref{5.28} follow from the requirement of symmetry with respect to indices $i_1, i_2$ in \eqref{4.58} and remove the extra component from $B_\mu^\pm$. This is the component {\bf 1} in the decomposition $(\sfrac12, \sfrac12)= {\bf 3}\oplus {\bf 1}$, when the group $\sSL(2,\C)$ is reduced to SU(2). The fields $B_c^\pm:=(-\im\ga^2)(B^\pm)^*\im\ga^2=(B_\mu^\pm)^*\ga^\mu$  are charge conjugate for the complex fields  \eqref{5.29}. The Proca field is called neutral (i.e. the particle coincides with antiparticle), if $B^-=B_c^+$. Then the field $B_\mu =B_\mu^++B_\mu^-$ is real.

\section{Quantum photons}

\subsection{Classical and quantum massless particles}

\noindent {\bf  Classical massless particles.} We considered massive spinning relativistic particles as points moving in phase space $T^*\R^{1,3}\times\C_L^2\times\C_R^2$. They have a velocity $v^\mu=p^\mu/m$ satisfying equation $\eta_{\mu\nu}v^\mu v^\nu =1$ and move in straight lines in $\R^{1,3}$ and in circles in spaces $\C_L^2$ and $\C_R^2$.
Massless particles also move in straight lines in spaces $\R^{1,3}$,
\begin{equation}\label{6.1}
x^\mu (\tau)=x^\mu + k^\mu\tau\ ,
\end{equation}
with a constant light-like velocity vector $k^\mu\in\R^{1,3}$ lying on a cone 
\begin{equation}\label{6.2}
\eta_{\mu\nu}k^\mu k^\nu =0\ .
\end{equation}
We will use the dimensionless velocity vector $k^\mu$ instead of the momentum 4-vector. The situation with spin degrees of freedom is more complicated than for the case of massive particles. This is due to zero length of the vector $k^\mu$.

Moving to spinor notation as in equations \eqref{4.24}-\eqref{4.28}, we obtain
\begin{equation}\label{6.3}
k^{\al\dal}=k_\mu\sigma^{\mu\al\dal},\quad \det(k^{\al\dal})=k_\mu k^\mu=0\ .
\end{equation}
Since the determinant of the matrix \eqref{6.3} is zero, there exist 2-component  spinors $k^\al$ and $k^\dal$ such that
\begin{equation}\label{6.4}
k^{\al\dal}=k^\al k^\dal\ .
\end{equation}
Due to the degeneracy of the Hermitian matrix  \eqref{6.4}, the symplectic two-form  \eqref{4.38} and the metric \eqref{4.41} become degenerate, which requires a modification of the relativistic Hamiltonian mechanics of partcles with spin that is presented in this paper. This is beyond the scope of this paper and will be discussed separately. However, at the first quantized level, the transition to massless particles is not difficult and we obtain the massless Klein-Gordon fields, two Weyl spinors as a massless Dirac spinor, and Maxwell fields as massless Proca fields.

\noindent {\bf  First quantized photons.} Note that the factor $1/m$ is introduced in the definition \eqref{5.29} to be able to consider zero mass limit in equations \eqref{5.28}-\eqref{5.30} with rescaled fields $\psi^\al_{\dot\beta}$. In this case for real fields $A_\mu = A_\mu^++A_\mu^-=\mathop{\lim\limits_{m\to 0}}B_\mu$ with $B_\mu^-=(B_\mu^+)^*$ we obtain the Maxwell equations
\begin{equation}\label{6.5}
\eta^{\lambda\sigma}k_\lambda k_\sigma A_\mu =0\und\eta^{\mu\nu}k_\mu A_\nu =0\ ,
\end{equation}
where $k^\mu$ is the null vector \eqref{6.2}-\eqref{6.4}. Earlier in this paper we showed that Proca's equations \eqref{5.28} for massive vector fields follow from the relativistic analogue of the Schr\"odinger equations \eqref{5.3} on
the extended phase space $T^*\R^{1,3}\times\C_L^2\times\C_R^2$. Therefore, Maxwell's equations \eqref{6.5} are the equations of the first quantized theory of massless particles of spin $s=1$. However, the real solutions of equations \eqref{6.5} are not photons.

We will define photons as the complexification of light-like velocities/momenta \eqref{6.2}-\eqref{6.4} and then will justify this definition. There are two possible complexifications of a real vector $k_\mu$ with a Hermitian matrix \eqref{6.4}, where $k^\dal =(k^\al)^*=\overline{k^\al}$. Namely, we have
\begin{equation}\label{6.6}
\begin{split}
\mbox{left-handed\ photons:}&\quad k^\al k^\dal\mapsto\psi_L\, \mu^\al k^\dal\ ,
\\[1pt]
\mbox{right-handed\ photons:}&\quad k^\al k^\dal\mapsto\psi_R\, k^\al \lambda^\dal\ ,
\end{split}
\end{equation}
where $\psi_L(k)$ and $\psi_R(k)$ are the wave functions of left- and right-circularly polarized photons. Antiphotons are given by complex conjugate vectors $\psi_L^*\,\bar\mu^\dal\, k^\al$ and $\psi_R^*\,k^\dal\,\bar\lambda^\al$.

Definition \eqref{6.6} is introduced for the following reasons. Penrose proposed to consider the complexified self-dual and anti-self-dual solutions $F^+$ and $F^-$ of Maxwell's equations as wave functions of photons \cite{Penrose1, Penrose2}. Note that if the 2-form $F=F^++F^-$ is real, then the fields $F^+$ and $F^-$ are complex conjugate to each other and therefore the condition $F^+=0$ or $F^-=0$ leads to $F=0$. That is why Penrose proposed to consider independent complex fields $F^-$ and $F^+$ to describe photons. The fields
$F^-$ and $F^+$ realize representations of the Lorentz group of types (1,0) and (0,1), so Penrose's definition diverges from the approach of quantum field theory (QFT). In QFT, a gauge potential $A=A_\mu\dd x^\mu$ with $F=\dd A$, corresponding to the representation of the Lorentz group of type $(\sfrac12, \sfrac12)$, is used to introduce photons. Our definition \eqref{6.6} brings Penrose's definition into line with the definition of QFT, since the space of right spinors $\lambda^\dal\in\C_R^2$ is associated with self-duality, and the space of left spinors $\mu^\al\in\C_L^2$ is associated with anti-self-duality, as follows from formulae \eqref{4.34} and \eqref{4.35}.

\subsection{Photons in position representation}

\noindent Let us introduce two complex isotropic vectors,
\begin{equation}\label{6.7}
A_\mu^L:\ \eta^{\mu\nu}A_\mu^LA_\nu^L=0\und
A_\mu^R:\ \eta^{\mu\nu}A_\mu^RA_\nu^R=0\ ,
\end{equation}
depending either on $k_\mu$ (momentum representation) or on $x^\mu$ (position representation). We will associate them with matrices
\begin{equation}\label{6.8}
\begin{split}
&\CA^L:=A_\mu^L\ga^\mu\Pi_L=\begin{pmatrix}0&0\\A^L&0\end{pmatrix}\ ,\quad 
\CA^R:=A_\mu^R\ga^\mu\Pi_R=\begin{pmatrix}0&A^R\\0&0\end{pmatrix}\ ,
\\[1pt]
&\CA:=\CA^L+\CA^R=\begin{pmatrix}0&A^R\\A^L&0\end{pmatrix}\und\CA_c:=(-\im\ga^2)\CA^*\im\ga^2=
\begin{pmatrix}0&\bar A^L\\\bar A^R&0\end{pmatrix}\ ,
\end{split}
\end{equation}
where
\begin{equation}\label{6.9}
A^R=A_\mu^R\sigma^\mu , \quad A^L=A_\mu^L\bar\sigma^\mu , \quad
\bar A^R=\bar A_\mu^R\bar\sigma^\mu\und\bar A^L=\bar A_\mu^L\sigma^\mu\ ,
\end{equation}
and the projectors $\Pi_L$ and $\Pi_R$ are given in \eqref{4.31}. In \eqref{6.8} the field $\CA_c$ is charge conjugate to the field $\CA$. Field $\CA$ defines photons, and field $\CA_c$ defines antiphotons. The bar over the vectors $\bar A_\mu^R$ and $\bar A_\mu^L$ in \eqref{6.9} denotes complex conjugation.

For the matrix field $\CA$ from \eqref{6.8}, we introduce a Dirac-type equation in position representation
\begin{equation}\label{6.10}
\ga^\mu\dpar_\mu\CA =0\ ,
\end{equation}
from which it follows that all components of this matrix satisfy the massless Klein-Gordon equation. Equation \eqref{6.10} is equivalent to the equations
\begin{equation}\label{6.11}
\begin{split}
\sigma^\mu\dpar_\mu A^L=0&\ \Rightarrow\ \eta^{\lambda\sigma}\dpar_\lambda\dpar_\sigma A_\mu^L=0,\ 
\eta^{\mu\nu}\dpar_\mu A_\nu^L=0\und\sigma^{\mu\nu}\dpar_\mu A_\nu^L=0   \ ,
\\[1pt]
\bar\sigma^\mu\dpar_\mu A^R=0&\ \Rightarrow\ \eta^{\lambda\sigma}\dpar_\lambda\dpar_\sigma A_\mu^R=0,\ 
\eta^{\mu\nu}\dpar_\mu A_\nu^R=0\und\bar\sigma^{\mu\nu}\dpar_\mu A_\nu^R=0 \ ,
\end{split}
\end{equation}
where the matrices $\sigma^{\mu\nu}$ and $\bar\sigma^{\mu\nu}$ are given in \eqref{4.34}.

The first two equations in \eqref{6.11} are Maxwell's equations \eqref{6.5} in the Lorenz gauge both for the field $A_\mu^R$ and the field $A_\mu^L$. The last equations in \eqref{6.11} mean that the complex gauge field $F_{\mu\nu}^R=\dpar_\mu A_\nu^R -\dpar_\nu A_\mu^R$ is self-dual, and the field $F_{\mu\nu}^L=\dpar_\mu A_\nu^L -\dpar_\nu A_\mu^L$ is anti-self-dual. Therefore, the solutions of equation \eqref{6.10} define complex self-dual and anti-self-dual solutions of Maxwell's equations. Thus, the complex vector fields $A_\mu^L$ and $A_\mu^R$ introduced by us are photons in the Penrose sense, but defined in terms of gauge potentials in accordance with QFT. Note that in the case of massless fields we should not speak about spin, but about helicity, which in the case under consideration coincides with chirality. The self-dual field $F^R$ has helicity $h=1$, and the anti-self-dual field $F^L$ has helicity $h=-1$.

\subsection{Photons in momentum representation}

\noindent Let us return to the momentum representation and write the fields $A^L$ and $A^R$ in spinor notation,
\begin{equation}\label{6.12}
A^L_{\dal\al}=A_\mu^L\bar\sigma^\mu_{\dal\al}\und
A^{R\al\dal}=A_\mu^R\sigma^{\mu\al\dal}\ .
\end{equation}
Equations \eqref{6.11} have the form
\begin{equation}\label{6.13}
k^{\al\dal}A^L_{\dal\beta}=0\und
k_{\dal\al}A^{R\al\dot\beta}=0\ .
\end{equation}
Recall that $k^{\al\dal}=k^{\al}k^{\dal}$, where $k^\al$ and $k^\dal$ are massless Weyl spinors.

We introduce massive Dirac spinors
\begin{equation}\label{6.14}
\Psi_+=\begin{pmatrix}\mu^\al\\\lambda_\dal\end{pmatrix}\und 
\Psi_-=\Psi_+^c=(-\im\ga^2)\Psi_+^*=\begin{pmatrix}\bar\lambda^\al\\\bar\mu_\dal\end{pmatrix}
\end{equation}
satisfying the Dirac equations
\begin{equation}\label{6.15}
(v_\mu\ga^\mu\mp 1)\Psi_\pm=0\ \Rightarrow\ \mu^\al=v^{\al\dal}\lambda_\dal\ \Rightarrow\ \mu^\al\bar\lambda_\al=v^{\al\dal}\bar\lambda_\al\lambda_\dal =1\ .
\end{equation}
Mass can be introduced as a multiplies $m$ in front of the velocity $v^\mu$ and its numerical value is arbitrary. We will define the velocity vector $v_\mu$ through spinors $\mu^\al , \lambda_\dal$ using formula
\begin{equation}\label{6.16}
v^{\al\dal}:=\mu^\al\bar\mu^\dal + \bar\lambda^\al\lambda^\dal \ \Rightarrow\  
v_{\al\dal}v^{\dal\beta}=\delta_\al^\beta\ \Leftrightarrow\ v_\mu v^\mu=1\ ,
\end{equation}
where we used the equalities $\mu^\al\bar\lambda_\al =\bar\mu^\dal\lambda_\dal =1$.
Recall that the Dirac equation contains twelve real degrees of freedom and their solutions are parametrized by eight real degrees of freedom. The solution can be parametrized either by the momentum $p_\mu =mv_\mu\in\R^{1,3}$ and one complex spinor $\lambda_\dal$ or by two complex spinors $\mu^\al, \lambda_\dal$, and then the momentum is expressed through them using formula \eqref{6.16}.

Using Dirac spinors \eqref{6.14}, solutions of equations \eqref{6.13} can be written as
\begin{equation}\label{6.17}
A_{\al\dal}^L=\psi_L(k)\mu_\al k_\dal\und A_{\al\dal}^R=\psi_R(k)k_\al \lambda_\dal\ ,
\end{equation}
proving the assertion \eqref{6.6} that photons are complexified light-like momenta. The momentum-dependent complex functions $\psi_L$ and $\psi_R$ are the wave functions of left- and right-polarized photons. Antiphotons are defined by complex conjugate vectors
\begin{equation}\label{6.18}
\bar A_{\al\dal}^R=\psi_R^*(k)\bar\lambda_\al k_\dal\und 
\bar A_{\al\dal}^L=\psi_L^*(k)k_\al\bar\mu_\dal\ ,
\end{equation}
which reflects zero quantum charge. The standard scalar product of fields $\bar A^L$ with $A^L$ and $\bar A^R$ with $A^R$ after normalization gives $\bar A_\mu^L A^{L\mu}=\psi_L^*\psi_L$ and $\bar A_\mu^R A^{R\mu}=\psi_R^*\psi_R$ and these expressions can be interpreted as probability densities. Thus, equations \eqref{6.7}-\eqref{6.15} define the relativistic quantum mechanics of photons.

From formulae \eqref{6.14}-\eqref{6.18} we see that the photon fields are composed of massless momentum spinors $(k^\al,k^\dal)$ and massive spinors $\Psi_+$ and $\Psi_-$, which may be interpreted as an electron and a positron. The explicit form of wave functions \eqref{6.17} can explain the processes of creation and annihilation of photons and electron-positron pairs. In addition, photons are complex-valued light-like momenta and the exchange of photons is an exchange of momenta. This can be seen especially clearly in the example of charged Weyl spinors interacting with the Maxwell fields,
\begin{equation}\label{6.19}
\begin{pmatrix}0&k^\al k^\dal\\k_\dal k_\al&0\end{pmatrix}
\begin{pmatrix}\psi^\al\\\chi_\dal\end{pmatrix}=0\quad\Rightarrow\quad
\begin{pmatrix}0&k^\al(k^\dal +\psi_R\lambda^\dal)\\k_\dal(k_\al+\psi_L\mu_\al)&0\end{pmatrix}
\begin{pmatrix}\tilde\psi^\al\\\tilde\chi_\dal\end{pmatrix}=0\ ,
\end{equation}
where the addition of photons changes the solution $(\psi^\al,\chi_\dal)=(k^\al,k_\dal)$ to the solution
\begin{equation}\label{6.20}
\tilde\psi^\al =k^\al + \psi_L^{}\mu^\al\und \tilde\chi_\dal =k_\dal +\psi_R^{}\lambda_\dal\ .
\end{equation}
In conclusion, we note that our discussion of quantum mechanics of photons is incomplete. A more detailed discussion requires a separate paper.

\section{Summary}

\noindent The aim of this paper was to consistently translate the concepts of quantum mechanics from the language of functional analysis (Hilbert spaces, operators, etc.) into the language of differential geometry. Such a translation was initiated in the geometric quantization approach \cite{Sour}-\cite{Wood}, where it was shown that the space of wave functions of non-relativistic quantum mechanics is the space of polarized sections of a complex line bundle over phase space. In this paper we have considered
\begin{itemize}
\item
a new approach to relativistic Hamiltonian mechanics and relativistic quantum mechanics,
\item
a new description of spin and antiparticles at the classical and quantum level,
\item
a description of first quantized photons as  complexified light-like momenta.
\end{itemize}
In the remainder of this section, we will provide a concise overview of the main poits of the paper.

\medskip

\noindent{\bf Charge conjugation.} In classical Hamiltonian mechanics, a particle is a point moving in phase space $X$ along a trajectory $x(\tau)\in X$ parametrized by $\tau\in\R$. We use this definition also for the relativistic case, when the particle moves in an extended phase space with internal degrees of freedom and a parameter $\tau$ on the trajectory. We define the
particle-antiparticle mapping as the mapping $\tau\mapsto -\tau$, which corresponds to a reversal of the particle's trajectory orientation. The map $\tau\mapsto -\tau$ is antilinear, it maps complex structures on the extended phase manifold $X$ and on any vector bundle over $X$ to conjugate complex structures, and holomorphic functions to antiholomorphic ones. Thus, the mapping $\tau\mapsto -\tau$ actually defines a charge conjugation operation $C$. We emphasize that $\tau$ is a scalar parameter not related to the coordinate time $x^0$. Note that the mapping $PT\!: T^*H_+^3\to T^*H_-^3$ at the classical and quantum levels  corresponds to the transition from a particle to an antiparticle and hence must be accompanied by the charge conjugation $C$. Therefore, the charge $q_\pm =\pm1$, which distinguishes particles from antiparticles, is included in all formulae and this eliminates non-physical states both at the classical and quantum levels.

\medskip

\noindent{\bf Intrinsic angular momentum.} It is usually assumed that the classical spin is a vector $S_a$ of fixed length in 3-dimensional space $\R^3$, which therefore defines a 2-dimensional sphere $\CPP^1\subset\R^3$ (see e.g. \cite{Sni}). Quantizing the phase space $\CPP^1$ yields the quantum spin space $\C^{n+1}$ and spin is a half integer number $s=\sfrac12 n$ parametrizing this representation of the group SU(2). In this description, nothing rotates anywhere, so it is usually asserted that spin is not related to the physical rotation of a particle, but rather it is an inherent quantum characteristic.

Note that in this scheme the spin vector $S_a$ is considered as a fundamental variable parametrizing the dual space $\rsu(2)^*{=}\R^3$ of the algebra su(2), and the sphere $\CPP^1$ is the coadjoint orbit of the group SU(2) acting in su(2)$^*$. In this case, on the space $\R^3$ one defines a degenerate two-form
\begin{equation}\label{7.1}
\omega_{\R^3}^{}=-\frac{s}{R^3}\,\veps_{abc}\,S^a\dd S^b\wedge\dd S^c\ \for\  R^2:=\delta_{ab}S^aS^b\ ,
\end{equation}
which induces on $\CPP^1$ a symplectic two-form
\begin{equation}\label{7.2}
\omega_{\CPP^1}^{}=\im 2s\frac{\dd z\wedge\dd\zb}{(1+z\zb)^2}\ \for\  z=\frac{S_1-\im S_2}{R+S_3}\ .
\end{equation}
Note that if $2s=n=1,2,...$  then the field $F_{\R^3}^{}=\im\omega_{\R^3}^{}$ is the field of the $n$-monopole solution of Yang-Mills equations on $\R^3$, inducing the curvature $F_{\CPP^1}^{}=\im\omega_{\CPP^1}^{}$ in the $n$-monopole bundle
\begin{equation}\label{7.3}
S^3/\Z_n\to\CPP^1
\end{equation}
discussed in \eqref{2.38} and this $F_{\CPP^1}^{}$ coincides with the curvature \eqref{2.52} in the associated bundle $\CO(-n)$ from \eqref{2.51}. Therefore, the assertion that $2s$ is any real number at the classical level, becoming discrete only at the quantum level, is not convincing. As soon as we assume that the particle moves in a space containing the sphere $\CPP^1$, the question arises of what is the geometry of this enveloping space. Formulae \eqref{7.1}-\eqref{7.3} indicate that this is the lens space \eqref{7.3} embedded in the orbifold $\C^2/\Z_n$ and the interger $n=2s$ is a part of the initial data defining the motion of a classical particle.

In view of all that has been said above, we have considered the coordinates $z_\al$ on the space $\C^2$ of the defining representation of the group SU(2) as fundamental spin variables. Spin vector $S_a=z\sigma_a z^\+$ is a quadratic combination of these variables $z=(z_0, z_1)$ similar to the usual angular momentum, which is a quadratic function of the coordinates $x^a$ and momentum $p_b$. Embedding the sphere $\CPP^1$ into the space ($\C^2$, U(1)) with a symplectic action of the group U(1) allows one to preserve information about the motion of the particle. The group U(1) acting on $\C^2$ is given by the Hamiltonian function $H_{int}^{}$, the constant value of which fixes the level surface $H_{int}^{}=1$ for the momentum map $\mu_{H_{int}}^{}\!:\ \C^2\to\R$ \cite{MW, Wood1, Wood, GS}. The space $\CPP^1$ is introduced as a K\"ahler quotient
\begin{equation}\label{7.4}
\CPP^1=\mu_{H_{int}}^{-1}(1)/\sU(1)=\sSU(2)/\sU(1)\cong(\C^2\setminus\{0\})/\C^*\ ,
\end{equation}
where $S^3\cong\sSU(2)$ is the level surface of $H_{int}^{}$ and $\C^*$ is the group of nonzero complex numbers. 
Note that the level surface $S^3$ is mapped to itself under the action of the group $\Z_n\subset\sU(1)$, so in  \eqref{7.4} we can consider the orbifold $(\C^2{\setminus}\{0\})/\Z_n\subset\C^2{\setminus}\{0\}$ and the lens space $S^3/\Z_n\subset S^3$, where $\Z_1$=Id. 
As a result, we obtain that a particle of spin $s=\sfrac12 n$ moves along a circle in the lens space $S^3/\Z_n$ embedded in the orbifold $\C^2/\Z_n$. 
In other words, it was shown that the quantum spin space $\C^{n+1}$ corresponds to the rotation of a classical particle in the orbifold $\C^2/\Z_n$ or, equivalently, to the $\Z_n$-invariant motion of a particle in $\C^2$.

\medskip

\noindent{\bf Wave function collapse.} When quantizing the spin space $\C^2$ we obtain the wave function
 \begin{equation}\label{7.5}
\psi (z) =\sum_{n=0}^\infty c_n\psi_n(z)\ ,
\end{equation}
where $\psi_n(z)$ corresponds to a particle moving in space $(\C^2, \sU(1), \Z_n)$ with $\Z_n\subset\sU(1)$. Let us apply the Copenhagen interpretation to the above functions $\psi$ and $\psi_n$. We have a massive particle whose spin $s$ is unknown, but it can be measured. During an observation, the particle interacts with a laboratory device and as a result of this the wave function {\it collapses}, $\psi\to\psi_n$, and we learn that the spin is $s=\sfrac12 n$. Thus, the spin of a particle must be described in probabilistic terms, since it is an observable quantity, just like energy. It is interesting that this is never done and the discussion is always about the wave function $\psi_n\in\C^{n+1}$ and not about
the superposition \eqref{7.5} of such functions. To fix $\psi_n$, we can put a factor $n$ before the symplectic structure $\omega_{int}^{}$ or fix it by specifying $\Z_n$-symmetry, but in any case this is some fixation of the initial condition. Recall that we call initial conditions not only the value of the wave function at the initial moment of time but also a specification of the geometry of the space on which it is defined. In the case under consideration, this is an indication of movement in the subspace $\C^2/\Z_n$ of $\C^2$. Therefore, the probabilistic nature of the spin value arises in the case where we do not know the initial data when specifying the motion of a particle in spin space $\C^2$.

Let us return to the function $\psi_n\in\C^{n+1}$ corresponding to the spin $s=\sfrac12 n$, which in turn is decomposed into a direct sum
\begin{equation}\label{7.6}
\psi_n(z) =\sum_{m=0}^n b_m\psi_{nm}(z)\ ,
\end{equation}
where the coefficients $b_m$ define the probability density of detecting the projection of the spin vector onto the 3rd axis equal to the number  $s_3=s-m$. It is curious that if the probabilistic interpretation of the value of spin $s$ is absent, then at the same time the probabilistic interpretation of its projection $s_3$ is generally accepted.
Namely, it is claimed that during measurement, the wave function $\psi_n$ collapses into $\psi_{nm}$, and the observable $s_3$ is fundamentally random. But is this true? Consideration of this paper suggests that the probabilities in this case are also arise due to incomplete knowledge of the initial data. Namely, the wave function $\psi_{nm}$ corresponds to the motion of a classical particle in space $\C/\Z_{n-m}\times\C/\Z_m\subset \C^2$, where $\C/\Z_k{:=}$point if $k$=0. This motion has initial data of the form
\begin{equation}\label{7.7}
(\C, \sU(1)_0, \Z_{n-m};\ \C, \sU(1)_1, \Z_{m})\ ,
\end{equation}
where two groups $\sU(1)_0$ and $\sU(1)_1$ are defined by functions $H_{int}^0=z_0\zb_{\dot 0}^{}$ and $H_{int}^1=z_1\zb_{\dot 1}^{}$.  This initial data can be specified as requirement of ordered $\Z_{n-m}\times  \Z_{m}$ symmetry imposed on the motion of a particle in $\C^2$. If we know these initial conditions \eqref{7.7}, then the quantum particle will be in state  $\psi_{nm}$, and the superposition of states \eqref{7.6} and the probabilities $|b_m|^2$ of detecting states
$\psi_{nm}$ arise from incomplete knowledge of the initial data of the particle's motion in spin space $\C^2$.  The situation looks similar to that which exists when throwing dice, where the probabilities of getting one side or another are not fundamentally non-deterministic.

\medskip

\noindent{\bf  Relativistic Hamiltonian mechanics.} We define relativistic Hamiltonian mechanics of scalar particles as a phase space $(T^*\R^{1,3}, \Omega_0)$ with a symplectic action (i.e. preserving $\Omega_0$) of a one-parameter Lie group $\CG_H$ generated by a Hamiltonian vector field $V_H$ associated with a Lorentz invariant function $H$ on $T^*\R^{1,3}$. The function $H$ defines a momentum map
\begin{equation}\label{7.8}
\mu_H^{}(\cdot , \cdot):\ T^*\R^{1,3}\to\R\with \mu_H^{}(x, p)=H(x,p)\ ,
\end{equation}
and the constant value $m>0$ of this function defines a hypersurface (a level set) in $T^*\R^{1,3}$,
\begin{equation}\label{7.9}
\mu_H^{-1}(m)=\{x,p\in T^*\R^{1,3}\mid     H(x, p)=m\}\ .
\end{equation}
Then the covariant phase space of scalar particles is given by a symplectic quotient of $T^*\R^{1,3}$ by the action of the group $\CG_H$,
\begin{equation}\label{7.10}
T^*\R^{1,3}//\CG_H :=\mu_H^{-1}(m)/\CG_H =Y_6\ .
\end{equation}
The action of elements $g=\exp(\tau V_H^{})$ of Lie group $\CG_H^{}$ on points from $Y_6$ (the space of initial data) generates the motion of a particle in the space $T^*\R^{1,3}$, where the Lorentz invariant parameter $\tau$ parametrizes motion along the orbits of the group $\CG_H^{}$.

We emphasize that the function $H$ is associated with the rest mass of the particle (rest energy), and not with its energy. The proposed formalism works for any Lorentz invariant function $H$. For example, for a particle in an external electromagnetic field, one should take
\begin{equation}\label{7.11}
H_{em}^{}=\sfrac1m\eta^{\mu\nu}(p_\mu + q_{\sf e}A_\mu)(p_\nu + q_{\sf e}A_\nu)\ ,
\end{equation}
where $q_{\sf e}$ is the electric charge of the particle. Note that the equation $H(x,p)=m$ in \eqref{7.9} replaces the energy-momentum relation, which is valid only for free particles.

For particles with spin, the phase space $T^*\R^{1,3}$ is extended to space $T^*\R^{1,3}\times\C_L^2\times\C_R^2$ and the sum  $H=H_0+H^L_{int}+H^R_{int}$ of three commuting Hamiltonian functions defines three momentum maps
\begin{equation}\label{7.12}
\mu_{H_0}^{}: T^*\R^{1,3}\to\R,\quad \mu_{H_{int}^L}^{}: \C_L^2\to\R\und \mu_{H_{int}^R}^{}: \C_R^2\to\R\ ,
\end{equation}
which lead to a symplectic reduction of the extended phase space to space $T^*H^3\times\CPP_L^1\times\CPP_R^1$. To define Bargmann-Wigner particles (including Dirac and Proca particles), an additional Hamiltonian function $H_D=\omega N_D$ is introduced, which carries out a Lorentz covariant reduction of the space $\C_L^2\times\C_R^2$ to $\C_+^2$ and the manifold $\CPP^1_L\times\CPP_R^1$ to $\CPP_+^1$.  All functions $H_0$, $H_{int}^{}$ and $H_D$ are defined so that they remain invariant under transformations of the Lorentz group O(1,3), and in particular do not change sign when $T^*H_+^3$ is mapped to $T^*H_-^3$.   The appearance of negative energies occurs precisely because of the ignoring of this invariance and the incorrect identification of energy $E$ with the component of momentum $p^0$. Energy is always positive at the classical and quantum level.

\medskip

\noindent{\bf Quantization.} We described quantization as a transition from the phase space $X$ to two complex line bundles $L_\C^\pm$ over $X$ with fixed curvature  $F^\pm_{\sf vac}=\pm\im\omega_X$, where $\omega_X$ is a symplectic 2-form on $X$. In the relativistic case, this transition is carried out in the same way as in the non-relativistic case and includes the mapping of the Lorentz invariant function $H$ on the extended phase space into the operator $\hat H$, as well as the assignment of evolution with respect to the parameter $\tau$ by the Schr\"odinger type equation
\begin{equation}\label{7.13}
\Jv\dpar_\tau\Psi =\hat H\Psi\for\Psi=\Psi_++\Psi_-\in L_\C^+\oplus L_\C^-\ ,
\end{equation}
where $\Jv =\im\qv =\pm\im$, when acting on the vectors $\Psi_\pm$. In fact, in the relativistic case, the operator $\hat H$ specifies the parameters of the particle (mass, spin, spin projection, charges) and the evolution of $\Psi$ in $\tau$ can refer to the change in their observed values.

In this paper we considered stationary states. We have shown that the eigenfunctions of the operator $\hat H$ in equation \eqref{7.13} are the Klein-Gordon ($s=0$), Dirac ($s=\sfrac12$), and Proca ($s=1$) fields for the zeroth, first, and second order terms in the expansion of the function $\Psi$ from \eqref{7.13} in the coordinates $Z_i$ of the spin space. For the homogeneous terms of the $n$-th order, equation \eqref{7.13} reduces to the Bargmann-Wigner equations.

\medskip

\noindent{\bf Photons.} We have shown that the Proca equations \eqref{5.28} for massive vector fields $B_\mu$ follow from the Schr\"odinger equation \eqref{5.3} on the extended phase space $T^*\R^{1,3}\times\C^4/\Z_2$ after restricting the spin space $\C^4$ onto the subspace $\C^2$ given by the equation \eqref{4.57}. Maxwell's equations \eqref{6.5} follow from these equations in the limit of zero mass. Penrose defined right- and left-handed photons as complex self-dual $F_{\mu\nu}^R$ and anti-self-dual $F_{\mu\nu}^L$ solutions of Maxwell's equations \cite{Penrose1, Penrose2}. To describe these fields in terms of gauge potential $A_\mu^R$ and $A_\mu^L$, we introduced a Dirac-type equation
\begin{equation}\label{7.14}
\ga^\mu\dpar_\mu\CA=0\ \for\ \CA=\begin{pmatrix}0&A^R\\A^L&0\end{pmatrix}\ ,
\end{equation}
where $A^R=A^R_\mu\sigma^\mu=(A^{R\al\dal})$ and $A^L=A^L_\mu\bar\sigma^\mu=(A^{L}_{\dal\al})$.
The solutions of these equations in the momentum representation are complexified light-like momenta of the form
\begin{equation}\label{7.15}
A^{R}_{\al\dal}=\psi_R(k)k_\al\lambda_\dal\und     A^{L}_{\al\dal}=\psi_L(k)\mu_\al k_\dal\ ,
\end{equation}
where $\psi_R(k)$ and $\psi_L(k)$ are the wave functions of the first quantized photons.

\bigskip

\noindent 
{\bf\large Acknowledgments}

\noindent
I am grateful to Tatiana Ivanova for stimulating discussions and remarks.

\bigskip


\end{document}